\documentclass[11pt, a4paper]{article}
\pdfoutput=1
\usepackage[utf8]{inputenc}
\usepackage[T1]{fontenc}
\usepackage{lmodern, textcomp}
\usepackage[american]{babel}
\usepackage{csquotes}

\usepackage{eurosym}
\usepackage[nottoc]{tocbibind}
\usepackage{authblk}
\usepackage{fancyhdr}
\usepackage{xspace}

\usepackage{graphicx}
\usepackage{subcaption}
\usepackage{float}
\usepackage{makecell}
\usepackage{multirow}
\usepackage{multicol}
\usepackage{marginnote}
\usepackage[multiple]{footmisc}

\usepackage[usenames, dvipsnames, svgnames, table]{xcolor}

\usepackage[backend=bibtex8, citestyle=numeric-comp, maxbibnames=99,
			sorting=none, sortcase=false, sortcites=true, giveninits=true]{biblatex}

\usepackage{amsmath, amsfonts, amssymb}
\usepackage{mathtools}
\usepackage{mathrsfs}
\usepackage{cancel}
\usepackage{braket}
\usepackage{tensor}
\usepackage{slashed}
\usepackage{siunitx}
\usepackage{listings}

\usepackage{stmaryrd}
\usepackage{caption}
\usepackage{relsize,exscale}
\usepackage{array}
\usepackage[toc,page]{appendix}

\usepackage{enumerate}

\DeclareSymbolFont{largesymbols}{OMX}{cmex}{m}{n}

\setcellgapes{1pt}
\makegapedcells
\newcolumntype{R}[1]{>{\raggedleft\arraybackslash }b{#1}}
\newcolumntype{L}[1]{>{\raggedright\arraybackslash }b{#1}}
\newcolumntype{C}[1]{>{\centering\arraybackslash }b{#1}}

\newtheorem{definition}{Definition}

\newtheorem{proposition}{Proposition}
\newtheorem{remark}{Remark}
\newtheorem{claim}{Claim}

\newcommand{\beq}{\begin{equation}}
\newcommand{\eeq}{\end{equation}}
\newcommand{\bea}{\begin{eqnarray}}
\newcommand{\eea}{\end{eqnarray}}
\definecolor{mygray}{gray}{0.3}

\newcommand{\bes}{\begin{eqnarray}}
\newcommand{\ees}{\end{eqnarray}}

\newcommand\restr[2]{{
  \left.\kern-\nulldelimiterspace 
  #1 
  \vphantom{\big|} 
  \right|_{#2} 
  }}

\usepackage{multicol}
\usepackage{amssymb}

\usepackage[heightrounded, top=4cm, bottom=4cm, left=3cm, right=3cm, headheight=14pt]{geometry}

\usepackage[pdftex, breaklinks=true, linktocpage,
	colorlinks=true, urlcolor=blue, linkcolor=blue, citecolor=red
]{hyperref}

\newcommand{\email}[1]{\href{mailto:#1}{\nolinkurl{#1}}}
\newcommand{\emailfoot}[1]{\thanks{\email{#1}}}

\usepackage{marginnote}
\newcounter{draftcommentcnt}
\NewDocumentCommand{\draftcomment}{s O{red} m}{%
	\def\margnote{\IfBooleanTF{#1}{\marginnote}{\marginpar}}%
	\stepcounter{draftcommentcnt}%
	\textcolor{#2}{#3}%
	\margnote{\textcolor{#2}{$\Leftarrow$ \arabic{draftcommentcnt}}}%
}

\newcommand{\preprint}[0]{MIT-CTP/5309}
\fancypagestyle{preprint}{
	\fancyhf{}

	\fancyhead[R]{\textsf{\preprint}}
}

\numberwithin{equation}{section}

\input{arxiv.def}

\hypersetup{
	pdftitle={Non-perturbative renormalization for the neural network-QFT correspondence},
}

\title{Non-perturbative renormalization for the neural network-QFT correspondence}

\author[1,2,3]{Harold Erbin\emailfoot{erbin@mit.edu}}
\author[3]{Vincent Lahoche\emailfoot{vincent.lahoche@cea.fr}}
\author[3,4]{Dine Ousmane Samary\emailfoot{dine.ousmanesamary@cipma.uac.bj}}

\affil[1]{%
	Center for Theoretical Physics, Massachusetts Institute of Technology
	\protect\\
	Cambridge, MA 02139, \textsc{Usa}
}

\affil[2]{%
	\textsc{Nsf Ai} Institute for Artificial Intelligence and Fundamental Interactions
}

\affil[3]{%
	Université Paris Saclay, \textsc{Cea}, \textsc{List}, Gif-sur-Yvette, F-91191, France
}

\affil[4]{%
	Faculté des Sciences et Techniques (ICMPA-UNESCO Chair)
	\protect\\
	Université d'Abomey-Calavi, 072 BP 50, Bénin
}

\begin{document}

\maketitle

\begin{abstract}
In a recent work~\cite{Halverson:2021:NeuralNetworksQuantum}, Halverson, Maiti and Stoner proposed a description of neural networks in terms of a Wilsonian effective field theory.
The infinite-width limit is mapped to a free field theory while finite $N$ corrections are taken into account by interactions (non-Gaussian terms in the action).
In this paper, we study two related aspects of this correspondence.
First, we comment on the concepts of locality and power-counting in this context.
Indeed, these usual space-time notions may not hold for neural networks (since inputs can be arbitrary), however, the renormalization group provides natural notions of locality and scaling.
Moreover, we comment on several subtleties, for example, that data components may not have a permutation symmetry: in that case, we argue that random tensor field theories could provide a natural generalization.
Second, we improve the perturbative Wilsonian renormalization from~\cite{Halverson:2021:NeuralNetworksQuantum} by providing an analysis in terms of the non-perturbative renormalization group using the Wetterich-Morris equation.
An important difference with usual non-perturbative RG analysis is that only the effective (IR) 2-point function is known, which requires setting the problem with care.
Our aim is to provide a useful formalism to investigate neural networks behavior beyond the large-width limit (i.e.~far from Gaussian limit) in a non-perturbative fashion.
A major result of our analysis is that changing the standard deviation of the neural network weight distribution can be interpreted as a renormalization flow in the space of networks.
We focus on translations invariant kernels and provide preliminary numerical results.
\end{abstract}

\thispagestyle{preprint}

\newpage

\hrule
\pdfbookmark[1]{\contentsname}{toc}
\tableofcontents
\bigskip
\hrule

\newpage

\section{Introduction and outline}

Deep learning and neural networks (NNs)~\cite{Goodfellow:2016:DeepLearning, Schmidhuber:2015:DeepLearningNeural} have experienced a rapid development in the last decade, with an ever-increasing number of remarkable applications.
In many cases, these systems outperform humans and ordinary algorithms.
However, there are still many challenges to be solved: in particular, most neural networks work as a black box and require a huge number of examples during the learning phases.
More generally, there is no complete theoretical understanding of why deep learning works so well and how to improve it further.
For example, it is not clear how training can be made more efficient and fast, how knowledge can be transferred to other tasks or how to choose hyperparameters systematically.
This lack of reliability poses, in certain cases, important ethical problems.
Indeed, with the growing use of AI for making decisions (for example, in banking, employment, medicine, military, etc.), it is crucial to be able to explain the choices of the AI in a transparent way~\cite{HighLevelExpertGrouponAI:2019:EthicsGuidelinesTrustworthy}.
Moreover, having a black box is also a drawback for scientific discovery since the goal of science is to interpret and explain, and knowledge can grow only from understanding~\cite{Roscher:2020:ExplainableMachineLearning}.
Our paper is part of the lively field of explainable AI~\cite{Weld:2018:ChallengeCraftingIntelligible, Gunning:2019:XAIExplainableArtificialIntelligence}, where physics do have a role to play~\cite{Zdeborova:2020:UnderstandingDeepLearning}.

A natural path for studying neural networks is provided by theoretical physics: it offers an array of tools useful to describe a wide range of complex systems~\cite{Saitta:2011:PhaseTransitionsMachine}.
In the recent years, evidence has accumulated~\cite{Bradde:2017:PCAMeetsRG, Beny:2018:InferringRelevantFeatures, Beny:2018:CoarsegrainedDistinguishabilityField, Lahoche:2021:FieldTheoreticalApproach-1, Lahoche:2021:FieldTheoreticalApproach-2,Lahoche:2020:SignalDetectionNearly, Lahoche:2020:GeneralizedScaleBehavior,Beny:2013:DeepLearningRenormalization,Mehta:2014:ExactMappingVariational, KochJanusz:2018:MutualInformationNeural, deMelloKoch:2020:DeepLearningRenormalization, Li:2018:NeuralNetworkRenormalization, deMelloKoch:2020:WhyUnsupervisedDeep, deMelloKoch:2020:ShortSightedDeep} in favor of a scenario involving a particular form of "coarse-graining", making contact with a familiar tool for physicists: the Wilsonian renormalization group (RG).

A macroscopic ideal gas is completely described by the ideal gas law, and a macroscopic fluid is well described by the Navier-Stokes equation.
Both equations ignore the microscopic atomic and molecular interactions and provide a coarse-grained description.
This idea was fully developed by Wilson's renormalization group, formalizing the general feature that long range behavior of physical systems does not require an understanding of the nature and interactions of its microscopic building blocks.
Through a very impressive argumentation, Wilson showed that it is possible to explain the apparent universality of physical systems near critical points from the observation that, up to the accuracy of physical predictions, the specific microscopic details can be absorbed through a few effective couplings, defining an effective large scale theory.
Despite the fact that the RG was born in the era of critical phenomena, it turned out to be a very general framework, largely responsible for the success of field theory descriptions of long range distance physics, both in condensed matter physics and in high energy physics~\cite{ZinnJustin:2002:QuantumFieldTheory,ZinnJustin:2007:PhaseTransitionsRenormalization,Weinberg:2005:QuantumTheoryFields-1,Weinberg:2005:QuantumTheoryFields-2}.

The ability of the RG to explain long range universality can be traced from information geometry~\cite{Amari:2016:InformationGeometryIts, Amari:2007:MethodsInformationGeometry, Balasubramanian:2015:RelativeEntropyProximity, Beny:2018:InferringRelevantFeatures, Beny:2018:CoarsegrainedDistinguishabilityField, Beny:2015:RenormalisationGroupStatistical, Beny:2014:RenormalisationInferenceProblem}.
The RG coarse-graining is performed on the eigenvalues of the (free) Fisher information metric, which is a local version of the Kullback-Leibler (KL) divergence $D_{\text{KL}}(p\vert\vert q)$ (or relative entropy) and which provides a reasonable measure of distinguishability between two probability distributions $p$ and $q$ \cite{Beny:2015:RenormalisationGroupStatistical}.
From coarse graining, and in absence of singular structures, KL divergence decreases, as well as distinguishability between distributions, as to become smaller than any experimental precision.
Beyond this limit, we cannot distinguish the two distributions, as different as they may have been originally.
The ability of the RG to extract the relevant features from a large set of interacting microscopic degrees of freedom is a compelling argument for a link with deep learning.
In fact, it is natural to expect a relation with any procedure able to extract relevant features from a massive data set, as it is the case, for instance, in principal component analysis (PCA), where some recent works stressed such a connection between signal detection and RG~\cite{Bradde:2017:PCAMeetsRG, Beny:2018:InferringRelevantFeatures, Beny:2018:CoarsegrainedDistinguishabilityField, Lahoche:2021:FieldTheoreticalApproach-1, Lahoche:2021:FieldTheoreticalApproach-2,Lahoche:2020:SignalDetectionNearly, Lahoche:2020:GeneralizedScaleBehavior}.

In this paper, we aim at developing further the correspondence between quantum field theory (QFT) and neural networks, called the NN-QFT correspondence~\cite{Halverson:2021:NeuralNetworksQuantum, Maiti:2021:SymmetryviaDualityInvariantNeural}.
The main objective is to provide a description of the NN behavior using the non-perturbative RG and the corresponding effective field theory.\footnote{In this paper, we will mostly use both “quantum field theory” (QFT) and “effective field theory” interchangeably, the context makes it clear if we speak of the microscopic (ultraviolet, UV) or effective theory (low-energy). We are working in Euclidean signature, in which case the term “statistical field theory” is sometimes used, to make clear that it describes thermal and not quantum fluctuations. However, we will use QFT for uniformity.}
This positions our paper in a growing tradition of papers describing how the behavior of NNs can be understood through a more and less sophisticated coarse-graining, which can itself be related to a RG.
Strong evidence in favor of a correspondence between RG and deep learning has been stressed for Restricted Boltzmann machines (RBM), whose architecture exhibits similarities with the Ising model (a theoretical model for ferromagnets)~\cite{Beny:2013:DeepLearningRenormalization,Mehta:2014:ExactMappingVariational, KochJanusz:2018:MutualInformationNeural, deMelloKoch:2020:DeepLearningRenormalization, Li:2018:NeuralNetworkRenormalization, deMelloKoch:2020:WhyUnsupervisedDeep, deMelloKoch:2020:ShortSightedDeep}.
Historically, the Ising model was precisely the conceptual cradle of the RG through Kadanoff's "block-spin" method, which can be viewed as an elementary version of the general Wilson coarse-graining~\cite{Kadanoff:1966:ScalingLawsIsing}.
The use of a field theoretical formalism is not a novelty as well~\cite{Halverson:2021:NeuralNetworksQuantum,Schoenholz:2017:CorrespondenceRandomNeural,Dyer:2019:AsymptoticsWideNetworks,Helias:2019:StatisticalFieldTheory,Yaida:2020:NonGaussianProcessesNeural,Bachtis:2021:QuantumFieldtheoreticMachine,Roberts:2021:PrinciplesDeepLearning}.
In fact, this is expected since physics has shown that field theories are a general feature for systems involving emergent collective dynamics.
For example, they appeared to provide a good understanding of the qualitative behavior of NNs through the spin-glass formalism~\cite{Nishimori:2001:StatisticalPhysicsSpin,Sherrington:1993:NeuralNetworksSpin}.

We follow the correspondence between NN and QFT pioneered by Halverson, Maiti and Stoner~\cite{Halverson:2021:NeuralNetworksQuantum}.
Its originality with respect to other approaches lies in the observation that, under very general conditions, NN with infinitely wide layers are described by a Gaussian process (GP) due to central limit theorem~\cite{Neal:1996:BayesianLearningNeural,Lee:2018:DeepNeuralNetworks,Matthews:2018:GaussianProcessBehaviour,Novak:2020:BayesianDeepConvolutional,GarrigaAlonso:2019:DeepConvolutionalNetworks,Yang:2020:ScalingLimitsWide,Yang:2021:TensorProgramsWide,Yang:2020:TensorProgramsII,Yang:2020:TensorProgramsIII}.
Realistic architectures never involve an infinite number of hyperparameters $N$, and their behavior fails to be well described by a GP.
However, this useful but purely theoretical limit allows approaching the non-Gaussian process (NGP) as a perturbation from the large $N$ limit, which is assumed to receive $1/N$ corrections which, for $N$ large enough, can be computed perturbatively.
In~\cite{Halverson:2021:NeuralNetworksQuantum, Maiti:2021:SymmetryviaDualityInvariantNeural}, the correspondence has been developed in the case of a fully connected network with a single hidden layer of width $N$.
They developed the field theoretical machinery necessary to describe neural networks (see Appendix \ref{app:NumSim} for a summary).
This includes computing correlation functions of outputs, obtained in QFT by constructing Green functions from Feynman rules in perturbation theory.
Effective interactions (also called couplings) can then be extracted by comparing the NN correlation functions and the QFT Green functions.
Finally, they introduced a RG flow from a cut-off on the volume of the input data, from the assumption that the effective field theory must be insensitive to the choice of the volume, up to a global rescaling of couplings entering in its definition.
In the QFT language, this corresponds to an infrared (IR, large volume) cut-off: a major difference in our paper is that we will use an ultraviolet (UV, data resolution) cut-off (see Section \ref{sec:data-space}).

The relation between different effective models can be translated locally through a set of $\beta$-functions which describe the evolution of couplings when the cut-off changes, with universal features of the theory emerging from the flow.
In this paper, we are aiming at proposing a non-perturbative formalism, based on the Morris-Wetterich equation~\cite{Wetterich:1991:AverageActionRenormalization,Wetterich:1993:ExactEvolutionEquation,Morris:1998:ElementsContinuousRenormalization,Bagnuls:2001:ExactRenormalizationGroupReview,Delamotte:2012:IntroductionNonperturbativeRenormalization}, to investigate the NN-QFT correspondence beyond the perturbative regime, i.e.\ beyond the large $N$ regime.
This means that our analysis does not requite the coupling constants to be small and that our equations are given in a $1/N$ expansion.
As mentioned above, our framework differs from the one used in~\cite{Halverson:2021:NeuralNetworksQuantum} in that we introduce a true partial integration of the degrees of freedom procedure, without any assumption on the expected large volume behavior of the corresponding effective field theory.
Among the major differences with respect to the situation with ordinary QFT, the full (effective or IR) $2$-point function is known theoretically, including non-Gaussian effects, whereas the free propagator (microscopic or UV) is not known.
This unconventional setting allows going beyond standard limitations of the non-perturbative framework, in particular to close the infinite hierarchical system of equation describing the RG flow and keeping the full momentum dependence of the correlation functions following the Blaizot-Mendez-Wschebor method~\cite{Blaizot:2006:NonPerturbativeRenormalization-1,Blaizot:2006:NonPerturbativeRenormalization-2,Blaizot:2007:NonPerturbativeRenormalizationGroup,Benitez:2012:NonperturbativeRenormalizationGroup}.

Another unconventional aspect concerns the notions of power-counting and locality, which are traditionally inherited from the background space-time (which we will call “data-space” in the case of neural networks).
In the case of QFT for neural networks, such a relation appears as an additional hypothesis that no experience motivates "a priori".
Other properties such as rotation and permutation invariances of the point components may not make sense for neural network data.
However, recent works in the context of background independent quantum gravity~\cite{Rivasseau:2011:RenormalizingGroupField, Rivasseau:2014:TensorTrackIII} have shown that the notions of scales and power-counting are more primitive than that of space-time, and such that locality can be derived from power-counting itself, ensuring moreover that the RG exists and is well-defined.
In this article, we will discuss how these ideas can be relevant for neural networks.

A RG can be then constructed by following the standard method, partially integrating on the degrees of freedom and starting with those associated with the highest scales (UV).
However, the situation for the NN-QFT correspondence is quite different compared to usual studies of the non-perturbative RG: indeed, we are able to solve exactly the $4$-point vertex function while keeping the full momentum-dependence without approximation on the $2$-point function (since it is already known exactly).
We can also solve almost exactly for the other momentum-dependent $n$-point vertex functions (when two momenta are equal and the others vanish, we can also find an exact solution without approximation).
In this paper, we consider two versions of the RG.
In a first approach, called \emph{passive}, the notion of scale is fixed by the resolution chosen to describe the data.
In that approach, the standard deviation of the hidden weights, $\sigma_W$, is viewed as a reference mass scale.
The resulting evolution equation provides an explicit realization of equivalence classes of networks having the same output (up to the machine precision), as the data is coarse-grained.
In a second approach, called \emph{active}, the RG flow is constructed by viewing $\sigma_W$ as a running scale.
In such a way, the equivalence class is between networks having the same output, keeping the data resolution fixed.
This implies that, for fixed $N$, neural networks with different $\sigma_W$ can be viewed as belonging to the same RG trajectory.
In particular, this implies that the renormalization flow can be used to make predictions for any $\sigma_W$ given the results for one of them.
We illustrate this by describing the behavior of the quartic coupling constant of the effective field theory and check numerically the flow equations.
In this paper, we focus on the analytic result and we plan to extend the numerical aspects in future works.


\medskip

\paragraph{Outline.}

In Section~\ref{sec2}, we discuss some general concepts about the field theories which may be used to describe neural networks.
In particular, we comment on the definition of the data-space, IR and UV regimes, (non-)locality and its consequences on scaling and power-counting.
At the end, we describe the passive and active points of views for the renormalization group.
In Section~\ref{sec3} and~\ref{sec4}, we derive the passive and active RG flow equations respectively.
In Appendix \ref{app:NumSim}, we review the numerical simulations from~\cite{Halverson:2021:NeuralNetworksQuantum} and provide some additional details.
Finally, Appendix~\ref{App1} contains the details of technical computations.

\section{NN-QFT, locality, scaling and RG}\label{sec2}

In this section, we present the framework of the NN-QFT correspondence proposed in~\cite{Halverson:2021:NeuralNetworksQuantum, Maiti:2021:SymmetryviaDualityInvariantNeural}. As explained in the introduction, we focus on the Gaussian network\footnote{The term “Gaussian” refers to the fact that the kernel is a Gaussian kernel, not that we have a Gaussian process in the infinite-width limit.} (or Gauss-net) which have a translation invariant kernel.
In this section, we first recall the main ideas of the correspondence (some numerical results from~\cite{Halverson:2021:NeuralNetworksQuantum} are reproduced in Appendix~\ref{app:NumSim}).

Then, we discuss the role played by non-local interactions.\footnotemark{}
\footnotetext{%
	They were not considered in the original version of~\cite{Halverson:2021:NeuralNetworksQuantum} but additional discussion has been added in a subsequent version during the preparation of this manuscript.
}
In particular, we describe the different ways to relax locality and how this naturally leads to break the rotation invariance of the data.
The most general QFT in the latter case are called random tensor field theories (or group field theories), which are generalization of random matrix field theories.

We are also revising the concept of power-counting, preferring a notion intrinsic to the renormalization group compared to the one used in~\cite{Halverson:2021:NeuralNetworksQuantum}, which is inherited from a background “data-space”.
In the latter case, a classical scale dimension is attributed to the data and dimensional analysis is performed by requiring that the action is dimensionless (such that its exponential can serve as a weight in the path integral).
However, it is not clear how to extend this notion in the presence of non-local interactions.
We introduce two notions of scales which emerge from the analysis: the first is attached to the data and called “working precision”, and the second is attached to the network and called the “observation scale”.
We consider two versions of the RG, flowing in these two parameter scales. We conclude the section with a short presentation of the Wetterich-Morris formalism~\cite{Wetterich:1991:AverageActionRenormalization,Wetterich:1993:ExactEvolutionEquation,Morris:1994:ExactRenormalisationGroup} for non-perturbative RG and a discussion about the RG version considered in the reference paper~\cite{Halverson:2021:NeuralNetworksQuantum}. Note that we voluntary use the same notations and conventions to make the comparison with their results easier.

\subsection{Correspondence between neural networks and quantum field theory (NN-QFT)}
\label{sec21}

In~\cite{Halverson:2021:NeuralNetworksQuantum}, the authors proposed a general QFT framework to describe the statistical behavior of neural networks (NN), working in the function-space rather than parameter-space (which can be viewed as a duality~\cite{Maiti:2021:SymmetryviaDualityInvariantNeural}).
The original motivation stems from the observation that neural networks in the infinite-width limit are described by a random Gaussian process~\cite{Neal:1996:BayesianLearningNeural}: the latter can also be described by a free (or Gaussian) quantum field theory.\footnote{We refer to~\cite{Halverson:2021:NeuralNetworksQuantum} for a gentle introduction to QFT with neural networks in mind.}
When the width is finite, the random process is not Gaussian and one can expect the NN to be mapped to an interacting field theory, which has been checked in~\cite{Halverson:2021:NeuralNetworksQuantum}.

\subsubsection{Neural network and experimental Green functions}
\label{sec:nn-exp}

We consider a fully connected neural network $f_{\theta,N}(x): f_{\theta,N}: \mathbb{R}^{d_{\text{in}}} \rightarrow \mathbb{R}^{d_{\text{out}}}$ with learnable parameters (weights and biases) $\theta = (W_0, b_0, W_1, b_1)$, a single hidden layer of width $N$, and an activation function $\sigma$:
\begin{equation}
f_{\theta,N}(x)=W_1(\sigma(W_0x+b_0))+b_1\,,
\end{equation}
where the weights $W_i$ and biases $b_i$ characterize the affine transformation of each layer and $\sigma$ acts element-wise.
The weights $W_0$ and $W_1$ follow centered Gaussian distributions $\mathcal{N}(0, \sigma_W^2 / d_{\text{in}})$ and $\mathcal{N}(0, \sigma_W^2/N)$ respectively, and both biases $b_0$ and $b_1$ are drawn from centered Gaussian distributions $\mathcal{N}(0, \sigma_b)$.
The input data $x$ is a $d_{\text{in}}$-dimensional vector, while we take $d_{\text{out}} = 1$ for the output data for simplicity.
As a consequence, $W_0$ is a $(d_{\text{in}}, N)$-matrix, $W_1$ a $(N, 1)$-matrix, $b_0$ a $N$-vector, and $b_1$ a scalar.
The Gauss-net activation is slightly peculiar because it acts as an exponential of the layer output normalized by the data of the previous layer:
\begin{equation}
\label{eq:activation}
x_1 \equiv \sigma(W_0 x + b_0) = \frac{ e^{W_0 x+b_0}}{\sqrt{\exp \bigg[2(\sigma_b^2+\frac{\sigma_W^2}{d_{\text{in}}} x^2)\bigg]}} \,.
\end{equation}
Finally, we stress that the NN is randomly initialized and that we will not consider the effect of training.

Information on the neural network can be extracted by considering correlations of the outputs: they are encoded by the “experimental” correlation (or Green) functions $G^{(n)}_{\text{exp}}$~\cite{Halverson:2021:NeuralNetworksQuantum}:
\begin{equation}
	G^{(n)}_{\text{exp}}(x_1, \ldots, x_n)
		= \langle f_{\theta,N}(x_1) \cdots f_{\theta,N}(x_n) \rangle,
	\label{eq:exp-green-n-points}
\end{equation}
where the statistical average\footnote{The notation $\langle \cdot \rangle$ should not be confused with the expectation value in QFT: we will always use it to denote the statistical average over a set of networks.} is taken over a large number of neural networks with identical $N$ and parameter distributions.
The numerical evaluation of these quantities is explained in Appendix~\ref{app:NumSim}.

\subsubsection{Large \texorpdfstring{$N$}{N}: free field theory}

Neural networks $f_{\theta,N}: \mathbb{R}^{d_{\text{in}}} \rightarrow \mathbb{R}^{d_{\text{out}}}$ with $N \to \infty$ are well-described statistically by a Gaussian distribution:
\begin{equation}
P[f] = \frac{1}{Z}\, \exp \left(-\frac{1}{2}\, \int_{(\mathbb{R}^{d_{\text{in}}})^2}\, dx dy \, f(x) \Xi (x,y) f(y) \right)\,, \label{freetheory}
\end{equation}
where the factor $Z$ ensures that the expression is normalized when integrating over the full functional space
\begin{equation}
	\int [df]\, P[f]= 1 \,,
\end{equation}
$[df]$ denoting the path integral measure in functional space and $\Xi(x, y)$ the kinetic operator (Gaussian kernel).
In general, we will omit the subscripts $(\theta, N)$ on NN function samples and write simply $f$.

The origin of this Gaussian behavior in the limit $N\to \infty$ can be traced from central limit theorem: since $f_\theta(x)$ is formally a sum of $N$ identically distributed random terms which self-average.
The function $f(x)$ splits into two contributions:
\begin{equation}
\label{eq:decomp-Wb}
f(x)=f_{W}(x)+f_b(x),
\end{equation}
where $f_b(x)\equiv b_1$ being essentially $N$ independent variables and following the Gaussian law $\mathcal{N}(0,\sigma_b)$, whereas $f_W(x)$ goes toward a Gaussian distribution only for large $N$. Formally, it reads as:
\begin{equation}
	f_W(x):= W_1 x_1\,. \label{defF}
\end{equation}
where $x_1$ is given by \eqref{eq:activation} (such that $f_W$ depends on $W_0, W_1$ and $b_0$).
As stated above, for large $N$, one expects that such a quantity self-averages around its mean, and thus that fluctuations are small:
\begin{equation}
\langle f_W\, f_W \rangle
\underset{N\to \infty}{\sim} \langle f_W \rangle \langle f_W \rangle
= 0\,,
\end{equation}
where the last equality follows from the assumptions that initial distributions for $\theta$ are centered and non-correlated.
Hence, the statistical properties of $f_W$ are essentially given by a centered Gaussian distribution, up to $1/N$ corrections. Obviously, the random nature of $f_W$ is inherited from the initial parameter distribution, however the asymptotic Gaussian behavior arises from the law of large numbers.

The $2$-point correlation (or Green) function
\begin{equation}
K(x, y)
\equiv \int [df]\, P[f] f(x) f(y) \,,
\end{equation}
is the inverse of the Gaussian kernel $ \Xi(x,y)$ which appears in the free action \eqref{freetheory}:
\begin{equation}
\int dz \, \Xi(x,z) K(z,y) = \delta(x-y)\,.
\end{equation}
However, according to \eqref{eq:decomp-Wb}, it is also possible to decompose $K$ as:
\begin{equation}
	K(x, y)
		= K_W(x, y) + K_b(x, y),
	\qquad
	K_b(x, y) = \sigma_b^2,
	\label{eq:kernel}
\end{equation}
where $K_W$ is the $2$-point function associated to $f_W$.
It corresponds to the Fisher information metric~\cite{Amari:2007:MethodsInformationGeometry,Amari:2016:InformationGeometryIts} in the information geometry language, and is fixed from the choice of the activation function.

In this paper, we essentially focus on translation invariant kernels $K_W(x,y)\equiv K_W(\vert x-y\vert)$, which is achieved by the \emph{Gauss-net} architecture \eqref{eq:activation}, corresponding to the kernel:
\begin{equation}
K_W(\vert x-y \vert) =\sigma_W^2 \,e^{- \frac{\sigma_W^2}{2d_{\text{in}}} \vert x-y\vert^2}\,, \label{exact2points}
\end{equation}
where $\vert x-y\vert:=\sqrt{\sum_i(x-y)_i^2}$ denotes the ordinary Euclidean distance between $x$ and $y$.

In field theory language, the kernel enters in the definition of the classical kinetic action\footnote{Note that in this paper we choose the subscript $\text{kin}$ for "kinetic", more familiar to physicist rather than $G$ for "Gaussian", used in the reference~\cite{Halverson:2021:NeuralNetworksQuantum}.} (i.e.~the log-likelihood in probability theory):
\begin{equation}
S_{\text{kin}}[f]:=\frac{1}{2}\, \int_{(\mathbb{R}^{d_{\text{in}}})^2}\, dx dy \, f(x) \Xi (x,y) f(y)\,,
\label{eq:free-action}
\end{equation}
the corresponding probability distribution being given by the exponential law $P[f]\propto e^{-S_{\text{kin}}[f]}$ in \eqref{freetheory}.
The $n$-point correlation (or Green) functions are defined as:
\begin{equation}
	G_0^{(n)}(x_1, \ldots, x_n)
		\equiv Z^{-1} \int [df]\, e^{-S_{\text{kin}}[f]} f(x_1) \cdots f(x_n).
	\label{eq:green-n-points}
\end{equation}
In the free theory, $G_0^{(n)}$ is completely determined in terms of $G_2(x, y) = K(x, y)$ through Wick's theorem, and vanishes for $n$ odd~\cite{Halverson:2021:NeuralNetworksQuantum}.
Hence, this implies that:
\begin{equation}
	G^{(n)}_{\text{exp}}(x_1, \ldots, x_n)
		\underset{N\to \infty}{\sim} G_0^{(n)}(x_1, \ldots, x_n).
\end{equation}

\subsubsection{Data-space and momentum space}
\label{sec:data-space}

In this subsection, we discuss some definitions related to the data-space\footnote{As we will see, its properties may be sufficiently different from the usual space of positions -- spacetime -- appearing in usual QFT to find another name.} corresponding to the neural network input $x$, and how they differ from~\cite{Halverson:2021:NeuralNetworksQuantum}.

Continuity and infinity in computer science exist only as idealizations.
First, any real number $x \in \mathbb R$ is represented numerically by a decimal number, bounded in precision by the number of bits used to encode it. For instance, if the maximal number of decimals is $n_0$, two numbers $x$ and $x+10^{-m}$ cannot be absolutely distinguished for $m>n_0$.
To be more realistic, we should see the data-space $\mathbb {R}^{d_{\text{in}}}$ as a lattice of step $a_0 = 10^{- n_0}$ rather than a continuum manifold.
The lattice spacing $a_0$ provides what physicists call an ultraviolet (UV) cut-off.
Varying this parameter amounts to changing the data resolution, in full similarity with spacetime resolution in usual QFT.
Second, computers cannot store an infinite amount of information.
For this reason, it cannot handle infinite numbers (except as special data types and formal rules) and it is necessary to restrict the data to a finite interval $x \in [- L/2, L/2]$.\footnote{In most of~\cite{Halverson:2021:NeuralNetworksQuantum}, the large-volume (what we call IR) cut-off in data-space $L$ is denoted as $2 \Lambda$. However, we keep this notation for the large-volume cut-off in momentum space.}
Hence, we consider the data-space to be a $(2 N_0)^{d_{\text{in}}}$ square lattice with spacing $a_0$, $N_0 \in \mathbb{N}$, and total hypervolume
\begin{equation}
	\mathcal{V}
		= L^{d_{\text{in}}},
	\qquad
	L := 2a_0N_0.
\end{equation}

It is generally more convenient to work in Fourier (or momentum) space.
The allowed momenta $p=(p_1,\ldots, p_{d_{\text{in}}})$ lie in the \emph{first Brillouin region}:
\begin{equation}
p_i=\frac{2\pi n_i}{L}\,, \qquad n_i \in \mathbb{Z}_{N_0}\,.
\end{equation}
Note that we assume periodic boundary conditions. This is not a problem for $L$ large enough; for small $L$,\footnote{We will precise “small with respect to what” in a moment.} we can simply repeat the data set a large number of times to obtain a large enough effective volume to make the boundary conditions irrelevant.

In the rest of this paper, we use the following definitions.
\begin{definition}
We call $(2N_0)^{d_{\text{in}}}$ the discrete volume and $a_0$ the working precision.
\end{definition}
Note that $(2N_0)^{d_{\text{in}}}$ also counts the number of states in the first Brillouin region. In this discrete setting, we can write the Fourier series of the network $f(x)$ (for $x\in \left(a\mathbb{Z}_{N_0}\right)^{d_{\text{in}}}$) as:
\begin{equation}
f(x)=:\frac{1}{N_0^{d_{\text{in}}}}\sum_p g(p) e^{ip x}\,, \label{Fourierseries}
\end{equation}
the basis functions $e^{ipx}$ being normalized such that $\sum_{x} e^{i(p_1-p_2) x} =N_0\delta_{p_1p_2}$. Note that \eqref{Fourierseries} holds for any discrete function on the lattice. In the continuum limit, for small $a_0$ and $N_0$ large such that $L$ remains fixed, discrete sums can be replaced by integrals. Moreover, for volume large enough, integrals becomes standard Fourier transform. We call this limit the \emph{thermodynamic limit} following the standard terminology in physics, and we focus on this regime in our investigations. Taking the Fourier transform of the $2$-point function:
\begin{equation}
\tilde{K}(p)= \frac{1}{(2\pi)^d} \int dx \, K(x) e^{-ipx}\,,
\end{equation}
where $px:= \sum_{i=1}^{d_{\text{in}}}\, p_i x_i$, we get for \eqref{exact2points}:\footnote{This type of kinetic term is reminiscent of $p$-adic string theory~\cite{Brekke:1993:PadicNumbersPhysics, Moeller:2002:DynamicsInfinitelyMany}.}
\begin{equation}
\tilde{K}(p)= (\sigma_W^2)^{1-d_{\text{in}}/2} \left(\frac{d_{\text{in}}}{2\pi} \right)^{d_{\text{in}}/2} e^{-\frac{d_{\text{in}}}{2\sigma_W^2}p^2}\,.
\label{eq:kernel-p}
\end{equation}
Note that in this continuum approximation, the Dirac delta $\delta(p)$ has to be understood as a shorthand notation for a Kronecker delta $(2\pi)^{-d_{\text{in}}} \mathcal{V} \delta_{p0}$.
Translation invariance is crucial to obtain a kernel \eqref{eq:kernel-p} which depends on a single momentum, and reflection invariance implies that it must be a function of $p^2$.
It would be interesting to understand how to generalize our computations for kernels which are not translation invariant~\cite{Halverson:2021:NeuralNetworksQuantum}.

For small $p^2$, we may expand $\tilde{K}(p)$ in power of $p^2$,
\begin{equation}
\tilde{K}(p)=(\sigma_W^2)^{1-d_{\text{in}}/2} \left(\frac{d_{\text{in}}}{2\pi} \right)^{d_{\text{in}}/2}-\frac{d_{\text{in}}}{2}(\sigma_W^2)^{-d_{\text{in}}/2} \left(\frac{d_{\text{in}}}{2\pi} \right)^{d_{\text{in}}/2} \,p^2+\mathcal{O}(p^4)\,.\label{expansionPropa}
\end{equation}
Up to $\mathcal{O}(p^4)$ corrections, the propagator looks like the canonical propagator of a free scalar field theory:
\begin{equation}
\tilde{K}(p)=\frac{1}{m^2_0+Z_0p^2+\mathcal{O}(p^4)}\,, \label{Klong}
\end{equation}
where:
\begin{align}
m^2_0=(\sigma_W^2)^{d_{\text{in}}/2-1} \left(\frac{d_{\text{in}}}{2\pi} \right)^{-d_{\text{in}}/2}\,, \quad Z_0=\frac{d_{\text{in}}}{2}(\sigma_W^2)^{d_{\text{in}}/2-2} \left(\frac{d_{\text{in}}}{2\pi} \right)^{-d_{\text{in}}/2}\,.
\end{align}
In the QFT terminology, $Z_0$ and $m^2_0$ are respectively the \emph{wave function renormalization} and \emph{bare mass}. One can rescale the field to to set $Z_0=1$, in which case the mass becomes:
\begin{equation}
\bar{m}_0^2
= Z_0^{-1} m_0^2
=\frac{2\sigma_W^2}{d_{\text{in}}}\,.
\end{equation}
We adopt the following definition:
\begin{definition}
The mass $\bar{m}_0^2$ defines the typical mass scale and its inverse defines the (IR) correlation length $\xi$, or the \emph{typical observation scale}
\begin{equation}
	\xi^2 := d_{\text{in}}/(2\sigma_W^2).
	\label{eq:observation-scale}
\end{equation}
\end{definition}
Note that the large volume limit is defined with respect to this correlation length, i.e.~$L\gg \xi$. Beside the existence of an intrinsic length scale, the system the propagator at large distance behaves like $(\bar{m}_0^2+p^2)^{-1}$.

Assigning the label $x$ (position space) to the original data and $p$ (momentum space) to the Fourier conjugate may seem arbitrary.
Indeed, while the machine precision provides a natural UV cut-off and an associated identification of the data-space as position space (since UV corresponds to small distances in that space), signals in Fourier space are also represented in the computer up to the machine precision.
However, in that case, using the machine precision as a UV cut-off would not match the usual intuition in QFT.
Given a translation-invariant kernel, another possibility is to identify the momentum space as the space where the propagator is diagonal, such that the propagator in position space depends on the distance $\vert x-x^\prime \vert$.

\subsubsection{Finite-\texorpdfstring{$N$}{N} corrections and interactions}
\label{sec:finite-N}

For a Gaussian process, as we have seen earlier, correlation functions $G^{(2n)}$ for $n>1$ can be decomposed as a sum of product of $2$-point functions thanks to the Wick theorem~\cite{ZinnJustin:2002:QuantumFieldTheory}. For $N$ large but finite, the distribution is not exactly Gaussian, and the correlation functions do not match with Gaussian predictions.

The deviation of the QFT and experimental correlation functions from the Gaussian case are denoted as:
\begin{equation}
	\Delta G^{(n)}_{\text{exp}}
		:= G^{(n)}_{\text{exp}} - G_0^{(n)},
	\qquad
	\Delta G^{(n)}
		:= G^{(n)} - G_0^{(n)}.
\end{equation}
Note that $G_0^{(n)}$ are still the large $N$ Green functions defined in \eqref{eq:green-n-points}.
Importantly, we identify the exact $2$-point function $G^{(2)}$ with the kernel $K$ which equals $G_0^{(2)}$.
Since it contains (quantum) corrections due to the interactions, the free $2$-point Green function computed from the kinetic term only is not known (in standard QFT, the converse is true, see Section \ref{secRenNN} for a discussion).
We will see in Section~\ref{secN} that connected functions $\Delta G^{(2n)}_c$ behaves as:
\begin{equation}
	\Delta G^{(2n)}_c = \mathcal{O}(1/N^{n-1}) \,,\qquad \forall\, n\,.
	\label{eq:decay-N-Gn}
\end{equation}
which has also been investigated analytically and numerically in~\cite{Halverson:2021:NeuralNetworksQuantum} (see also Appendix~\ref{app:NumSim}).\footnote{The original version~\cite{Halverson:2021:NeuralNetworksQuantum} did not contain this discussion which has been added while the present manuscript was in preparation.}
This scaling is consistent with the fact that the exact $2$-point function is independent of~$N$.

Qualitatively, this is reminiscent of what happens for the Ising model in large dimension. The local magnetization self-averages because the number of closest neighbors is large, and the statistical properties remain (quasi)-Gaussian. For space dimension $d$ large but finite, thermodynamical quantities can be computed as power series in $1/d$, which do not affect universal quantities, as soon as $d>4$. For $d<4$ however, the decoupling of physical scales breaks down and the Gaussian approximation is not suitable~\cite{ZinnJustin:2007:PhaseTransitionsRenormalization}.

The same scenario is expected to be true for neural networks. For finite $N$, the distribution does not obey Wick theorem, and correlations functions receive contributions which do not reduce to products of $2$-point functions, and the classical action must include non-Gaussian contributions, i.e.~products of $f$ of degrees higher than $2$. However, as soon as $N$ remains large enough, deviations from the Gaussian behavior are expected to remain small. In the classical action, these corrections materialize as product of $m$ fields ($m>2$), that we call \emph{interactions}:
\begin{equation}
	S[f]
		= S'_{\text{kin}}[f] + S_{\text{int}}[f],
\end{equation}
where $S'_{\text{kin}}[f]$ is a new free action of the form \eqref{eq:free-action}.
We follow the orthodox assumption in field theory that $S$ is polynomial, and we denote generally as \emph{couplings} the monomials.
The correlation functions are computed using \eqref{eq:green-n-points} by replacing $S_{\text{kin}}[f]$ with $S[f]$.
But, since the interacting action $S_{\text{int}}[f]$ is built from cubic and higher powers of $f$, this generally prevents from computing the path integral exactly, and one has to resort to a perturbative expansion encoded in terms of Feynman graphs~\cite{Halverson:2021:NeuralNetworksQuantum}.
The form of the interactions is discussed in the next subsection.
For the rest of this paper, we discard the contribution $f_b$ in \eqref{eq:decomp-Wb} from our analysis, and omit the subscript $W$.

\subsection{Locality, scaling(s), and power-counting}
\label{sec:locality-scaling}

In this subsection, we precise the class of interactions which are assumed to suitably reproduce the non-Gaussian properties of correlations. Moreover, we discuss the scaling behaviors, especially relevant for the renormalization group investigations in the next section.

\subsubsection{The theory space}

The set of allowed couplings defines the \emph{theory space}. They are generally guided by physical arguments, the symmetries of the system, and fundamental assumptions about the physical laws. This is especially the case for fundamental physics, where the expected properties of the space-time background play a key role. In turn, the structure of space-time is itself a consequence of the interactions between physical matter.\footnote{This point of view is named "relational" in physics, and is essentially the one used in quantum theory of gravity~\cite{Rovelli:2007:QuantumGravity}} Indeed, if we are able to say that something is “here”, this is because we can interact with this thing. In words, a statement such that “the field must interact locally” is physically equivalent to “locality is defined by the interactions of fields”. In other words, space-time in physics is more that a set of $d_{\text{in}}$ coordinates $x\in \mathbb{R}^{d_{\text{in}}}$. It is equipped with a group structure, the Poincaré group, which dictates how the coordinates can be transformed from one to the other.
As the history of the relativity theory shows~\cite{Deruelle:2018:RelativityModernPhysics, Wald:1984:GeneralRelativity, Rovelli:2007:QuantumGravity}, these properties are essentially consequences of interactions between light and matter.

In the QFT framework, the role of the background space is played by $\mathbb{R}^{d_{\text{in}}}$.
In~\cite{Halverson:2021:NeuralNetworksQuantum}, the authors adopt a conservative approach for most of their analysis, building the couplings as products of fields at the same point $x\in \mathbb{R}^{d_{\text{in}}}$:
\begin{equation}
S_{\text{int}}= \sum_n g_n\, \int_{\mathbb{R}^{d_{\text{in}}}} dx \, \big( f(x) \big)^n\,. \label{localint}
\end{equation}
However, this makes various assumptions which may not be valid for a general neural network QFT.
For this reason, we will make them explicit and explain how to gradually lift them to consider the most general QFT.
Deciding which assumptions to use should be dictated by numerical evidences: in particular, it was found in~\cite{Halverson:2021:NeuralNetworksQuantum} that \eqref{localint} is sufficient for the activation functions and range of input parameters considered there (see Appendix~\ref{app:NumSim} for more details).
This approach can be considered as neural network phenomenology, in the sense that we are writing a model to match observations, but we can also use this model to check theoretical facts such as dualities~\cite{Betzler:2020:ConnectingDualitiesMachine,Krippendorf:2020:DetectingSymmetriesNeural,Maiti:2021:SymmetryviaDualityInvariantNeural}.

The first assumption is locality of the interaction: the fields appearing in the monomial $f(x)^n$ can be taken at different points (for simplicity, we consider a single coupling in $S_{\text{int}}$):
\begin{equation}
	\label{eq:coupling-nonlocal-cst}
	S_{\text{int}}
		= g \int d x_1 \cdots d x_n \, f(x_1) \cdots f(x_n).
\end{equation}
This breaks locality because fields at different points in space(time) can interact together.
In fact, since $g$ is a constant, this happens for arbitrarily large distances.
Note that this preserves translation invariance $x_i \to x_i + a$.

The next natural step is to replace $g$ by a \emph{coupling function}, i.e.~a function of space but independent of the field.
Going back to \eqref{localint} where all fields are at the same points, we can write a local action with a coupling function:
\begin{equation}
	S_{\text{int}}
		= \int dx \, g(x) \, \big( f(x) \big)^n\,.
\end{equation}
It was argued in~\cite{Halverson:2021:NeuralNetworksQuantum} from technical naturalness that $g(x)$ must be approximately constant since a coupling function $g(x)$ breaks the translation invariance of the action.
However, this is correct only when assuming locality of the action: replacing $g$ by a coupling function in \eqref{eq:coupling-nonlocal-cst} gives the non-local action~\cite{Weinberg:2005:QuantumTheoryFields-2}:
\begin{equation}
	S_{\text{int}}
		= \int d x_1 \cdots d x_n \, g(x_1, \ldots, x_n) f(x_1) \cdots f(x_n)\,.
\end{equation}
However, translation invariance can be preserved if $g$ depends only on the distances between the points:
\begin{equation}
	g(x_1, \ldots, x_n)
		= g(x_1 - x_2, \ldots, x_1 - x_n, \ldots x_{n-1} - x_n)\,.
\end{equation}
Moreover, having a coupling functions gives more control on the interaction region, for example, by restricting the non-locality to a small region.
For instance, we can set $g(x_1, \ldots x_n) = 0$ if $|x_i - x_j| > \ell$ for any pair $(i, j)$.
This allows representing non-locality by derivatives in momentum space and to show that they are subleading in the deep IR.
Simple non-local models of this form have been considered in~\cite{Halverson:2021:NeuralNetworksQuantum}.

A special type of such a non-local interaction is obtained by smearing the fields (only in the interactions): in \eqref{localint}, we can replace the field $f(x)$ by another $\tilde f(x)$ given by a convolution with a kernel $\kappa(x, y)$~\cite{Efimov:1968:ClassRelativisticInvariant,Alebastrov:1974:CausalityQuantumField,Tomboulis:2015:NonlocalQuasilocalField}:
\begin{equation}
	\tilde f(x)
		:= \int d y \, \kappa(x, y) f(y)
\end{equation}
such that
\begin{equation}
	S_{\text{int}}
		= g \int dx \, \big( \tilde f(x) \big)^n
		= g \int dx dy_1 \cdots dy_n \,
			\kappa(x, y_1) \cdots \kappa(x, y_n) \,
			f(y_1) \cdots f(y_n)\,.
\end{equation}
In order for this to make sense, the Fourier transform of the kernel $\kappa(x, y)$ must be an entire analytic function (with rapid decay if one wants to ensure UV finiteness).
This corresponds to a coupling function
\begin{equation}
	g(x_1, \ldots, x_n)
		= g \int dx \, \prod_{i=1}^n \kappa(x, x_1)\,.
\end{equation}
Smeared fields naturally appear in string theory and are responsible for its well-behaved UV behavior~\cite{deLacroix:2017:ClosedSuperstringField, Erbin:2021:StringFieldTheory}.
In fact, for the Gauss-net, rescaling the field $f$ to remove the exponential from the kinetic term \eqref{eq:kernel-p} is equivalent to smearing the field (as pointed out earlier by comparing with the $p$-adic string~\cite{Brekke:1993:PadicNumbersPhysics, Moeller:2002:DynamicsInfinitelyMany}).

There is a final assumption in all the previous interactions we wrote: that all components of $x$ (which is a $d_{\text{in}}$-dimensional vector) are homogeneous.
First, this means that coordinates can be added/subtracted to each other.
Second, this also implies that the role played by the $i$-th coordinate can be played by the $j$-th, or by any linear combination of the coordinates.
Physically, this means that the previous interactions have a $O(d_{\text{in}})$ symmetry (the Euclidean rotation group, or the Lorentz group in Lorentzian signature).
Together with translations (if present\footnote{This is a property of the kernel, but usual activation functions do not seem to provide it~\cite{Halverson:2021:NeuralNetworksQuantum}.}) $x_i \to x_i + a$, this builds the $d_{\text{in}}$-dimensional Euclidean group $\mathrm{Is}(d_{\text{in}})= \mathbb{R}^{d_{\text{in}}} \rtimes O(d_{\text{in}})$ (or Poincaré group in Lorentzian signature), which leaves the Euclidean distance invariant.
This is why we often write $f(x)$ instead of $f(x_1, \ldots, x_{d_{\text{in}}})$ where $x = (x_1, \ldots, x_{d_{\text{in}}})$.

However, it is not clear a priori that the data-space possesses this symmetry: it may not be possible to exchange two data components or even consider linear combinations if the components are not homogeneous.
Despite the fact that the free theory supports such a symmetry, the rotational invariance has no meaning for a neural network in general. Moreover, symmetries of the free theory can be broken by interactions, which are necessary to fully characterize the system.
Conditions under which input and output symmetries can be present has been analyzed in~\cite{Maiti:2021:SymmetryviaDualityInvariantNeural}.
In general, one can start by assuming no symmetry in order to describe the most general model, and then adapt to what the numerical experiments are indicating.

Hence, we need to consider fields for which each component is independent: this amounts to interpreting $f(x)$ as a field over $d_{\text{in}}$ independent copies of $\mathbb{R}$, meaning that each of the $d_{\text{in}}$ components is independent and cannot be transformed into the others.
Given that there are several fields, this means that the $i$-th component of a given point can be inserted only in the $i$-th argument of a field, however, it is not necessary to use all components of a single point in a single field.
Obviously, this expression is non-local because the field is evaluated for components corresponding to different points.
For example, for $d_{\text{in}} = 3$, one can write the following cubic interaction:
\begin{equation}
	S_{\text{int}}
		= g \int d x d y d z \, f(x_1, y_2, z_3) f(y_1, z_2, x_3) f(z_1, x_2, y_3)\,,
\end{equation}
where $x_i$, $y_i$ and $z_i$ are the components of the $3$-dimensional points $x$, $y$ and $z$.
Note that nothing prevents to use only two points, for example setting $y = z$ and integrating only over $x$ and $y$, or more generally to repeat the same component in any number of fields (for an early example, see~\cite{Godfrey:1991:SimplicialQuantumGravity}).

Such general theories are too wild and it is hard to make sense of them.
A controllable subclass is provided by random tensor field theories~\cite{Gurau:2016:RandomTensors}.
In this case, the fields are tensors, each component of the positions being seen as a (continuous) index and indices can be contracted pairwise only (which is achieved by integrating over the component, since the index is continuous), such that a given component can appear at most twice.
An intuitive way to represent it is to assign a \emph{color} to each component, and Feynman diagrams can be written in terms of strand graphs (generalizing ribbon graphs from matrix models).
For instance, a possible quartic interaction for $d_{\text{in}} = 3$ is:
\begin{equation}
	S_{\text{int}}
		= g \int d^3x d^3y f(x_1,x_2,x_3) f(x_1,y_2,y_3) f(y_1,y_2,y_3) f(y_1,x_2,x_3)\,.
	\label{examplemelon}
\end{equation}
We will see in the next section that tensor field theories are particularly interesting in the RG approach because, under some additional conditions, they possess a natural background-independent power-counting (see next subsection).

We conclude this section by clarifying a subtlety concerning QFT in curved spaces.
In this case, the Euclidean (or Poincaré) group is not a global symmetry (symmetries of the action are given by the isometry group of the background space) and one may ask what is the difference with tensor field theories.
The point is that this group is still a local symmetry (general relativity can be seen as gauging the Poincaré group) such that the properties discussed above continue to hold.
Indeed, one can always consider the tangent space associated to a point: since it is isomorphic to flat space, it means that the coordinates are still homogeneous.

\subsubsection{\texorpdfstring{$\Lambda$}{Lambda}-scaling and power-counting}\label{LambdaScal}

We are aiming to construct a field theory which admits a well-defined RG flow. In standard QFT, the rigorous construction of such a flow requires essentially three basics ingredients: 1) a scale decomposition, 2) a locality principle, 3) a power-counting.

The scale decomposition is the first ingredient to construct slices, and then to define a partial integration procedure. Power-counting and locality, in turn, are essential to understand the notions of effective couplings, i.e.~how Feynman graphs can be replaced by an effective vertex together with a slice-dependent coupling. As long as we are endowed with $\mathbb {R}^{d_ {in}}$ as a background space, all of these notions are obvious. Scale decomposition is intuitively related with the notion of metric distance, locality and non-locality are defined with respect to the background itself, and power-counting is related to dimensionality as well. Indeed, the existence of an extrinsic length scale and the requirement that the classical action $S$ is dimensionless, to give meaning to the exponential $e^{-S}$, allow fixing the dimensions (in terms of the length scale unit) of the couplings appearing in the classical action. This is the choice made in~\cite{Halverson:2021:NeuralNetworksQuantum} done. Assuming $[dx]_x=1$, where $[Q]_x$ denotes the dimension of the quantity $Q$ in units of $x$, they were able to fix the dimension of couplings like \eqref{localint},
\begin{equation}
[\delta S_{\text{in}}^{(n)}]_x=0 \,\quad \Leftrightarrow \quad [g]_x=-d_{\text{in}}-\frac{n [K]_x}{2}\,. \label{dimensioncan}
\end{equation}
We call it \emph{$\Lambda$-scaling} such a scaling, for some reference scale $\Lambda$ having ($x$)-dimension $1$. However, from the discussion above, we can be a little puzzled by the fact of assigning a physical dimension to the variable $x$, and to truly view $\mathbb {R}^{d_{in}}$ as a background space. Rather, we adopt the minimal point of view considering it only as a configuration space, without dimension. Sacrificing the background space then makes the issues of scale, locality and power-counting less intuitive. The discussion in Section~\ref{sec:data-space} shows that the theory has a canonical notion of scale, given by the Fourier modes (spectrum) of the propagator. Regarding the notions of locality and power-counting, the difficulty is quite similar to that encountered in canonical approaches to quantum gravity, where space-time and background metric disappear~\cite{Rivasseau:2011:RenormalizingGroupField}.
In this context, a clever solution was found, which in some sense defines the power-counting from a locality principle, starting from the observation that standard locality in field theory can be algebraically translated as the ability of connected Feynman diagrams to be contracted to a point. Locality can then be defined algebraically from the requirement that, at least for some leading order sector, such a contraction procedure exists. A recent example, arising from quantum gravity models is provided by tensorial field theories~\cite{Geloun:2013:Renormalizable4DimensionalTensor,Rivasseau:2014:TensorTrackIII,Lahoche:2018:NonperturbativeRenormalizationGroup}. In these theories, interactions are non-local in the usual sense (from the point of view of the configuration space) but, for some of them, the only divergences come from a sub-family of Feynman diagrams (in general the so-called \emph{melonic} diagrams), which is contractible to an elementary vertex compatible with some internal symmetry defining the tensorial interactions themselves. Interactions having these properties are then said to be local. In turn, graphs admitting such a contraction property have been shown to admit a well-defined power-counting. The reason for this is that, to be well-defined, a power-counting requires that the existence of a family of Feynman graphs having the same behavior with respect to some cut-off $\Lambda$. If, order by order in the perturbative series, quantum corrections have different scaling behavior with respect to $\Lambda$, no power-counting exists. The existence of a contraction procedure allows defining the relative scaling of the various terms entering in the classical action with respect to $\Lambda$, such that there exists non-vanishing leading sectors of the perturbative expansion which have the same behavior with respect to $\Lambda$.

Let us illustrate heuristically on a simple example how contractibility and power-counting allows fixing the scaling dimension of couplings. Consider the following classical action:
\begin{equation}
S[\phi]= \int dx \left[ \frac{1}{2} \phi(x) (-\Delta+m^2)\phi(x)+ g \phi^4(x)\right]\,, \label{freefield}
\end{equation}
describing the scalar field $\phi: \mathbb{R}^d\to \mathbb{R}$, and where $\Delta$ denotes the standard Laplacian. It is moreover local in the usual sense. For $g$ small enough, quantum corrections can be computed using standard perturbation theory. The first contribution to the effective mass $\delta^{(1)} m^2$ arise from the following integral in Fourier space (the symmetry factors are irrelevant for our discussion):
\begin{equation}
\delta^{(1)} m^2 \propto g\int \frac{dp}{p^2+m^2} \propto g \Lambda^{d-2}\,,
\end{equation}
for some cut-off $\Lambda$ for large momenta. In the same way, the first correction for $g$, say $\delta^{(2)} g$ involves the following integral:
\begin{equation}
\delta^{(2)} g \propto g^2 \int \frac{dp}{(p^2+m^2)^2}\propto g^2 \Lambda^{d-4}\,, \label{scalg}
\end{equation}
the upper index referring to the number of vertices involved in the Feynman diagram. Now, to obtain a well-defined power-counting, the correction for $g$ has to scale with $\Lambda$ in the same way as $g$ itself. This is solved by $g\sim \Lambda^{4-d}$, and we say that the $\Lambda$-scaling of $g$ is $[g]_{\Lambda}=4-d$. This moreover implies $\delta m^2 \sim \Lambda^2$, and thus $[m^2]_{\Lambda} =2$. Now, we have to check that it is coherent to all orders of the perturbative expansion. To this end, let us consider a Feynman graph $\mathcal{G}_V$, of order $V$ contributing to the perturbative expansion through the amplitude $\mathcal{A}_{\mathcal{G}_V} \sim g^V\Lambda^{\omega(\mathcal{G}_V)}$. Contracting along a spanning tree $\mathcal{T}_V\subset \mathcal{G}_V$, we reduce the original number of propagator edges $L$ to $L-V+1$, and the resulting graph looks like an effective (local) vertex, having $L-V+1$ loops of length one (tadpoles). Each tadpole behaves like $\int dp/(p^2+m^2)$, and thus scales as $\Lambda^{d-2}$. The divergent degree for the contracted graph $\mathcal{G}_V\backslash\mathcal{T}_V$ is therefore:
\begin{equation}
\omega(\mathcal{G}_V\backslash\mathcal{T}_V)=(d-2)(L-V+1)\,.
\end{equation}
Because the contraction procedure removes $V-1$ propagator edges, it increases $\omega(\mathcal{G}_V)$ by $2(V-1)$: $\omega(\mathcal{G}_V\backslash\mathcal{T}_V)=\omega(\mathcal{G}_V)+2(V-1)$. Finally, because the interaction is quartic, we have the relation $2L=4V-N$, $N$ being the number of external edges. Finally, we get:
\begin{equation}
\omega(\mathcal{G}_V)=(d-4)V+2+(d-2)\left(1-\frac{N}{2} \right)\,.
\end{equation}
Each vertex contributes a factor $\Lambda^{d-4}$, and the scaling $g\sim \Lambda^{4-d}$ ensures that all the quantum corrections have the same scaling. Moreover, setting $N=2$, we get $\omega=2$, in agreement with the one-loop scaling dimension for mass.

Obviously, because this theory is local in the usual sense, the derived scaling dimensions are exactly the same as the one derived from the standard dimensional analysis of the classical action. The two methods however do not coincide for non-local interactions such that \eqref{examplemelon}.
We argue that this more abstract way to think about locality, scaling and power-counting is more appropriate in a context where the construction of the theory space is not guided by experimental evidences, such that it seems more appropriate to work from the outset within a sufficiently broad framework to accommodate future developments in formalism. However, the exploration of these aspects for NNs is beyond the scope of this paper, since standard locality seems to hold for the Gauss-net kernel~\cite{Halverson:2021:NeuralNetworksQuantum}.


\subsubsection{\texorpdfstring{$N$}{N}-scaling}\label{secN}

There exists another scaling dimension, called $N$-\emph{scaling}, associated to the behavior of correlation functions with respect to the width $N$ of the hidden layer. The Gaussian universality for large $N$ ensures that the couplings $g_n$ behave as $g_n \sim N^{-\alpha(n)}$ for some positive function $\alpha(n)$.

The computation can be done by returning to the definition \eqref{defF} and using the fact that $W_1$ follows a centered Gaussian distribution with variance $\sigma_W^2/N$. For instance, we find:
\begin{equation}
\langle f(x) f(y) \rangle =\sum_{i,j=1}^N \langle W_1^{(ik)} W_1^{(jk))} \rangle \langle x_1^{(i)} y_1^{(j)} \rangle = \frac{\sigma_W^2}{N} \sum_i \, \langle x_1^{(i)} y_1^{(i)} \rangle\,,
\end{equation}
which is of order $1$.
The computation of higher correlation functions can be done using a similar strategy, from the assumptions that the $x_1^{(i)}$ with different index $i$ are statistically independent variables.
This in particular ensures that:
\begin{equation}
\langle x_1^{(i)} x_1^{(i)} x_1^{(j)} x_1^{(j)} \rangle \sim \langle x_1^{(i)} x_1^{(i)} \rangle \langle x_1^{(j)} x_1^{(j)} \rangle\,,\label{decomp}
\end{equation}
from this observation, a tedious calculation which is given in~\cite{Halverson:2021:NeuralNetworksQuantum} shows that the connected $4$-point function $G^{(4)}_{c}(x_1,x_2,x_3,x_4)$ has to scale as $1/N$, and more generally that $\alpha(n)=n/2-1$.

This analytic result also shows the limitations of the approach. Indeed, we expect that a more fundamental method will be able to predict the weights of the interactions. Moreover, the derivation assumes the relation \eqref{decomp}, and thus the independence of the $x_1^{(i)}$ having different indices $i$, but such an assumption seems to be in conflict with an interaction such that \eqref{localint}, which morally must introduce couplings mixing different outputs from the definition \eqref{defF} of $f_W$. One may expect that these difficulties could be solved by working with a random vector of size $N$, with components $\varphi_i(x)$ rather than with the function $f_W(x)$, defining it as an observable $f_W(x):= \langle \varphi_i \rangle$ i.e.~the vacuum of the corresponding theory. However, the construction of such a theory is going beyond the scope of this paper, and we plan to investigate it in a forthcoming work.

\subsection{Renormalization group} \label{sectionren}

\subsubsection{The Wilson approach}\label{Wilson}

The RG is probably one of the most important concepts discovered in physics during the last century and forms together with field theory the reference framework of modern physics, from condensed matter to high energies. Pioneered in the works of Wilson and Kadanoff~\cite{Polchinski:1984:RenormalizationEffectiveLagrangians,Kadanoff:1966:ScalingLawsIsing,Wilson:1971:RenormalizationGroupCritical-1,Wilson:1971:RenormalizationGroupCritical-2}, RG is based on the idea of organizing the theory according to length scales, integrating out short distance degrees of freedom following a recursive procedure called \emph{coarse-graining} and providing an effective description for the long distance degrees of freedom, through an effective action where microscopic interactions are hidden in effective interactions. Note that RG is in fact a semi-group, which is non-invertible. Thus at each step, information is lost, and RG can be viewed as a systematic procedure to extract large scale relevant features.

To illustrate the physics underlying the Wilson procedure, and before making contact with NN field theory, let us consider a physical system made of a single real scalar field $\phi$ whose configuration probability follows the exponential form $p[\phi]=e^{-S[\phi]}$, for some classical action $S[\phi]$. To have a concrete example in mind, we can take for $\phi$ the real field described by the classical action \eqref{freefield}. All the statistical properties of the distribution can be derived from the generating functional (partition function):
\begin{equation}
Z[j]=\int [d\phi] \, e^{-S[\phi]+\int dx \, j(x)\phi(x)}\,.\label{partition1}
\end{equation}
This integral being over all configurations for $\phi(x)$, all the degrees of freedom are integrated out in one step. Equation \eqref{partition1} provides a canonical definition of what is microscopic and what is macroscopic, two limits that we conventionally call  ultraviolet (UV) and infrared (IR):
\begin{enumerate}
	\item In the UV limit, no fluctuations are integrated out. The field configurations are therefore fixed from the extrema of the classical action $S$.
	\item In the IR limit, all fluctuations are integrated out. The configurations are fixed by a new action $\Gamma$, called \emph{effective action}.
\end{enumerate}
The effective action $\Gamma$ is in turn defined as the Legendre transform of the free energy $\mathcal{W}[j]:=\ln Z[j]$,
\begin{equation}
\mathcal{W}[j]+\Gamma[\Psi]=\int dx\, j(x) \Psi(x)\,, \label{defGamma}
\end{equation}
the classical field $\Psi$ being defined as $\Psi(x):=\delta \mathcal{W}/\delta j(x)$.

\begin{figure}
\begin{center}
\includegraphics[scale=0.8]{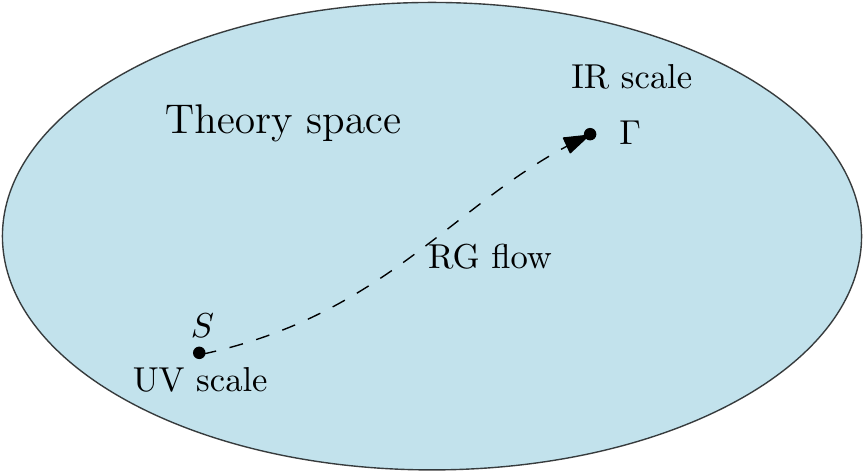}
\end{center}
\caption{The RG trajectory into the theory space, from UV to IR physics. } \label{FigGR}
\end{figure}

The RG is nothing but a path between these two boundaries. It is constructed by partially integrating out the degrees of freedom building the field $\phi$. Note that such a partial integration procedure is never arbitrary, and the Wilson RG assumes the existence of a canonical slicing $s=\{s_1,s_2,\cdots ,s_{\infty}\}$\footnote{The notation $s_{\infty}$ simply denotes the last slice.} in the configuration space of elementary degrees of freedom, allowing to integrate partially following a preferred order.
In general, this slicing is provided by the spectral distribution $\mu(E)$, $E\in \mathbb{R}$ of the UV $2$-point function for an exponential family like \eqref{partition1}: $s_i \subset \mu(E)$.
In fact, the $2$-point function can be identified with the Fisher information metric along the constrained space with fixed couplings.
This gives a connection between RG and information geometry \cite{Beny:2014:RenormalisationInferenceProblem} because of the regularity property of the Fisher metric, and in absence of singular structures, distance between probability distributions has to be reduced with coarse-graining, explaining the power of RG to discuss of universality in physics~\cite{ZinnJustin:2007:PhaseTransitionsRenormalization}. Integrating all the degrees of freedom in the first slice $s_1$ leads to an effective model with classical action $S^{\prime}$ which defines a new effective physics where effects coming from degrees of freedom in the first slice are hidden in effective interactions. Now, integrating the slice $s_2$, we obtain a new classical action $S^{\prime \prime}$ and so on. Such a partial integration (up to a global rescaling of fields to reach a fixed point) is called a \emph{RG transformation}, and the chain of RG transformations describes a "move" in the interior of the theory space:
\begin{equation}
S\to S^{\prime}\to S^{\prime\prime} \to \cdots \,,
\end{equation}
bounded by UV and IR effective physics (Figure \ref{FigGR}).

Let us illustrate how that works on the concrete example of the scalar field $\phi$ described by action \eqref{freefield}. In that case, $\mu(E)$ corresponds to the spectrum of the Laplacian $\Delta$, whose eigenmodes are Fourier modes, and $E\equiv p$. Assuming continuity of the spectrum, we can consider infinitesimal coarse-graining, integrating out slices of infinitesimal thickness. This leads to a differential equation describing how the couplings change as the reference scale changes. Formally, this can be done as follows. We assume the existence of an upper bound for $p$, say $\Lambda$, and we call $\mu_\Lambda(p)$ the spectrum of the free $2$-point function with cut-off $\Lambda$, $K_{\Lambda}(p)$. As the cut-off $\Lambda$ moves, degrees of freedom are added or removed from the spectrum. Thus, let us consider the bare action “at scale $\Lambda$”:
\begin{equation}
S_\Lambda[\phi]=\frac{1}{2} \int \mu(p) dp \, \phi(p) K_{\Lambda}^{-1}(p) \phi(-p)+\mathcal{V}_\Lambda[\phi]\,,
\end{equation}
where $\mathcal{V}[\phi]$ includes interactions following our definition of section \ref{sec21}. Now let us consider the running cut-off $\Lambda(s)=s\Lambda$, for $s\in [0,1]$, which interpolate between the UV scale $s=1$, and IR scale $s=0$. If $K_{\Lambda(s)}(E)$ is at least $\mathcal{C}^{(1)}$ in $s$, we can consider the variation at first order from $s$ to $s'=s+\delta$:
\begin{equation}
K_{\Lambda(s)}(p)= K_{\Lambda(s^\prime)}(p)+ \Delta_{\Lambda(s)}(p) \delta\,.
\end{equation}
The following decomposition can be translated as a partial integration from the original partition function using the functional identity:
\begin{equation}
\int [d\phi d\chi] \, e^{-\tilde{S}_{\Lambda (s^\prime)}[\phi,\chi]}= \left( \frac{\det{\Delta_{\Lambda(s)}}\det{K_{\Lambda(s^\prime)}}}{\det{K_{\Lambda(s)}}}\right)^{1/2}\, \int [d\phi] \, e^{-S_{\Lambda (s)}[\phi]}\,, \label{id1}
\end{equation}
where:
\begin{equation}
\tilde{S}_{\Lambda (s^\prime)}[\phi,\chi]=\frac{1}{2} \int \mu(p) dp \left( \phi(p) K_{\Lambda(s^\prime)}^{-1}(p) \phi(-p)+\chi(p)\Delta^{-1}_{\Lambda(s)}\delta^{-1} \chi(-p) \right)+\mathcal{V}_{\Lambda(s)}[\phi+\chi]\,,
\end{equation}
and $\chi$ denotes the degrees of freedom integrated out. Indeed, defining:
\begin{equation}
e^{-\tilde{\mathcal{V}}_{\Lambda(s^\prime)}[\phi]}:=(\det{\Delta_{\Lambda(s)}})^{-1/2} \int [d\chi] e^{- \frac{1}{2} \int \mu(p) dp \, \chi(p)\Delta^{-1}_{\Lambda(s)}\delta^{-1} \chi(-p)- \mathcal{V}_{\Lambda(s)}[\phi+\chi]}\label{RGtransform}
\end{equation}
and:
\begin{equation}
S_{\Lambda(s^\prime)}[\phi]=\frac{1}{2} \int \mu(p) dp \, \phi(p) K_{\Lambda(s^\prime)}^{-1}(p) \phi(-p)+\tilde{\mathcal{V}}_{\Lambda(s^\prime)}[\phi]\,,
\end{equation}
we show that the identity \eqref{id1} can be rewritten as:
\begin{equation}
\int [d\phi] \, e^{-S_{\Lambda(s^\prime)}[\phi]}= \left(\frac{\det{K_{\Lambda(s^\prime)}}}{\det{K_{\Lambda(s)}}}\right)^{1/2} \int [d\phi] \, e^{-S_{\Lambda(s)}[\phi]}\,.
\end{equation}
The classical action at scale $\Lambda(s^\prime)$ formally looks like the action at the scale $\Lambda(s)$.
What is different between them is the interaction, which comes at scale $\Lambda(s^\prime)$ from a partial integration over the field $\chi$. The transformation \eqref{RGtransform} can be translated in a differential equation for $\delta$ small enough. Indeed, in this limit, the modes $\chi$ have a large mass, and can be treated perturbatively. Thus, expanding $\mathcal{V}_{\Lambda(s)}[\phi+\chi]$ in powers of $\chi$ and keeping only terms of order $2$, we get Polchinski's equation~\cite{Polchinski:1984:RenormalizationEffectiveLagrangians}:
\begin{equation}
\Lambda \frac{d \tilde{\mathcal{V}}_{\Lambda}}{d\Lambda}\bigg|_{\Lambda=\Lambda(s)}=-\frac{1}{2} \int dp  \mu(p) \, \Delta_{\Lambda(s)}(p) \Big( \frac{\delta^2 \tilde{\mathcal{V}}_{\Lambda(s)} }{\delta \phi(p)\delta\phi(-p)} -\frac{\delta \tilde{\mathcal{V}}_{\Lambda(s)}}{\delta \phi(p)} \frac{\delta \tilde{\mathcal{V}}_{\Lambda(s)}}{\delta \phi(-p)}  \Big)\,. \label{flowPol}
\end{equation}
This equation is formally “exact”. However, this has the reputation to be very hard to solve for many reasons. The first one is that it takes place in a functional space of infinite dimension. If we decide to work in a reduced phase space, taking into account only the most relevant interactions, difficulties appear, instabilities with respect to the considered truncation appear as soon as we try to get beyond the perturbative sector, which is precisely what we are aiming at in this paper.
For this reason, and as it is the case for the largest part of non-perturbative investigations in the literature~\cite{Delamotte:2012:IntroductionNonperturbativeRenormalization,Dupuis:2021:NonperturbativeFunctionalRenormalization,Wetterich:1991:AverageActionRenormalization,Wetterich:1993:ExactEvolutionEquation,Morris:1994:DerivativeExpansionExact,Morris:1994:ExactRenormalisationGroup,Berges:2002:NonPerturbativeRenormalizationFlow}, we will prefer to use the Wetterich formalism, better for dealing with non-perturbative approximations. We will discuss this method in the next section.

\subsubsection{Renormalization group(s) for the NN-QFT}\label{secRenNN}

The analogy between NNs and RG is evident: both are aiming at extracting relevant features from a massive number of degrees of freedom. RG shows that microscopic details can be ignored to describe long distance physics, and that microscopic theories can be indistinguishable from their common large distance properties. Extracting regularities from large sets of data is exactly what machine learning does; and, as we recalled in the introduction, the question of the relevance of the renormalization group in artificial intelligence is growing in the literature~\cite{Beny:2013:DeepLearningRenormalization,Li:2018:NeuralNetworkRenormalization,deMelloKoch:2020:DeepLearningRenormalization,KochJanusz:2018:MutualInformationNeural,Mehta:2014:ExactMappingVariational}. However, the effective field theory that we presented in the first part offers a new framework to discuss aspects related to the renormalization group in the study of the behavior of NNs~\cite{Halverson:2021:NeuralNetworksQuantum}. As the previous section stressed out, the field theory that we consider exhibits strong similarities with theories usually considered by physicists: the long distance (i.e.~large volume, small momenta) limit \eqref{Klong} of the free propagator being the same as for the usual scalar field $\phi$ described by the action \eqref{freefield}. This formal similarity will serve as a guide in the construction of the renormalization group, and it is very tempting to carry out a coarse-graining in momenta, exactly as for the scalar field $\phi$ in the previous section. We will discuss two different coarse-graining strategies which we call respectively  \emph{passive} and \emph{active} RGs. But before we go into them in detail, let's make a few general remarks about what distinguishes this NN field theory from ordinary theories.

\begin{figure}
\begin{center}
\includegraphics[scale=0.8]{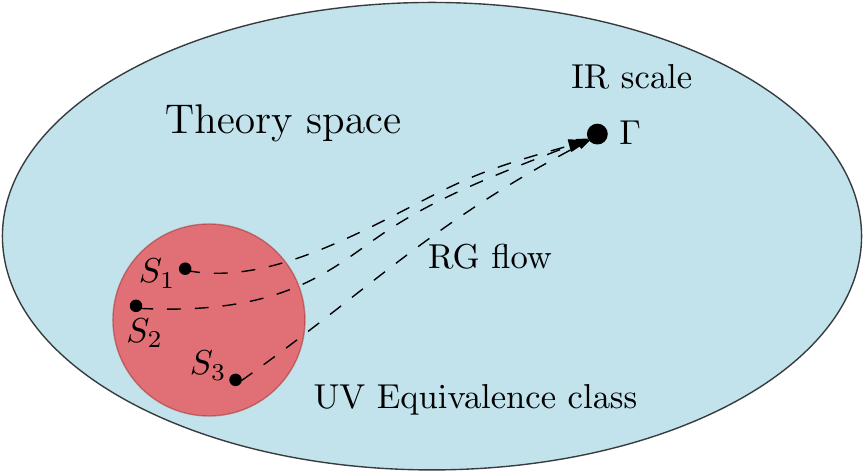}
\end{center}
\caption{Behavior of the RG flow with different initial conditions. The red region corresponds to initial conditions for all microscopic actions whose RG flows are experimentally indistinguishable in the deep IR regime, and corresponds to the same effective physics described by $\Gamma$.} \label{figequiv}
\end{figure}

In the standard scenario, what is known is the UV theory i.e.~the classical action. This action itself is viewed as an effective description, valid at some fundamental scale and ignoring the details about nature and physics of microscopic degrees of freedom underlying the physical world. The choice of the classical action is constrained by predictivity (which promotes just-renormalizable theories), consistency with quantum effects (compensation of anomalies in gauge theories, for instance), and the effective structures at the scale at which the theory is defined, which generally implies some symmetries (rotation, reflection, gauge invariance...). In this respect, the RG aims to provide an approximation of the exact quantum theory, and to compare it with experiments.
The case of the field theory that we consider differs from this general picture in its relations between UV and IR scales. The propagator \eqref{exact2points} is exact and defined in the deep IR. From a RG point of view, the knowledge of this propagator takes into account all the fluctuations at all scales. But, for finite $N$, the knowledge of the $2$-point functions is not sufficient to reproduce higher correlations functions, and non-Gaussian interactions are required in the classical action to reproduce experimental correlation functions. But, due to these interactions, the flow of the different ingredients entering in the definition of the classical action becomes non-trivial, with the consequence that both $S_{\text{kin}}$ and $S_{\text{int}}$ in the UV are unknown. Thus, in some sense, the situation is the inverse of what we do in ordinary field theory: we have to infer the form of the UV theory (or more likely a class of UV theory) from the knowledge of only a part of the IR theory. By construction, such an inference cannot lead to a single solution, but a class of solutions which have to satisfy the following requirements:
\begin{enumerate}
	\item reproduce the exact $2$-point functions up to the experimental precision;
	\item reproduce the deviations from Wick's theorem, due to interactions, and which are less and less perturbative as $N$ becomes small, once again up to irrelevant corrections with respect to the experimental precision.
\end{enumerate}
Any measurement in physics comes with a finite precision: hence, two effective descriptions are considered to be equivalent and sufficient to describe something if the predictions agree up to the experiment precision. The precision is also finite in numerical simulations, and this explains why that we are able to infer only an equivalent class of models rather than a point in the theory space. In the first section, we showed that the relative relevance of the interactions is not the same such that irrelevant interactions contribute below the machine precision threshold, meaning that we have no way to distinguish between several initial conditions whose trajectories are sufficiently close in the IR (see Figure \ref{figequiv}).
This argument allows working, in a first approximation, within a finite subspace of the full theory space, focusing on interactions having the largest canonical dimension.

\begin{figure}
	\begin{center}
	\includegraphics[scale=0.6]{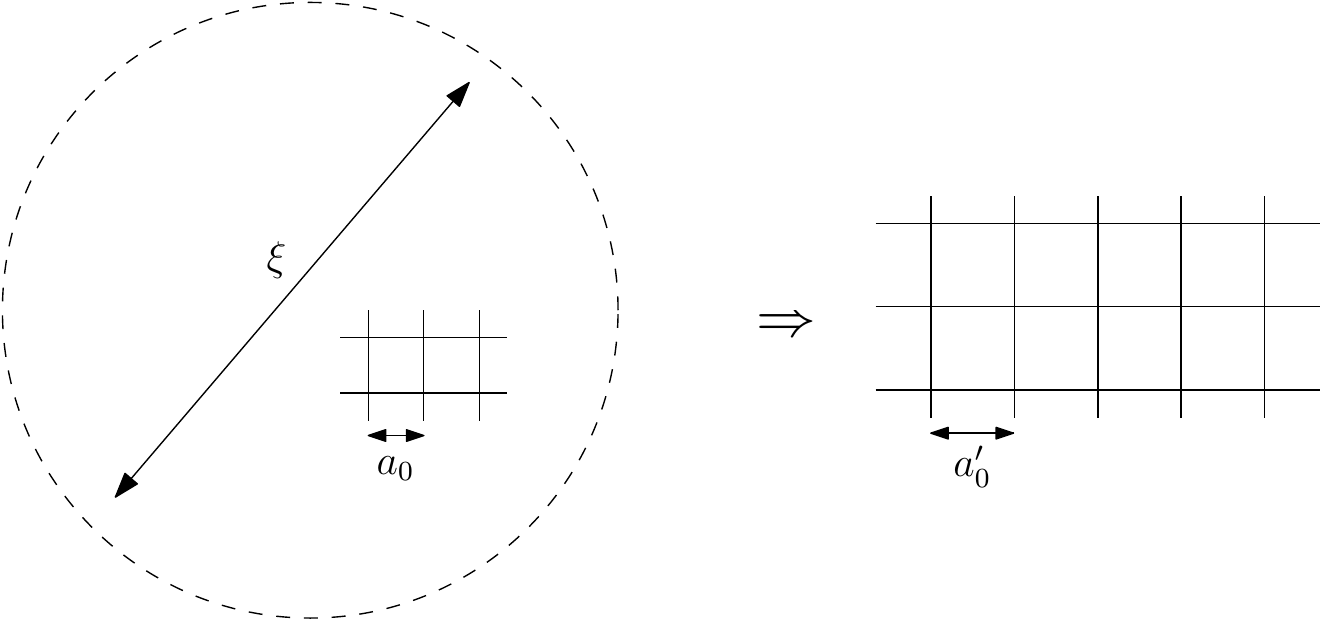}
	\caption{A passive change of the grid scale, provided by a dilatation of working precision from $a_0$ to $a_0^\prime$. } \label{figGrid}
	\end{center}
\end{figure}

\paragraph{Passive RG.} Because of the existence of an intrinsic length scale $\xi$ defined in \eqref{eq:observation-scale}, we can think to partially integrate microscopic degrees of freedom with respect to this length scale to construct a proper RG flow following standard field theory. In this picture, what is playing the role of a microscopic scale is the working precision (see Section~\ref{sec:data-space}), which introduces a cut-off in momentum integration $\Lambda=1/a_0$.
We can then construct a coarse graining procedure from grid size dilatation (see Figure \ref{figGrid}).

Note that from such a procedure, one needs to have $\xi \gg a_0$. Because the maximal value for $p$ is $p_\infty= 2\pi/a_0$, we can show that it implies $p_\infty \xi \gg 1$, which invalidates the expansion \eqref{expansionPropa}. However, it may happen that such an expansion holds in a sufficiently large domain. A necessary condition is that the expansion \eqref{expansionPropa} holds for the smallest (nonzero) momentum $p_0=2\pi/a_0N_0$, implying:
\begin{equation}
\xi \ll a_0 N_0 \equiv L\,,
\end{equation}
which is the condition defining the large volume limit.

A dilatation procedure as described in Figure \ref{figGrid} induces a renormalization group by partial integration of momenta into the windows $\sim ]1/a_0^\prime, 1/a_0]$. The existence of two complementary limits $\xi \ll L$ and $\xi \gg a_0$ is reminiscent of a crossover scale behavior, between a deep ultraviolet (UV) limit $p\sim 1/a_0$ and a deep infrared limit ($p\sim 1/L$), which we will study separately in the next section. Note that such a crossover scale appears generally in situations where two very different mass scales appear, ensuring decoupling\footnote{Such a decoupling is at the origin of the large mass expansion in field theory.} of some effects associated to the larger one when experiments focus on the first one~\cite{Collins:1986:Renormalization}.
Here, what plays the role of a large mass is the inverse of the typical observation scale $\xi$; in the very large mass limit, $p_\infty \ll (\xi)^{-1}$, and the IR sector recovers all the physics. In the opposite limit, $p_0 \gg (\xi)^{-1}$, everything is UV, and an expansion such that \eqref{expansionPropa} does not hold. In other words, for $p\ll (\xi)^{-1}$, one expects that quantum effects are suppressed with powers of $(\xi)^{-1}$. This observation can be a source of improvement for approximations used to solve the RG flow equation \eqref{Wett} in the next section. In particular, we understand that contributions coming from higher couplings will tend to stay small if they are at the transition scale $(\xi)^{-1}$. Section \ref{sec3} is devoted to this RG strategy.

\begin{figure}
\begin{center}
\includegraphics[scale=0.6]{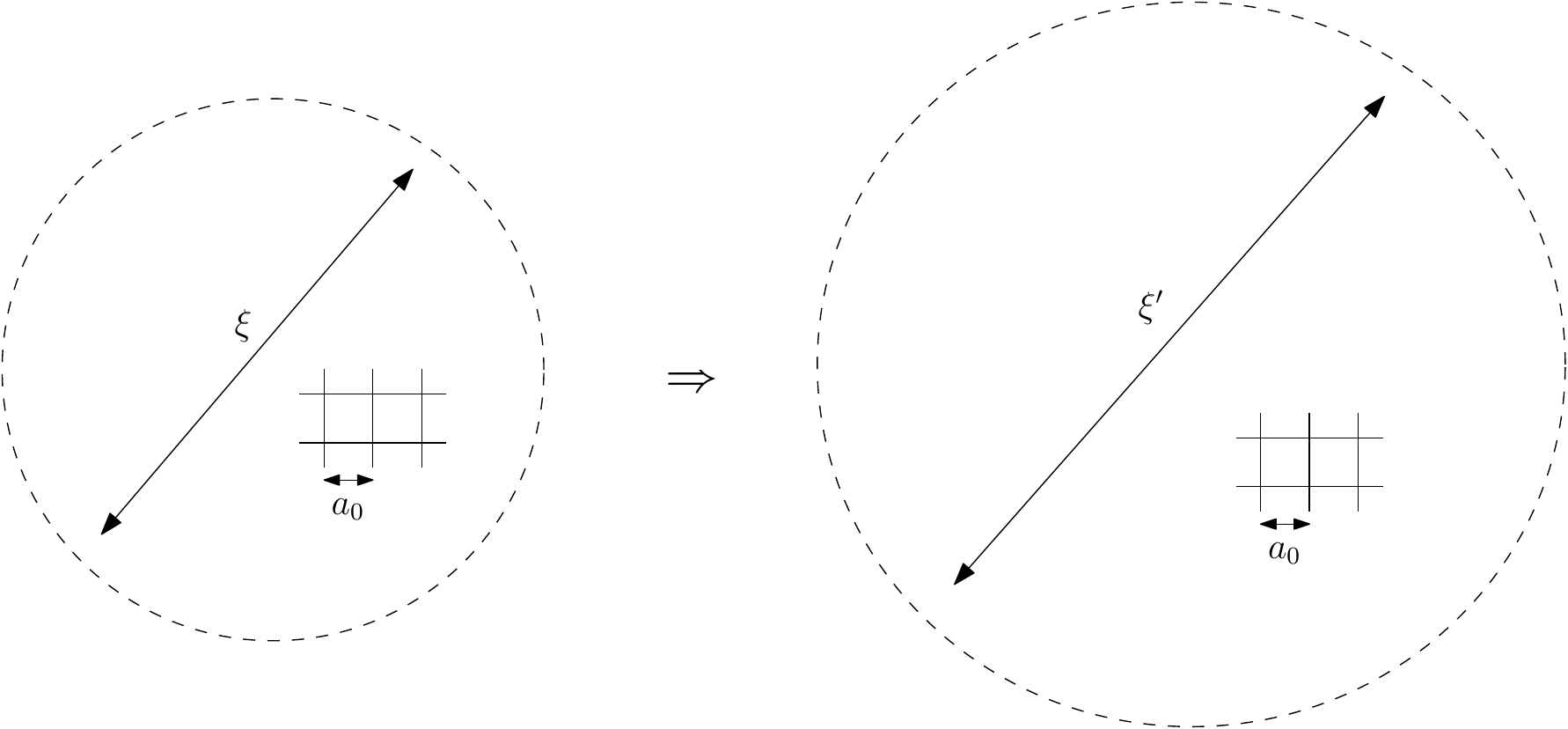}
\end{center}
\caption{An active change of the observation scale from $\xi$ to $\xi^\prime$, without dilatation of working precision.} \label{figGrid2}
\end{figure}

\paragraph{Active RG.} In the process described above, the observation scale $\xi$ is kept fixed and the working precision is changed. Conversely, we can keep the data (i.e.~the working precision) fixed and change the observation scale. If the first version is essentially \emph{passive} with respect to the NNs (i.e.~the latter is not changed), this strategy is, in contrast, \emph{active} (see Figure \ref{figGrid2}).
Indeed, remembering the expression \eqref{eq:observation-scale}, $\xi$ is completely determined in terms of $\sigma_W$, the standard deviation of the weight distribution.
Hence, flowing in the observation scale is equivalent to changing the weight standard deviation, and thus the neural network.

Physically, if we think of a thermodynamic system like a ferromagnet, such a strategy is equivalent to turning the thermostat's knob to lower the temperature towards the critical regime. This alternative point of view is the subject of the section \ref{sec4}.

\begin{remark}
	The active RG is closer to the RG version considered in the reference~\cite{Halverson:2021:NeuralNetworksQuantum} than the passive scheme, as the flow equations derived in section \ref{sec4} show explicitly. However, despite this formal contact, our approach differs by its very construction. While, from their point of view, the RG is the mathematical explanation of a principle of invariance with respect to a certain volume “cut-off”, our RG is the result of a procedure of partial integration of the degrees of freedom of the field.

	Indeed, the RG flow is usually performed with respect to a UV cut-off (spacetime /data-space resolution) and not an IR cut-off (volume).
	In~\cite{Halverson:2021:NeuralNetworksQuantum}, the large volume cut-off was introduced because the $2$-point function diverges at large distance (at least, for the ReLU-net), which reminds the short-distance divergence of the canonical propagator in particle QFT.
	Moreover, one can ask whether the data-space should be identified with the position or momentum space in usual spacetime QFT, and in principle this could depend on the problem.
	From our arguments in Section~\ref{sec:data-space}, it seems more natural to identify the data-space with the position space (except in the case where the data is already the Fourier transform of a space(time) process).
	IR divergences are also present in particle QFT, and they are cured following different methods according to their origin.
	The first case are massless particles, for which a refined definition of amplitudes is needed~\cite{Kulish:1970:AsymptoticConditionsInfrared,Weinberg:2005:QuantumTheoryFields-1}.
	An infrared cut-off (such as a mass) can be introduced at intermediate stages to regulate the integral, but it is not a renormalization parameter.
	In practice, the divergences of the ReLU-net arise from a similar origin (singularity in the propagator for large distance / zero-momentum).
	Second, IR divergences appear for internal on-shell propagators: they translate the fact that quantum effects shift the vacuum and masses of the fields.
	Resummation of quantum effects through renormalization leads to finite results~\cite{deLacroix:2017:ClosedSuperstringField,Weinberg:2005:QuantumTheoryFields-1}.
	Third, some quantities can diverge for infinite volume for example when studying phase transition: in that case, the usual method is to study the theory with different values of a volume cut-off and to extrapolate to infinite volume (thermodynamic limit)~\cite{Orth:2004:VolumeDependenceLight}.
	However, this is \emph{not} a renormalization flow.
	For these reasons, we take a more conservative approach and identify small resolution in data space with the UV limit, and perform the RG flow for the associated cut-off.

	It is also noted that in~\cite[sec.~4.3]{Halverson:2021:NeuralNetworksQuantum} that the Gauss-net does not require renormalization because the $2$-point function is exponentially decaying with the distance, such that all integrals are convergent.
	In fact, the previous paragraph shows that renormalization is still needed in this case because its role is not only to handle properly (spurious) UV divergences, but also to take into account quantum effects (some of which lead to IR divergences).
	Said another way, renormalization provides a mapping between the bare and physical parameters (at a given energy scale) there is always a renormalization flow in the space of couplings.
	Indeed, the bare parameters describe the properties of the fields without interactions: they are not physical because fields do not live in isolation and any measurement implies an interaction.
	A famous example of a perfectly finite theory but which has an infinite number of finite (such that predictivity is not lost) counter-terms and non-trivial RG flow (with the so-called stub length) is string field theory~\cite{Sonoda:1990:ClosedStringField,Sen:2016:WilsonianEffectiveAction,Erbin:2021:StringFieldTheory}.
\end{remark}

\section{Flowing through NN-QFT theory space: the passive RG}\label{sec3}

In this section, we show how the passive RG within Wetterich formalism allows predicting the behavior of correlation functions for a fully connected NN with a single hidden layer. We start with a short presentation of the Wetterich formalism, before turning on to applications. We will consider separately two different regimes, the deep IR regime $k \ll (\xi)^{-1}$ where the effective propagator can be suitably approximated with an ordinary Laplacian $\sim (-\Delta+m^2)^{-1}$, and the UV regime $k \sim (\xi)^{-1}$, where the propagator follows the exponential law $\sim e^{-\Delta/m^2}/m^2$.

\subsection{Wetterich formalism}
\label{sec:wetterich}

In section \ref{Wilson}, we provided a formal introduction to Wilson's ideas for the RG. In this section, we present another incarnation, the so-called Wetterich formalism~\cite{Wetterich:1991:AverageActionRenormalization,Wetterich:1993:ExactEvolutionEquation,Delamotte:2012:IntroductionNonperturbativeRenormalization}, which focuses on the effective action for integrated degrees of freedom rather than on the effective classical action for the remaining degrees of freedom, as it is the case in \eqref{flowPol}. We focus on the passive RG as presented in the section \ref{secRenNN}. Let $\Lambda=1/a_0$ be some reference working precision and $k\in [0,\Lambda]$. Assuming that we performed partial integration up to the scale $k$, we denote as $\Gamma_k$ the effective action for those averaged degrees of freedom. Obviously, it must satisfy the boundary conditions:
\begin{enumerate}
\item $\Gamma_{k=\Lambda}=S$, no fluctuations are integrated out, and the effective action reduces to the classical action.
\item $\Gamma_{k=0}=\Gamma$, all fluctuations are integrated out and we recover the full effective action $\Gamma$ defined in \eqref{defGamma}.
\end{enumerate}
The Wetterich formalism aims to construct a smooth interpolation between these two limits. To this end, it is convenient to modify the classical action with a scale dependent mass term $\Delta S_k$, which reads in momentum space:
\begin{equation}
\Delta S_k=\frac{1}{2}\int  \, f(-p) r_k(p^2) f(p)\,.
\end{equation}
The substitution $S\to S+\Delta S_k$ defines a $k$-dependent partition function $Z_k$ through the definition \eqref{partition1}. The shape of the scale-dependent mass $r_k(p^2)$ is designed to freeze low momenta modes $p^2<k^2$, decoupling them from long distance physics whereas high energy modes $p^2>k^2$ remain essentially unaffected. Moreover, in order to recover the full classical action $\Gamma$ for $k=0$, $r_k(p^2)$ has to vanish in that limit. In the same way, it has to become very large in the opposite limit, for $k\to \Lambda$, in order to satisfy the UV boundary condition $\Gamma_{k\to \Lambda} \to S$ (all the fluctuations are frozen).
The interpolating functional $\Gamma_k$ is defined as:
\begin{equation}
\Gamma_k[\Psi]+\ln Z_k[j]=\int dx  j(x)\,\Psi(x)- \frac{1}{2}\int  \, \Psi(-p) r_k(p^2) \Psi(p)\,.\label{defGammak}
\end{equation}
As $k$ varies from $k$ to $k-\delta k$, effective couplings involved in the effective action change. To obtain the differential equation governing the behavior of $\Gamma_k$, as $k$ varies, we can differentiate the definition \eqref{defGammak} with respect to $k$. After a tedious calculation whose details can be found in~\cite{Delamotte:2012:IntroductionNonperturbativeRenormalization}, we get the following functional equation:
\begin{equation}
\frac{d}{dk}\Gamma_k= \frac{1}{2}\, \int \frac{dp}{(2\pi)^{d_{\text{in}}}}  \, \frac{d r_k}{dk}(p^2)\,(\Gamma^{(2)}_k+r_k)^{-1}(p,-p)\,, \label{Wett}
\end{equation}
where $\Gamma^{(n)}_k$ denotes the $n$-th functional derivative with respect to the classical field $\Psi(x):=\delta \ln Z_k/\delta j(x)$. This equation, up to the formal character of its derivation, is as exact as equation \eqref{flowPol} is. It defines a trajectory through a functional space and is as hard to solve as the equation \eqref{flowPol}.
Approximations are required to make the underlying physics tractable. The standard strategy, called \emph{truncation}, is to identify a relevant finite-dimensional subspace of the full theory space, and to project the flow equation \eqref{Wett} onto it. Working with equation \eqref{Wett} has the great advantage that this projection procedure does not require to assume that couplings are small, and thus allows investigating approximate but non-perturbative solutions of the RG flow.

\subsection{Local potential approximation in the deep IR}

The local potential approximation (LPA) is one of the most popular approximation procedures~\cite{Delamotte:2012:IntroductionNonperturbativeRenormalization} to solve the exact RG flow equation \eqref{Wett}. This approximation focuses on the region of the full theory space spanned by local interactions in the sense of \eqref{localint}. For the investigations in this section, we assume $p^2 \ll 2\sigma_W^2/d_{\text{in}}$, which is our reference mass scale. This implies:
\begin{equation}
k \ll \xi^{-1}\,,
\end{equation}
which defines the IR regime (see Section \ref{secRenNN}). Note that due to the scaling behavior of derivative contributions, one expects that the validity of this description survives in the weak UV regime: $1/a_0 \gg k \gg (\xi)^{-1}$ due to the \emph{large river effect} which states, that in a suitable vicinity of the Gaussian fixed point and in the absence of singularities along the flow, the latter projects itself into the subspace spanned by the most relevant couplings~\cite{Bagnuls:2001:ExactRenormalizationGroup}.

\subsubsection{Symmetric phase}

To begin, we focus on the simplest truncation around sixtic interactions, discarding from our analysis contributions arising from higher couplings. This is equivalent to setting:
\begin{equation}
\Gamma_k^{(2n)}[\Psi=0]=0 \,, \quad \text{for}\, n>3\,.\label{truncationcutoff}
\end{equation}
where, to avoid confusion with the example given in Section \ref{sectionren}, we denote as $\Psi$ the classical field. Note that such an expansion around $\Psi=0$ is named symmetric phase expansion, and we call symmetric phase the domain of the full phase space where it remains valid. It may happen that such an expansion breaks down, in cases where $\Psi=0$ becomes an unstable vacuum. This is the case when phase transitions are encountered. In this section, we focus on the symmetric phase, and discuss more elaborate formalisms in the next section. Approximation \eqref{truncationcutoff} ensures that we keep effects up to order $1/N^2$.
\medskip

To be more concrete, we assume that $\Gamma_k[\Psi]$ can be decomposed as a sum of two contributions:
\begin{equation}
\Gamma_k[\Psi]= \Gamma_{k,\text{kin}}[\Psi]+U_k[\Psi]\,,
\end{equation}
where:
\begin{enumerate}
\item The kinetic contribution $\Gamma_{k,\text{kin}}[\Psi]$ keeps all the quadratic terms in $\Gamma_k[\Psi]$.

\item The effective potential $U_k[\Psi]$ gathers the non-Gaussian contributions in the expansion of $\Gamma_k[\Psi]$.
\end{enumerate}

Without lost of generality, the kinetic contribution can be written as:
\begin{equation}
\Gamma_{k,\text{kin}}[\Psi]= \frac{1}{2}\sum_p \, \Psi(p) \mathcal{K}_k(p^2) \Psi(-p)\,.
\end{equation}
The kernel $\mathcal{K}_k(p^2)$ is a priori difficult to track. Fortunately, because we are aiming to deal with IR effects, the momentum $p$ is expected to be small, justifying to expand $\mathcal{K}(p)$ in power of $p^2$:
\begin{equation}
\mathcal{K}_k(p)= \mathcal{K}_k(0)+p^2 \mathcal{K}_k^\prime(0)+\mathcal{O}(p^4)\,.
\end{equation}
The first term of this expansion define the \emph{running mass}, and we denote it as $m^2(k)$. In the same way the second term of the expansion is called \emph{running wave function renormalization}, and we denote it as $Z(k)$. Nerveless, it is easy to check that in the symmetric phase $Z(k)$ does not depend on the running scale $k$ (see below). Thus we must have $Z(k)=Z_0=1$. This scheme defines the \emph{derivative expansion}~\cite{Morris:1994:ExactRenormalisationGroup,Morris:1994:DerivativeExpansionExact,Morris:1998:ElementsContinuousRenormalization,Balog:2019:ConvergenceNonPerturbativeApproximations}, and for this section we focus on the two first terms:
\begin{equation}
\Gamma_{k,\text{kin}}[\Psi]= \frac{1}{2} \sum_{p} \, \Psi(p) (p^2+m^2(k))\Psi(-p)\,. \label{kinetic}
\end{equation}
To keep only effects up to order $1/N^2$, we consider the following truncation for the effective potential:
\begin{equation}
U_k[\Psi]= \frac{u_4}{4!} A_4[\Psi]+\frac{u_6}{6!} A_6[\Psi]\,,\label{truncationdequal1}
\end{equation}
where:
\begin{equation}
A_n[\Psi]:=\sum_{\{p_i\}} \delta_{0,\sum_{i=1}^n p_i} \prod_{j=1}^n \Psi(p_j)\,.
\end{equation}
This form follows from the expression of a local interaction of order $n$ in position space: the Fourier transformation to momentum space introduces one momentum $p_j$ for each field together with a delta function for momentum conservation, and a sum (since the $p_j$ take discrete values) over each value of the momentum.
The final piece is the regulator $r_k$. From the choice of the kinetic truncation \eqref{kinetic}, it is suitable to use the modified version of the standard optimized Litim's regulator~\cite{Litim:2001:OptimisedRenormalisationGroup}:
\begin{equation}
r_k(p^2):= (k^2-p^2)\theta(k^2-p^2)\,,
\end{equation}
for which analytic computations are possible.
In this equation, $\theta$ is the step function such that $\theta(x > 0) = 1$ and $\theta(x < 0) = 0$.
The flow equations can be deduced from the exact RG equation \eqref{Wett} by taking successive derivatives with respect to the classical field $\Psi$. Taking the second derivative gives the flow equation for $\Gamma_{k}^{(2)}(\vec{p}_1, \vec{p}_2)$.
\begin{equation}
k \frac{d}{dk}\Gamma_{k}^{(2)}(p_1, p_2)= -\frac{1}{2} \sum_{\{p_\alpha\}, \alpha=0,3,4}\,\left(k\frac{d}{dk}r_k(p_0^2) \right) G_k(p_0, p_3) \Gamma_k^{(4)}(p_1, p_2,p_3,p_4) G_k(p_4, p_0)\,,\label{flowmass}
\end{equation}
where, on the RHS, functions are computed for $\Psi=0$. From the truncation \eqref{kinetic}, we must have:
\begin{equation}
G_k(p,p^\prime):= \frac{\delta_{p,-p^\prime}}{p^2+m^2(k)+r_k(p^2)}\,.
\end{equation}
The fourth derivative $\Gamma_k^{(4)}(p_1, p_2,p_3,p_4)$ can be easily computed from the truncation \eqref{truncationdequal1}, leading to:
\begin{equation}
\Gamma_k^{(4)}(p_1, p_2,p_3,p_4)=u_4 \delta_{0,p_1+ p_2+p_3+p_4}\,, \label{rencondGamma4}
\end{equation}
replacing $\Psi=0$ at the end of the computation.
Thus, setting $p_1=0$ on both sides of equation \eqref{flowmass}, we get after some calculations\footnote{Details are given on Appendix \ref{App1}.}:
\begin{equation}
k \frac{d}{dk}m_k^2= - \frac{u_4}{(k^2+m^2(k))^2} \mathrm{Vol}(k) \,,
\end{equation}
where
\begin{equation}
\mathrm{Vol}(k):=\left(k^2\sum_p \theta(k^2-p^2) \right)\,.
\end{equation}

\begin{remark}
In equation \eqref{flowmass}, the only dependence on the external momenta $p_1$ and $p_2$ on left-hand side is through the conservation delta $\delta_{p_1,-p_2}$ arising from the structure of the four-point function vertex $\Gamma_k^{(4)}$. Thus, the field strength $Z(k)$, whose flow equation could be deduced by taking derivatives on both sides of equation \eqref{flowmass} with respect to $p_1^2$, vanish identically.
\end{remark}

\begin{figure}
\begin{center}
\includegraphics[scale=0.7]{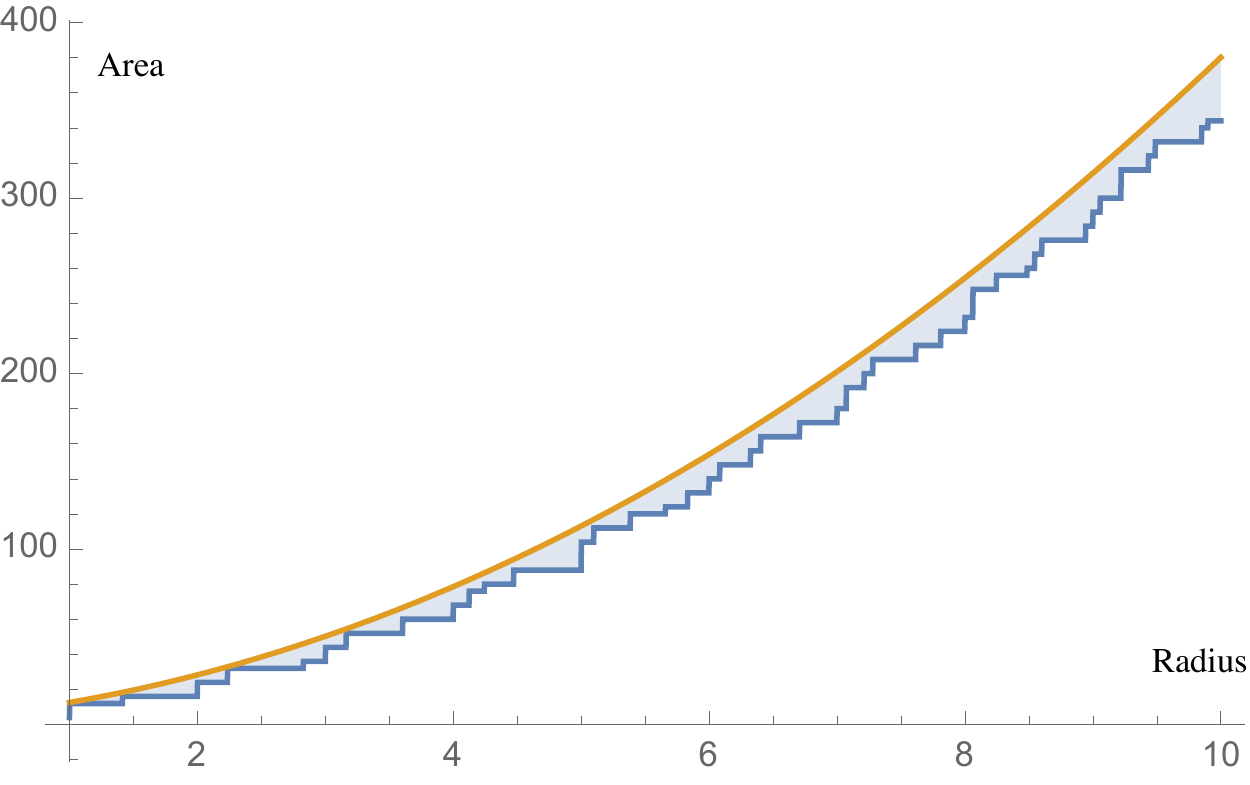}
\end{center}
\caption{The discrete volume in two dimension (blue curve) versus the continuous version given by $\pi (R+1)^2$ (brown curve).}\label{figcontinuous}
\end{figure}

In the same way, taking the fourth and sixth derivatives with respect to $M$ of the flow equation \eqref{Wett}, and from the condition \eqref{rencondGamma4}, we get schematically:
\begin{equation}
k\frac{d}{dk}{\Gamma}_k^{(4)}= -\frac{1}{2}\tilde{G}_k \Gamma_k^{(6)} G_k+ 3\tilde{G}_k \Gamma_k^{(4)}G_k\Gamma_k^{(4)}G_k\,,\label{eqlocal4}
\end{equation}
and:
\begin{equation}
k\frac{d}{dk}{\Gamma}_k^{(6)}=-\frac{1}{2}\tilde{G}_k \Gamma_k^{(8)} G_k+15\tilde{G}_k \Gamma_k^{(6)}G_k\Gamma_k^{(4)}G_k-45\tilde{G}_k \Gamma_k^{(4)}G_k\Gamma_k^{(4)}G_k\Gamma_k^{(4)}G_k\,.\label{eqlocal6}
\end{equation}
A tedious calculation leads to:
\begin{equation}
k\frac{du_4}{dk}=-\frac{u_6}{(k^2+m^2(k))^2}\mathrm{Vol}(k)+ \frac{6u_4^2}{(k^2+m^2(k))^3}\mathrm{Vol}(k) \,,
\end{equation}
and
\begin{equation}
k\frac{du_6}{dk}=\frac{30 u_4u_6}{(k^2+m^2(k))^2}\mathrm{Vol}(k)-\frac{90 u_4^3}{(k^2+m^2(k))^3}\mathrm{Vol}(k)\,.
\end{equation}
These equations illustrate how the scaling can be fixed without assuming any background dimension, as discussed in section \ref{LambdaScal}. Indeed, a moment of reflection shows that the argument below equation \eqref{scalg} about the existence of a non-trivial expansion is equivalent to the statement that a global rescaling of all couplings must exist such that the flow equations become an autonomous system. For $k$ large enough, the sum in $\mathrm{Vol}(k)$ can be well approximated by an integral, and:\footnote{$\sum_p\to \frac{1}{(2\pi)^{d_{\text{in}}}}\int dp$.}
\begin{equation}
\mathrm{Vol}(k) \underset{k \gg 1}{\sim} \frac{1}{(2\pi)^{d_{\text{in}}}}\frac{\pi^{d_{\text{in}}/2}k^{d_{\text{in}}+2}}{\Gamma(d_{\text{in}}/2+1)}=:K_{d_{\text{in}}}k^{d_{\text{in}}+2}  \,. \label{volk}
\end{equation}
One expects that such an approximation remains valid for $k^2 \gg 4\pi^2/L^2$, with $L$ being large (see Figure \ref{figcontinuous}). Thus, defining,
\begin{equation}
\bar{u}_2(k)=k^{-2} m^2(k)\,,\quad \bar{u}_{2n}=k^{-2n+(n-1)d_{\text{in}}}u_{2n}\,,
\end{equation}
we get the autonomous system ($\beta_{2n}:=k d\bar{u}_{2n}/dk$):
\begin{align}
\beta_2&=-2\bar{u}_2-\frac{K_{d_{\text{in}}}\bar{u}_4}{(1+\bar{u}_2)^2}\,,\\
\beta_4&=-(4-d_{\text{in}})\bar{u}_4- \frac{ K_{d_{\text{in}}}\bar{u}_6}{(1+\bar{u}_2)^2}+ \frac{6 K_{d_{\text{in}}}\bar{u}_4^2}{(1+\bar{u}_2)^3}\,,\\
\beta_6&=-(6-2d_{\text{in}})\bar{u}_6+ \frac{30 K_{d_{\text{in}}} \bar{u}_4\bar{u}_6}{(1+\bar{u}_2)^3}-\frac{90 K_{d_{\text{in}}} \bar{u}_4^3}{(1+\bar{u}_2)^4}\,.
\end{align}

\begin{figure}
\begin{center}
\includegraphics[scale=0.5]{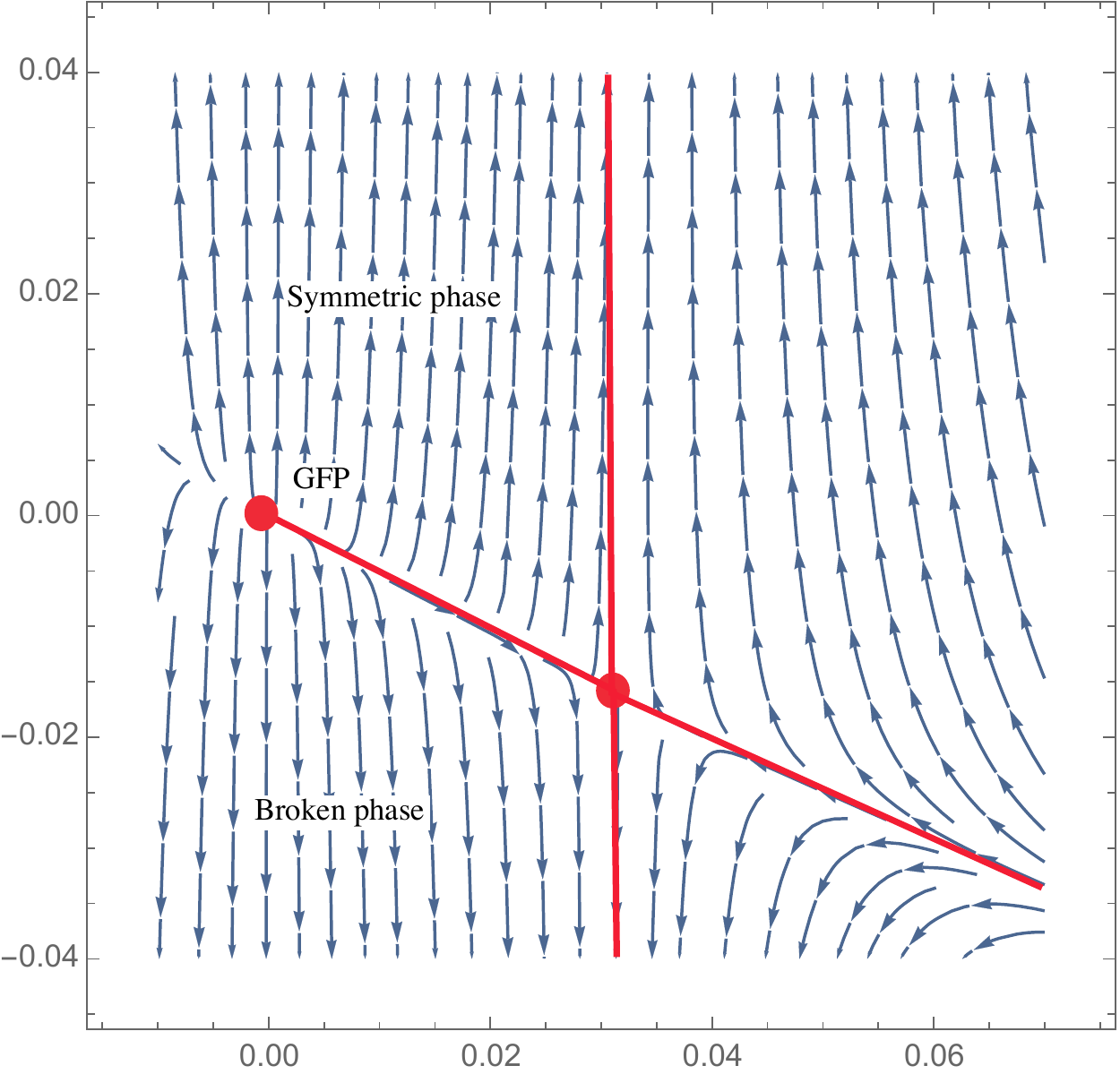}
\qquad
\includegraphics[scale=0.5]{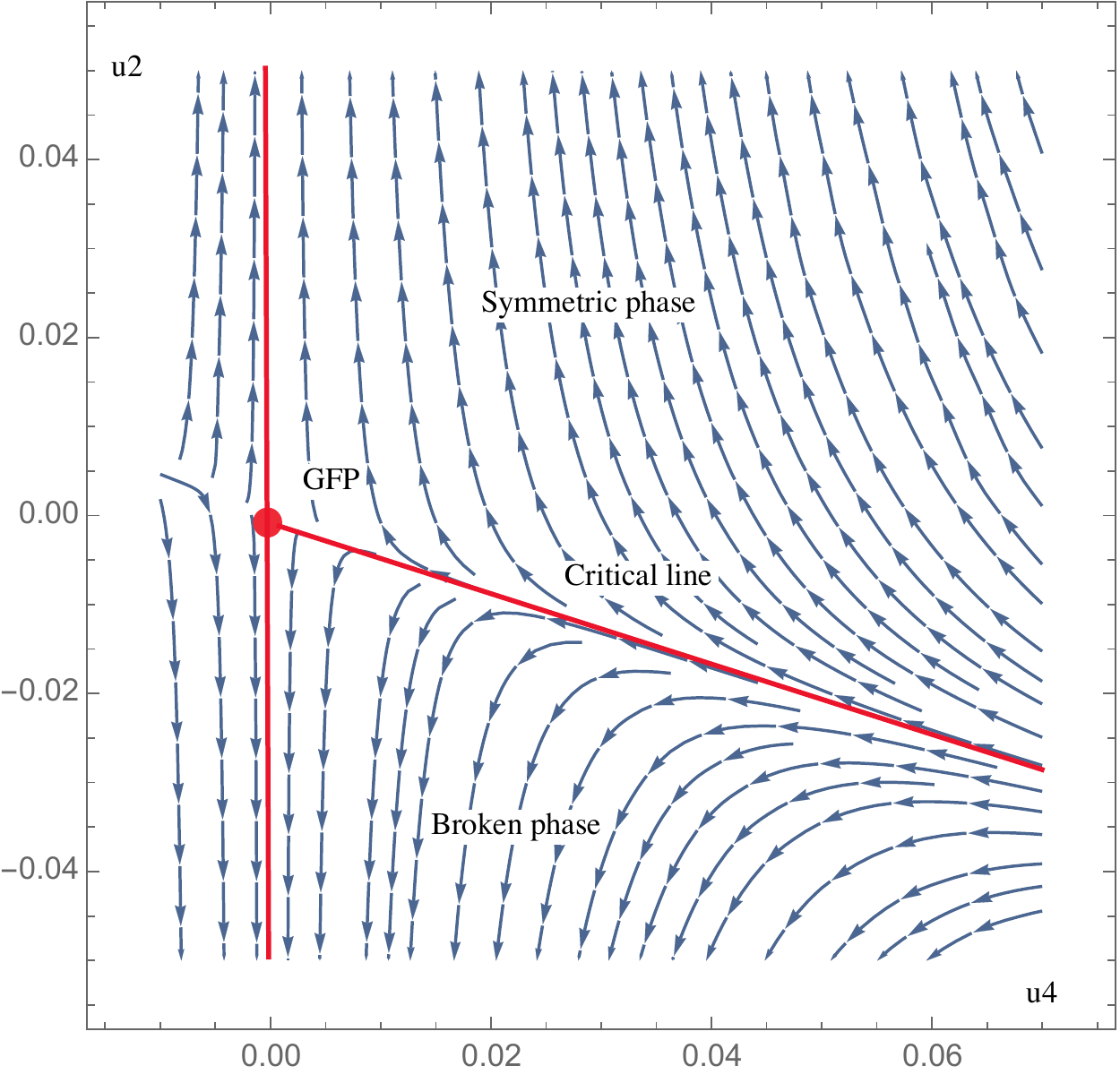}
\end{center}
\caption{Typical behavior of the RG flow for $d<4$ (on left) and $d>4$ (on right). In the first case, the flow is repelled from the Gaussian fixed point GFP) and there exists an IR fixed point; with one attractive and one repulsive direction. The integral curve of the attractive direction connects it with the Gaussian fixed point, and defines the critical line, splitting the RG trajectories in two families, going toward positive and negative mass respectively. For the second case, there are no fixed points, and the transition line between symmetric and broken phase is controlled by the Gaussian fixed point itself.}\label{flow}
\end{figure}

\begin{figure}[htp]
	\centering

	\subcaptionbox{$(u_2, u_4)$}[\linewidth]{%
		\includegraphics[scale=0.5]{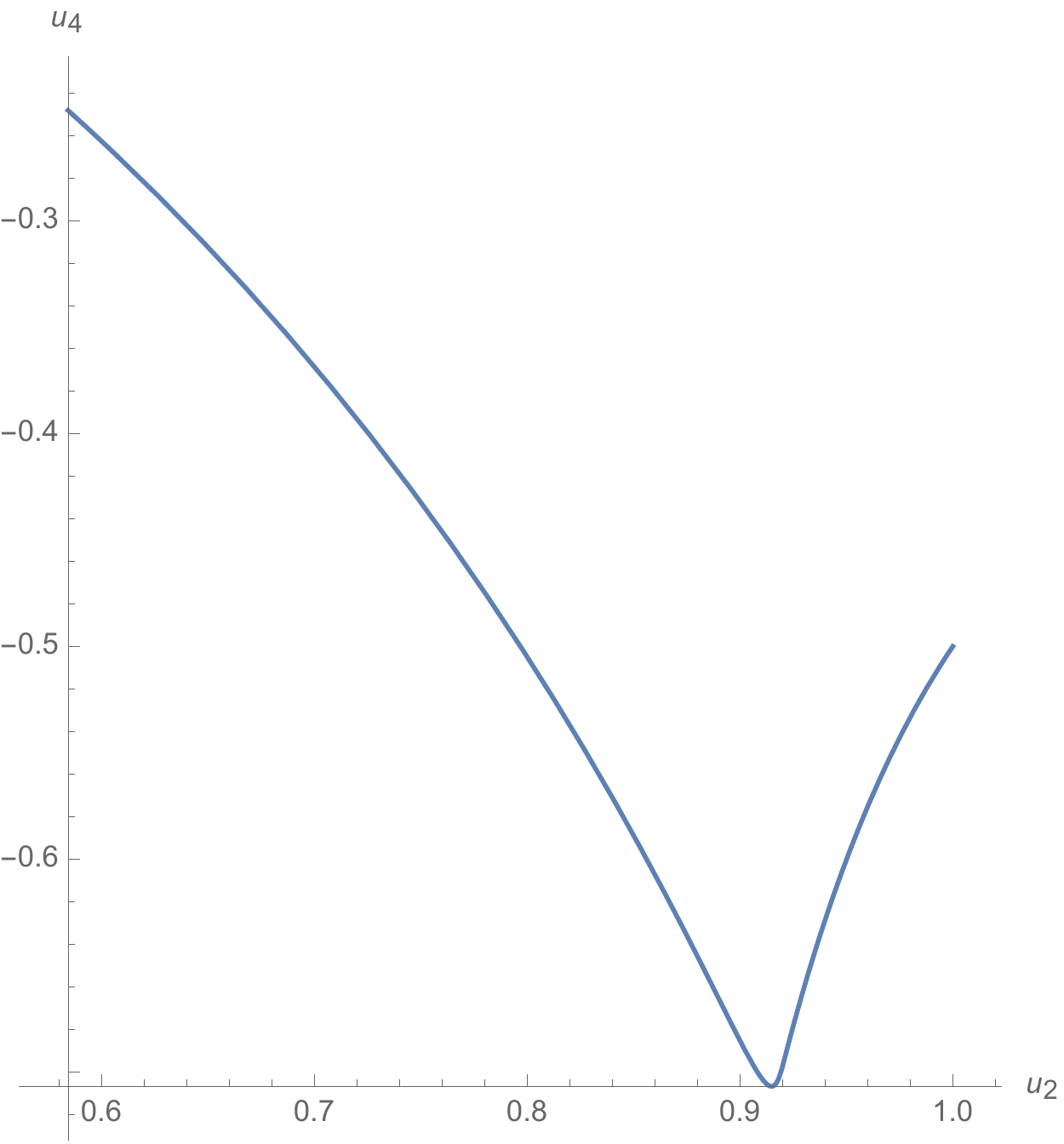}
	}

	\bigskip
	\subcaptionbox{$(u_2, u_6)$}[\linewidth]{%
		\includegraphics[scale=0.6]{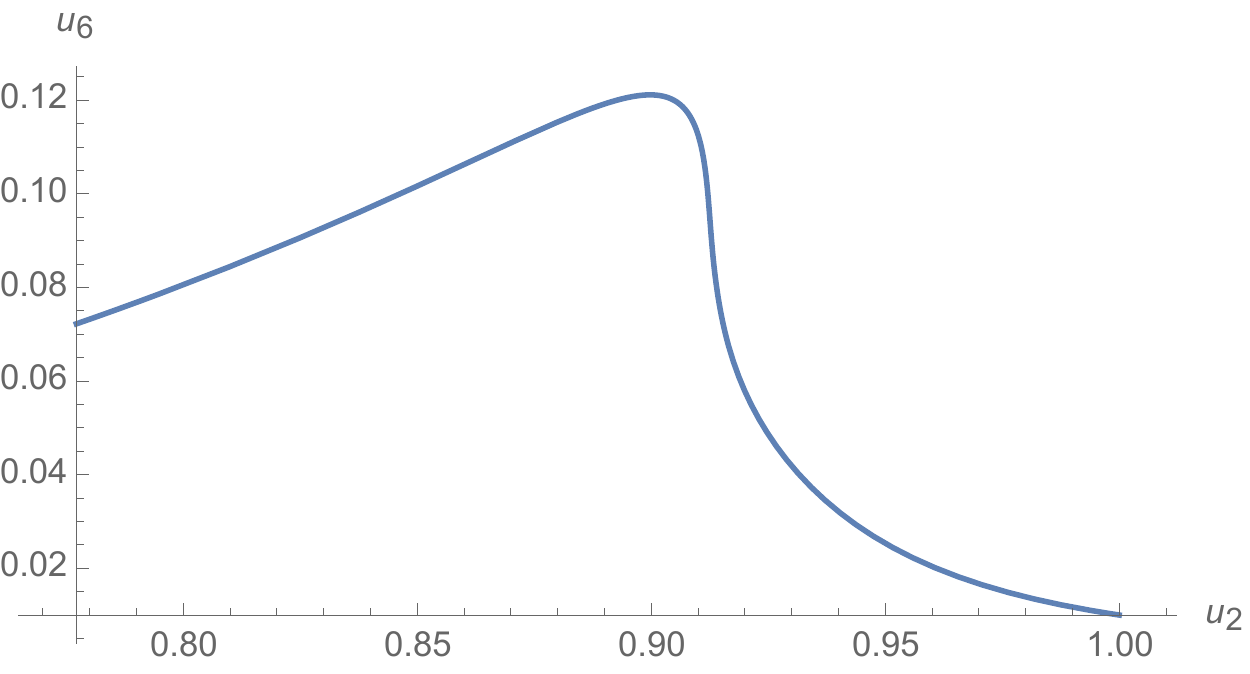}
	}

	\bigskip

	\subcaptionbox{$(u_4, u_6)$}[\linewidth]{%
		\includegraphics[scale=0.6]{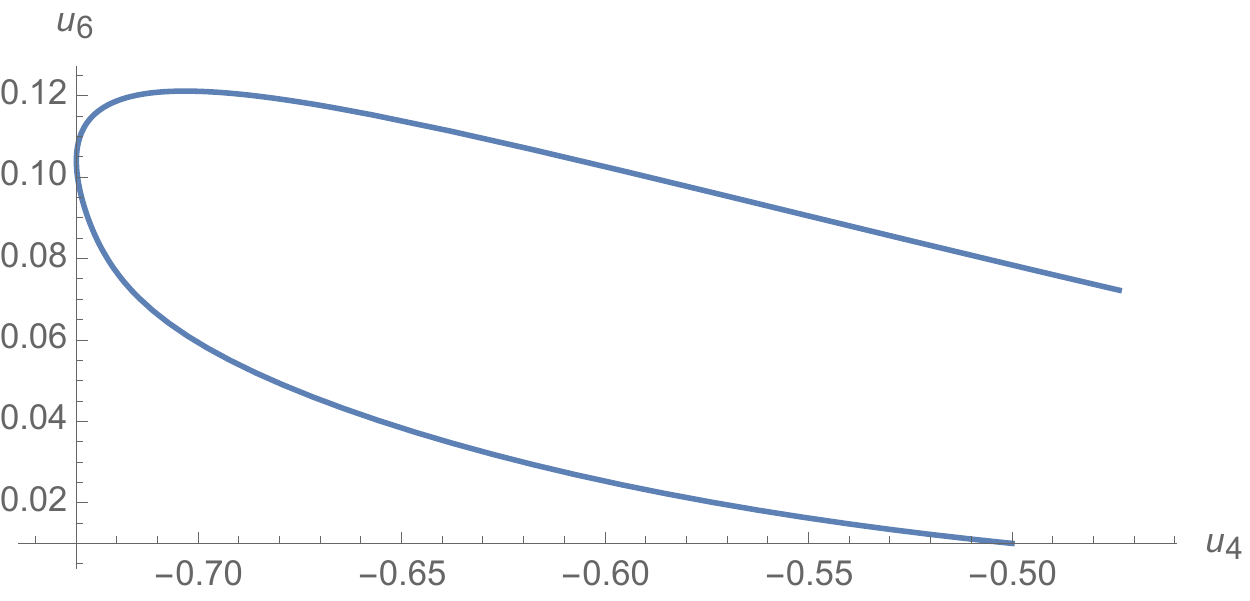}
	}

	\caption{Solution to the passive flow equations for $u_2 = 1$, $u_4 = - 0.5$, $u_6 = 0.01$.}
	\label{fig:flow-passive}
\end{figure}

From these equations, it is obvious that the behavior of the flow depends on the dimension $d_{\text{in}}$. For instance, for $d_{\text{in}}>4$, all the couplings are \emph{irrelevant} and trajectories return toward the Gaussian region, the $\bar{u}_2$ axis being the only direction of instability. In contrast, for $d_{\text{in}}<4$, some couplings become \emph{relevant}, and trajectories are repelled from the Gaussian region. $u_4$ is the first one to become relevant, for $d_{\text{in}}>3$; for $d_{\text{in}}<3$, $u_6$ becomes relevant as well. Figure \ref{flow} illustrates the behavior of the RG flow for several dimensions.
We have integrated numerically the flow equations for $u_2 = 1$, $u_4 = - 0.5$, $u_6 = 0.01$ in Figure~\ref{fig:flow-passive}.

\subsubsection{Beyond the symmetric phase}

In this section, we consider another approximation scheme for the effective potential $U_k$. Focusing on the IR regime, we assume that $\Psi(p)$ essentially reduces to its zero component (the macroscopic field):
\begin{equation}
\Psi(p) \sim \Psi_0 \delta_{0p}\,,
\end{equation}
and, defining $\chi:=\Psi_0^2/2$, we expand the effective potential per unit volume in power series around $\chi = \kappa(k)$:
\begin{equation}
\mathcal{U}_k[\Psi_0]=\frac{1}{2!} u_4(k) \bigg(\chi-\kappa(k)\bigg)^2+\frac{1}{3!} u_6(k) \bigg(\chi-\kappa(k)\bigg)^3+\cdots\,,
\end{equation}
Within this parametrization, we identify directly $\kappa$ with the (non-zero) vacuum, which runs with the scale $k$. The two-point function $\Gamma^{(2)}_k$ is moreover defined as:
\begin{equation}
\Gamma^{(2)}_k(p,p^\prime)=\left(Z(k)p^2+\frac{\partial^2 \mathcal{U}_k}{\partial \Psi_0^2 }\right) \delta(p+p^\prime)\,. \label{formulagood}
\end{equation}
Note that we introduced the field strength renormalization $Z(k)$ because its own flow is nonzero as soon as $\kappa\neq 0$, i.e.~broken phase effect introduces an anomalous dimension. As a technical device, we move the mass contribution in the effective potential. For a uniform field configuration, we must have:
\begin{equation}
\Gamma_k[\Psi_{\text{uniform}}]= \mathcal{V} \mathcal{U}_k[\Psi_{\text{uniform}}]\,.
\end{equation}
Therefore, taking the derivative with respect to $t:=\ln(k/\Lambda)$ ($\dot X:= k dX/dk$) and writing $\mathcal{U}_k^\prime (\chi) = \partial \mathcal{U}_k / \partial \Psi_0$, we get from \eqref{Wett}:
\begin{equation}
\dot{\mathcal{U}}_k[\Psi_0]=\frac{1}{2} \int \frac{dp}{(2\pi)^{d_{\text{in}}}} \, \dot{r}_k(p^2) \left( \frac{1}{\Gamma^{(2)}_k+r_k} \right)(p,-p)\,.\label{exactRGEbis}
\end{equation}
or, using the definition \eqref{formulagood}:
\begin{equation}
\dot{\mathcal{U}}_k[\chi]=\frac{1}{2}\, \int \frac{dp}{(2\pi)^{d_{\text{in}}}} \,  \dot{r}_k(p^2) \left( \frac{1}{Z(k)p^2+r_k(p^2)+\mathcal{U}_k^\prime (\chi)+ 2\chi \mathcal{U}_k^{\prime \prime}(\chi)} \right)\,.
\end{equation}
As in the previous section, we use the Litim regulator\footnote{Which is optimal in the following sense: the functional renormalization group (FRG) equations are defined only if the effective propagator $G$ is well-defined; that is, if $G^{-1}$ has no zero modes (and therefore do not develop IR divergences). This can be achieved by demanding that $G^{-1}$ has a sufficiently large gap, i.e.~a sufficiently large minimum. See~\cite{Litim:2001:OptimisedRenormalisationGroup} for more details.}, but modify it to deal with the running field strength $Z(k)$:
\begin{equation}
r_k(p^2)=Z(k)(k^2-p^2)\theta(k^2-p^2)\,,\label{Litimmod}
\end{equation}
leading straightforwardly to:
\begin{equation}
\dot{\mathcal{U}}_k[\chi]= \frac{\mathrm{Vol}(k)}{Z(k)k^2+\mathcal{U}_k^\prime (\chi)+ 2\chi \mathcal{U}_k^{\prime \prime}(\chi)}\,.
\end{equation}
For $k$ large enough, we may use the same integral approximation as for \eqref{volk}, but taking into account that $K_{d_{\text{in}}}$ must now depend on the anomalous dimension $\eta_k$ because of the factor $Z(k)$ in \eqref{Litimmod}:
\begin{equation}
\eta_k:= k\frac{d}{dk}\ln(Z(k))\,,
\end{equation}
such that:
\begin{equation}
\dot{\mathcal{U}}_k[\chi]= K_{d_{\text{in}}}(\eta_k) k^{d_{\text{in}}+2}\, \frac{1}{Z(k)k^2+\mathcal{U}_k^\prime (\chi)+ 2\chi \mathcal{U}_k^{\prime \prime}(\chi)}\,.\label{equationPotential}
\end{equation}
As in the previous section, we introduce dimensionless quantities (labeled with overlines) as:
\begin{equation}
\bar{\mathcal{U}}_k=(k^{2})^{-d_{\text{in}}/2} \, \mathcal{U}_k\,,\qquad \bar{\chi}=Z(k)(k^2)^{1-d_{\text{in}}/2} \chi\,.
\end{equation}
Note that all these changes of variable make sense from the requirement that all the terms in the potential must have the same dimension (the dimensions of $g$ and $h$ have been fixed). The derivative on the RHS in equation \eqref{equationPotential} is taken at $\chi$ fixed. Therefore, we have:
\begin{equation}
\dot{\mathcal{U}}_k[\chi]=(k^{2})^{d_{\text{in}}/2} \left[ \dot{\bar{\mathcal{U}}}_k[\bar{\chi}]+(d_{\text{in}}+\eta_k)\bar{\mathcal{U}}_k[\bar{\chi}] -(d_{\text{in}}-2) \bar{\chi} \frac{\partial}{\partial \bar{\chi}} \bar{\mathcal{U}}_k[\bar{\chi}]     \right]\,, \label{eqLagrange}
\end{equation}
where in the RHS the derivative is taken with $\bar\chi$ fixed. We obtain:
\begin{equation}
\dot{\bar{\mathcal{U}}}_k[\bar{\chi}]=-(d_{\text{in}}+\eta_k)\bar{\mathcal{U}}_k[\bar{\chi}] +(d_{\text{in}}-2) \bar{\chi} \frac{\partial}{\partial \bar{\chi}} \bar{\mathcal{U}}_k[\bar{\chi}] + \, \frac{K_{d_{\text{in}}}(\eta_k)}{1+\bar{\mathcal{U}}_k^\prime (\bar\chi)+ 2\bar{\chi} \bar{\mathcal{U}}_k^{\prime \prime}(\bar{\chi})}\,.
\end{equation}
The flow equations can be deduced from the normalization conditions at scale $k$:
\begin{equation}
\frac{\partial \mathcal{U}_k}{\partial \chi}\bigg\vert_{\chi=\kappa}=0\,, \qquad \frac{\partial^n \mathcal{U}_k}{\partial \chi^n}\bigg\vert_{\chi=\kappa}=u_{2n}\,. \label{rencouplings}
\end{equation}
Hence, because $\dot{\bar{\mathcal{U}}}_k[\bar{\chi}=\bar{\kappa}]=-\bar{g}\, \dot{\bar{\kappa}}$, we obtain for $\bar{\kappa}$,
\begin{align}
\beta_\kappa=-(d_{\text{in}}-2-\eta_k) \bar{\kappa} +K_{d_{\text{in}}}\,\frac{3+2\bar{\kappa} \frac{\bar{u}_6}{\bar{u}_4}}{(1+ 2\bar{\kappa} \bar{u}_4)^2}\,,
\end{align}
and after a tedious calculation, we obtain for $u_4$ and $u_6$:
\begin{align}
\beta_4&=-(4-d_{\text{in}}+2\eta_k) \bar{u}_4+(d_{\text{in}}-2) \bar{\kappa} \bar{u}_6-\,\frac{5K_{d_{\text{in}}}\bar{u}_6}{(1+2\bar{\kappa} \bar{u}_4)^2}+2K_{d_{\text{in}}}\,\frac{(3\bar{u}_4+2\bar{\kappa} \bar{u}_6)^2}{(1+ 2\bar{\kappa} \bar{u}_4)^3}\,,\\
\beta_6&=-(6-2d_{\text{in}}+3\eta_k) \bar{u}_6-6K_{d_{\text{in}}}\, \frac{(3\bar{u}_4+2\bar{\kappa}\bar{u}_6)^3}{(1+2\bar{\kappa} \bar{u}_4)^4}+20 K_{d_{\text{in}}}\bar{u}_6\,\frac{3\bar{u}_4+2\bar{\kappa}\bar{u}_6}{(1+ 2\bar{\kappa} \bar{u}_4)^3}.
\end{align}
The computation of the anomalous dimension is long and provided in Appendix \ref{App1}. The result is:
\begin{proposition}\label{prop1}
For the Litim's regulator \eqref{Litimmod}, the anomalous dimension $\eta_k$ in the local potential approximation with kinetic truncation up to order $p^2$ is given by:
\begin{equation}
\eta_k=-H(d_{\text{in}})\frac{(3 \bar{u}_4 \sqrt{2\bar{\kappa}} + \bar{u}_6 (2\bar{\kappa})^{3/2})^2}{(1+2\bar{\kappa} \bar{u}_4)^4}\,,
\end{equation}
where:
\begin{equation}
H(d_{\text{in}}):=\frac{\pi^{\frac{d_{\text{in}}+1}{2}}}{d_{\text{in}}+1}\frac{\Gamma\left(\frac{d_{\text{in}}}{2}\right)}{\Gamma \left(\frac{d_{\text{in}}+1}{2}\right)\Gamma \left(\frac{d_{\text{in}}-1}{2}\right)}\,.
\end{equation}
\end{proposition}

\subsection{\texorpdfstring{$1/a_0> k \gtrsim \xi^{-1}$: deep UV regime}{Deep UV regime}}\label{secUV}

In this section, we are aiming to discuss the deep UV regime $1/a_0> k \gtrsim (\xi)^{-1}$, where the expansion \eqref{expansionPropa} is not valid. For this regime, the derivative expansion breaks down as well, and a local approximation for interactions is no longer justified. We present a method, inspired from the Blaizot-Mendez-Wschebor (BMW) formalism~\cite{Blaizot:2006:NonPerturbativeRenormalization-1,Blaizot:2006:NonPerturbativeRenormalization-2,Blaizot:2007:NonPerturbativeRenormalizationGroup,Benitez:2012:NonperturbativeRenormalizationGroup}, which considerably improves the accuracy of truncations in regimes where the momentum-dependence of vertex functions is as relevant as the purely local ones, i.e.~relevant enough to invalidate the derivative expansion.
In this section, we summarize the essential results through three compact statements, in order to focus on the results, and leave the technical details in Appendix~\ref{App1}.

The procedure that we propose is based on the following three approximations (see also~\cite{Blaizot:2006:NonPerturbativeRenormalization-1,Blaizot:2006:NonPerturbativeRenormalization-2}):
\begin{enumerate}
\item We parametrize the $2$-point function $\Gamma_k^{(2)}(p,p^\prime)$ with a single parameter, the running mass $m^2(k)$, such that:
\begin{equation}
\Gamma_k^{(2)}(p,p^\prime):= \delta(p+p^\prime)\,m^2(k) \exp{\frac{p^2}{m^2}}\,,
\end{equation}
such that $\Gamma_k^{(2)}(p,p^\prime)$ reduces to the exact $2$-point function in the deep IR for $m^2(k=0)=\sigma_W^2/2 d_{\text{in}}$.

\item We assume that vertices are slowly varying with respect to the momenta $q$ running through the effective loops in the flow equation. The allowed windows of momenta being such that $q^2\lesssim k^2$, for $k$ small enough with respect to other momenta, we require:
\begin{equation}
\Gamma^{(n)}(p_1,p_2,..., p_{n-1}+q,p_n-q, \Psi_0)\approx \Gamma^{(n)}(p_1,p_2,..., p_{n-1},p_n, \Psi_0)\,,
\end{equation}
for some vacuum $\Psi_0$.

\item The third approximation is about the propagator entering in the flow equation. For $q$ in the windows of momenta allowed by $\partial_k r_k(q^2)$, we must have:
\begin{equation}
G_k((p+q)^2)\approx G_k(q^2)\theta \left(1-\alpha \,\frac{p^2}{k^2} \right)\,,
\end{equation}
where $\theta$ is the Heaviside step function and $\alpha$ a positive number, expected to be of order $1$.
\end{enumerate}
To complete these approximations we need to choose a suitable regulator. In principle, we could always use the Litim regulator (or any other regulator used in the literature). However, due to the parameterization of the phase space that we have chosen, and in particular the expression of the $2$-point function, this regulator loses its crucial advantage which consists in freezing all the fluctuations below the scale $k$. The Litim optimal condition is moreover expected to be a relevant constraint to define a regulator, especially in the symmetric phase, and we have the following statement:
\begin{claim}\label{prop2}
The scale dependent mass:
\begin{equation}
r_k(p^2)= m^2(k)\left(\exp\frac{k^2}{m^2(k)}-\exp\frac{p^2}{m^2(k)} \right)\theta(k^2-p^2)\,, \label{regulatorsuper}
\end{equation}
satisfies all the requirements for a regulator as soon as $m^2(k)>0$, freezes out all fluctuations with momentum $q^2<k^2$, and is optimal in Litim's sense.
\end{claim}
The physical discussion motivating this choice being a little technical, we provide it in Appendix \ref{App1}. Note that Litim's condition, which relies on the existence of an optimized gap for the inverse $2$-point function $\Gamma^{(2)}_k+r_k(p^2)$, is not an absolute criterion in regard to the reliability of the results. Indeed, some choices are expected to provide an optimal bound for the gap, which may have an influence on the computation of physical quantities like critical exponents~\cite{Canet:2003:OptimizationDerivativeExpansion,Lahoche:2020:RevisitedFunctionalRenormalization,Lahoche:2020:ReliabilityLocalTruncations}. Working within the set of “optimized regulators” in Litim's sense, we may complete the optimization argument with a principle of minimal sensitivity~\cite{Canet:2003:OptimizationDerivativeExpansion,Ball:1995:SchemeIndependenceExact,Comellas:1998:PolchinskiEquationReparameterization}, requiring that physical quantity has to be stationary with respect to some parameters spanning a family of regulators. This can be done for instance by replacing $r_k \to \beta r_k$, and to vary the physical quantities with respect to $\beta$. This will be the only optimization scheme that we will discuss in this paper.
\medskip

Within these approximations, the equation for the $2$-point function reads:
\begin{align*}
\dot{\Gamma}^{(2)}_k(p,-p)&=\int \frac{dq}{(2\pi)^{d_{\text{in}}}} \dot{r}_k(q) \bigg[ G(q^2) \Gamma^{(3)}_k(p,0,-p)G((q+p)^2)\Gamma^{(3)}_k(-p,p,0)G(q^2)\\
&\qquad -\frac{1}{2} G(q^2) \Gamma^{(4)}_k(p,-p,0,0)G(q^2) \bigg]\,,
\end{align*}
which, from the observation that $\Gamma_k^{(n+1)}(p_1,\cdots, p_n,0)\equiv \partial \Gamma_k^{(n)}(p_1,\cdots, p_n)/\partial \Psi_0$, leads to a closed equation for $\Gamma_k^{(2)}$:
\begin{equation}
\dot{\Gamma}^{(2)}_k(p,-p)=\int \frac{dq}{(2\pi)^{d_{\text{in}}}} \dot{r}_k(q) G^2(q^2) \bigg[ \left( \frac{\partial \Gamma^{(2)}_k(p,-p)}{\partial \Psi_0} \right)^2 G((p+q)^2)-\frac{1}{2}\frac{\partial^2\Gamma_k^{(2)}}{\partial \Psi_0^2} \bigg]\,.
\end{equation}
This is the standard BMW strategy. Our approach will however be a little different. First, we work in the symmetric phase $\Psi_0=0$. Second, we exploit the fact that the $2$-point function in our parametrization depends only on a single parameter (the mass) to close the hierarchy around the $6$-point function, thus removing the need for the usual assumption of a proportionality relation between the $6$ and $4$ points contributions in the flow equation of $\Gamma_k{(4)}$ (see~\cite{Blaizot:2006:NonPerturbativeRenormalization-1}). On the contrary, we will be able to deduce an expression for the $6$-point function from the knowledge of the $4$-point function, itself deduced from the flow equation of the $2$-point function. The only relevant parameters at sufficiently large times being the local parameters, $u_{2n}$, whose flow equations are deduced from the derivative expansion. The derivation of these equations being technical, we provide it in Appendix \ref{App1}, summarizing them in the following statement:

\begin{proposition}\label{prop3}
Truncating around sixtic interactions (i.e.~up to $\mathcal{O}(1/N^2)$ effects) in the deep UV, neglecting the momentum dependence of effective vertices in the computation of effective loops and for external momenta large enough, the flow equations for the local couplings $\bar{u}_2$, $\bar{u}_4$ and $\bar{u}_6$ are:
\begin{align}
\beta_2&=-2\bar{u}_2+2K_{d_{\text{in}}} \bar{u}_4 \,\frac{(1-\bar{u}_2)^2+\bar{u}_2(\bar{u}_2F_0-F_1)}{2\bar{u}_2^3e^{1/\bar{u}_2}-K_{d_{\text{in}}}\bar{u}_4 \left((1-\bar{u}_2)+(\bar{u}_2F_0-F_1)\right)}\\
\beta_4&=-(4-d_{\text{in}})\bar{u}_4-L_k(\bar{u}_2) \left(\frac{\bar{u}_6 e^{-2/\bar{u}_2}}{\bar{u}_2^2}-\frac{6\bar{u}_4^2e^{-3/\bar{u}_2}}{\bar{u}_2^3}\right)\,,\\
\beta_6&=-(6-2d_{\text{in}})\bar{u}_6-L_k(\bar{u}_2) \left(\frac{90 \bar{u}_4^3 e^{-4/\bar{u}_2}}{\bar{u}_2^4} -\frac{30 \bar{u}_6\bar{u}_4e^{-3/\bar{u}_2}}{\bar{u}_2^3}\right)\,,
\end{align}
where:
\begin{equation}
L_k(\bar{u}_2):=e^{1/\bar{u}_2}K_{d_{\text{in}}} \left[\frac{\beta_2+2\bar{u}_2}{2}\left(F_0-\frac{F_1}{\bar{u}_2}\right)-\left(1+  \frac{\beta_2+2\bar{u}_2}{2} \right)\left(1-\frac{1}{\bar{u}_2}\right)\right]\,,
\end{equation}
and $F_n(\bar{u}_2):=  d_{\text{in}}\int_{0}^{1} r^{d_{\text{in}}+2n-1} e^{\frac{r^2-1}{\bar{u}_2}}dr$. Furthermore, defining the minimal dimensionless vertex functions as ($x:=p/k$):
\begin{equation}
\Gamma_k^{(4)}(p,-p,0,0)=:k^{4-d_{\text{in}}}\bar{f}_k(x,0)\,,\qquad \Gamma_k^{(6)}(p,-p,0,0,0,0)=:k^{6-2d_{\text{in}}}\bar{h}_k(x) \,,
\end{equation}
we have, for $p$ large enough:
\begin{equation}
\bar{f}_k(x,0)= \frac{\bar{u}_2(\bar{u}_2-x^2)(2\bar{u}_2+\beta_2) e^{(1+x^2)/\bar{u}_2}}{K_{d_{\text{in}}} \left[\frac{\beta_2+2\bar{u}_2}{2}\left(F_0-\frac{F_1}{\bar{u}_2}\right)-\left(1+  \frac{\beta_2+2\bar{u}_2}{2} \right)\left(1-\frac{1}{\bar{u}_2}\right)\right]}\,,
\end{equation}
and:
\begin{equation}
\bar{h}_k(x)\approx \left( \frac{6(\bar{u}_4) e^{-1/\bar{u}_2}}{\bar{u}_2}-\frac{\bar{u}_2^2(4-d_{\text{in}})}{L_k(\bar{u}_2)} e^{2/\bar{u}_2}\right) \bar{f}_k(x,0)+e^{2/\bar{u}_2} \bar{u}_2^2 \frac{2x^2 \bar{f}^\prime_k(x,0)-\dot{\bar{f}}_k(x,0)}{L_k(\bar{u}_2)}\,.
\end{equation}
\end{proposition}
It is moreover interesting to note that for external momenta large enough, the knowledge of $\bar{f}_k(x,0)$ allows reconstructing the $4$-point function. Once again, we put the proof in Appendix \ref{App1}, and summarize the result in a compact statement:
\begin{claim}\label{prop4}
For momenta large enough ($p_i^2\sim (\xi)^{-2}(k)$), the $4$-point vertex $\Gamma_k^{(4)}(p_1,p_2,p_3,p_4)$ can be suitably approximated as:
\begin{equation}
\Gamma_k^{(4)}(p_1,p_2,p_3,p_4)\approx \delta \left(\sum_{i=1}^4 p_i\right)  \sum_{j=2}^4 \gamma_k(p_1+p_j)\,,
\end{equation}
with
\begin{equation}
\gamma_k(p):=\frac{1}{2} k^{4-d_{\text{in}}}\bar{f}_k\left(\frac{p}{k},0\right)-\frac{1}{6}u_4\,.
\end{equation}
\end{claim}

\section{Flowing though the neural network space: the active RG}\label{sec4}

In this section, following the discussion of section \ref{secRenNN}, we consider the active RG, viewing the networks parameters $2 \sigma_W^2 / d_{\text{in}}$ as an UV cut-off rather than a running mass. First, we derive the corresponding flow equation. We show that $n$-point functions exhibit a purely scaling behavior, and that the corresponding $\beta$-functions reduce to the linear dimensional contributions. As discussed in section \ref{secRenNN}, this RG is formally the same as the one used in~\cite{Halverson:2021:NeuralNetworksQuantum}, up to the lack of explicit scaling for mass, explaining why our scaling dimensions are different. Second, we investigate the content of the flow equations that we obtained. As pointed out in the section \ref{secRenNN}, the major advantage of this approach is to avoid introducing a working precision, or any special structure regarding the nature of the data.

Let us consider a network $(\sigma_W,\sigma_b)$, and  define $\Lambda^2:= 2 \sigma_W^2 / d_{\text{in}}$. Within this suggestive notation, the exact propagator looks like a UV regularized free propagator:
\begin{equation}
K_{\Lambda}(p^2):= \frac{e^{-p^2/\Lambda^2}}{\Lambda^2}\,,
\end{equation}
$\Lambda$ playing the role of a UV cut-off, which suppresses large momenta. The question is therefore: what happen if we smoothly change the parameter $\sigma_W$?\footnotemark{}
\footnotetext{%
	In fact, this situation is familiar in string field theory, where $\Lambda$ is called the stub parameter~\cite{Brustein:1991:RenormalizationGroupEquation,Brustein:1992:SpacetimeWorldsheetRenormalization,Sen:2016:WilsonianEffectiveAction,Erbin:2021:QFTStubs}.
}%
Formally, this is equivalent to moving the UV cut-off, which can be translated as a chain of equivalence relations between classical actions through the differential equation \eqref{flowPol}, all of them having the same long distance physics. One can think for the evolution equation of the classical action to something like equation \eqref{flowPol}, i.e.
\begin{equation}
\Lambda \frac{d \tilde{\mathcal{V}}_{\Lambda}}{d\Lambda}=-\frac{1}{2} \int \frac{dp}{(2\pi)^{d_{\text{in}}}}  \Lambda \frac{dK_{\Lambda}}{d\Lambda}(p^2) \Big( \frac{\delta^2 \tilde{\mathcal{V}}_{\Lambda} }{\delta \phi(p)\delta\phi(-p)} -\frac{\delta \tilde{\mathcal{V}}_{\Lambda(s)}}{\delta \phi(p)} \frac{\delta \tilde{\mathcal{V}}_{\Lambda(s)}}{\delta \phi(-p)}  \Big)\,. \label{flowPol2}
\end{equation}
Such an equation however assumes that $K_{\Lambda}$ is the free propagator. Rather, in our construction, it has to be understood as the effective propagator, taking into account fluctuations. Therefore, we have to construct a coarse graining with a fixed shape of the effective propagator, the corresponding free propagator remaining unknown. There is a pragmatic way to do this. We formally introduce a regulator $\Delta S_k$ in the classical action. This leads to the Wetterich equation \eqref{Wett}, but with the additional constraint that:
\begin{equation}
\Gamma_k^{(2)}(p_1,p_2)+r_k(p^2)\delta(p_1+p_2)\equiv k^2\exp \Big( \frac{p_1^2}{k^2}\Big)\delta(p_1+p_2)\,. \label{Twopointcondition}
\end{equation}
This equation simply means that we relate the running scale $k$ to the standard deviation of the neural network weights as
\begin{equation}
	\label{eq:relation-k-sigmaw}
	k^2
		= \frac{2 \sigma_W^2}{d_{\text{in}}}\,,
\end{equation}
and that we keep the shape of the $2$-point function fixed along the RG trajectory (if it exists), fixing $\Gamma_k^{(2)}(p,-p)$ as soon as $r_k(p^2)$ is given. Let us show how this condition allows closing the hierarchical flow equations. Let us consider the flow equation \eqref{flowmassNew} for $\Gamma_{k}^{(2)}(p_1, p_2)$. Neglecting the momentum dependence of the effective vertex $\Gamma_k^{(4)}(p_1, p_2,q,-q)$ with respect to the momenta "$q$" running through the effective loop following the discussion of Section \ref{secUV}, we get:
\begin{equation}
\dot{\Gamma}_{k}^{(2)}(p,-p)\approx  -\frac{1}{2} \Gamma_k^{(4)}(p, -p,0,0) \int \frac{dq}{(2\pi)^{d_{\text{in}}}}\dot{r}_k(q^2)  G_k^2(q^2) \,.\label{flowmassNew}
\end{equation}

\begin{remark}
Note that this approach implicitly assumes $k$ is small enough to justify the replacement: $ \Gamma_k^{(4)}(p, -p,q,-q)\to  \Gamma_k^{(4)}(p, -p,0,0)$
in \eqref{flowmassNew}. Hence, the resulting flow equations are expected to be exact for reference scales in the IR.
\end{remark}

Because the LHS can be explicitly computed from \eqref{Twopointcondition}, we therefore obtain:
\begin{equation}
\Gamma_k^{(4)}(p, -p,0,0) \approx-2k^4 \frac{2(k^2-p^2)\exp \Big( \frac{p^2}{k^2}\Big)-\dot{r}_k(p^2)}{\int \frac{dq}{(2\pi)^{d_{\text{in}}}}\dot{r}_k(q^2)  \exp \Big(\frac{-2q^2}{k^2}\Big)} \delta(0)\,.\label{fourpointtrue}
\end{equation}
In the same way, the flow equation for $\Gamma_k^{(4)}(p, -p,0,0)$ allows in principle to compute $\Gamma_k^{(6)}(p, -p,0,0,0,0)$ within the same approximation. Let us illustrate how this works. Let us consider a given network defining the “fundamental scale” $k\equiv \Lambda_0$. We can measure the $4$-point function at zero momentum $\Gamma_{k=\Lambda_0}^{(4)}(0, 0,0,0)\equiv u_4(\Lambda_0) \delta(0)$. This condition in turn fixes the value of $\dot{r}_k(0)$. For instance, let us consider the following explicit example, working with the slightly modified Litim's regulator:
\begin{equation}
r_k(p^2)=\alpha(k^2-p^2) \theta(k^2-p^2)\,.
\end{equation}
Straightforwardly, we have $r_k(0)=\alpha k^2$ and $\dot{r}_k(0)=2\alpha k^2$, and the previous equality reads as:
\begin{equation}
u_4(\Lambda_0)=\Lambda_0^{4-d_{\text{in}}} \left( -2\frac{1-\alpha}{\alpha} \frac{1}{I_2}\right)\,,
\end{equation}
where we introduced the dimensionless variable $x := q/k$, and
\begin{equation}
I_2:=\int \frac{dx}{(2\pi)^{d_{\text{in}}}}\theta(1-x^2) e^{-2x^2}\,.
\end{equation}
Introducing the dimensionless coupling $\bar{u}_4:=\Lambda_0^{d_{\text{in}}-4} u_4$, and solving on $\alpha$, we thus obtain:
\begin{equation}
\alpha= \frac{1}{1-\frac{\bar{u}_4I_2}{2}}\,.\label{condalpha}
\end{equation}
Because $u_4=\mathcal{O}(1/N)$, and $I_2$ is a pure number, $\alpha$ is close to $1$ for large $N$. However, $\alpha$ increases as $N$ decreases, and for $\bar{u}_4\sim 2/I_2$ the approximation breaks down. Note that we could expect this not to be a limitation of the approach in itself, but a limitation of the Litim regulator, however, a moment of reflection shows that such a singular behavior is in fact very general, and independent on the choice of the regulator. Under the condition \eqref{condalpha}, the problem \eqref{fourpointtrue} is well posed but trivial: it reduces to a pure scaling behavior. Indeed, given \eqref{fourpointtrue}, we have:
\begin{equation}
u_4(k)=k^{4-d_{\text{in}}} \left( -2\frac{1-\alpha}{\alpha} \frac{1}{I_2}\right)=\Big(\frac{k}{\Lambda_0}\Big)^{4-d_{\text{in}}} u_4(\Lambda_0)\,. \label{dimensional}
\end{equation}
The flow is entirely fixed by dimensional analysis, and the flow equation for $u_4$ reduces to its linear contribution:
\begin{equation}
k \frac{du_4}{dk}=(4-d_{\text{in}}){u}_4\,.
\end{equation}
In turn, this equation determines $\Gamma_k^{(6)} \sim (\Gamma_k^{(4)})^2\sim \mathcal{O}(1/N^2)$. The effective loop behaves like $k^{6-2d_{\text{in}}}$, times a factor which is $k$ independent.
Hence, we deduce:
\begin{equation}
k \frac{du_6}{dk}=(6-2d_{\text{in}}) {u}_6\,,
\end{equation}
meaning that $u_6$ follows a purely scaling behavior as well.

\begin{figure}[p]
	\centering

	\subcaptionbox{$N = 2$}[.45\linewidth]{%
		\includegraphics[width=\linewidth]{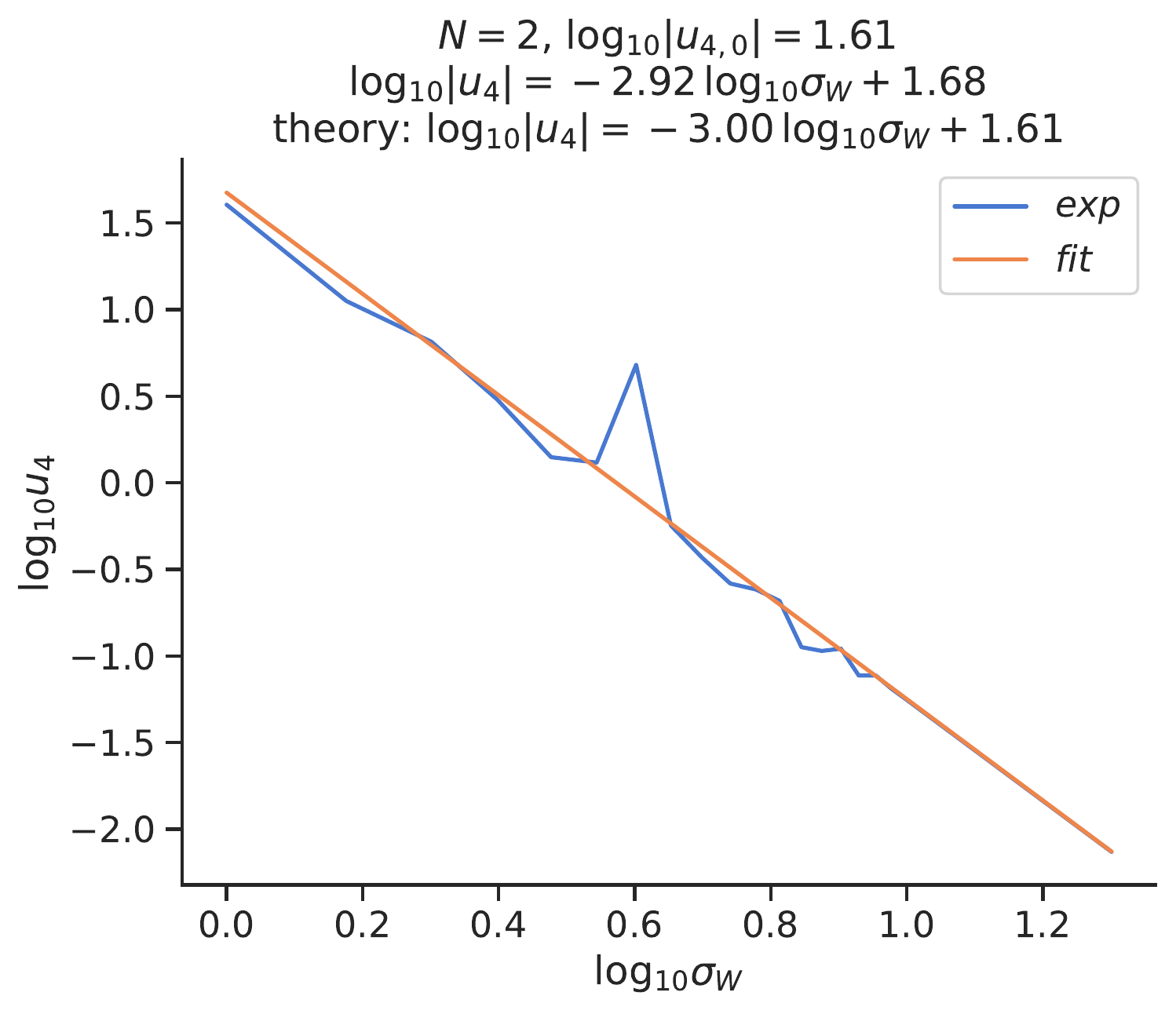}
	}
	\qquad
	\subcaptionbox{$N = 3$}[.45\linewidth]{%
		\includegraphics[width=\linewidth]{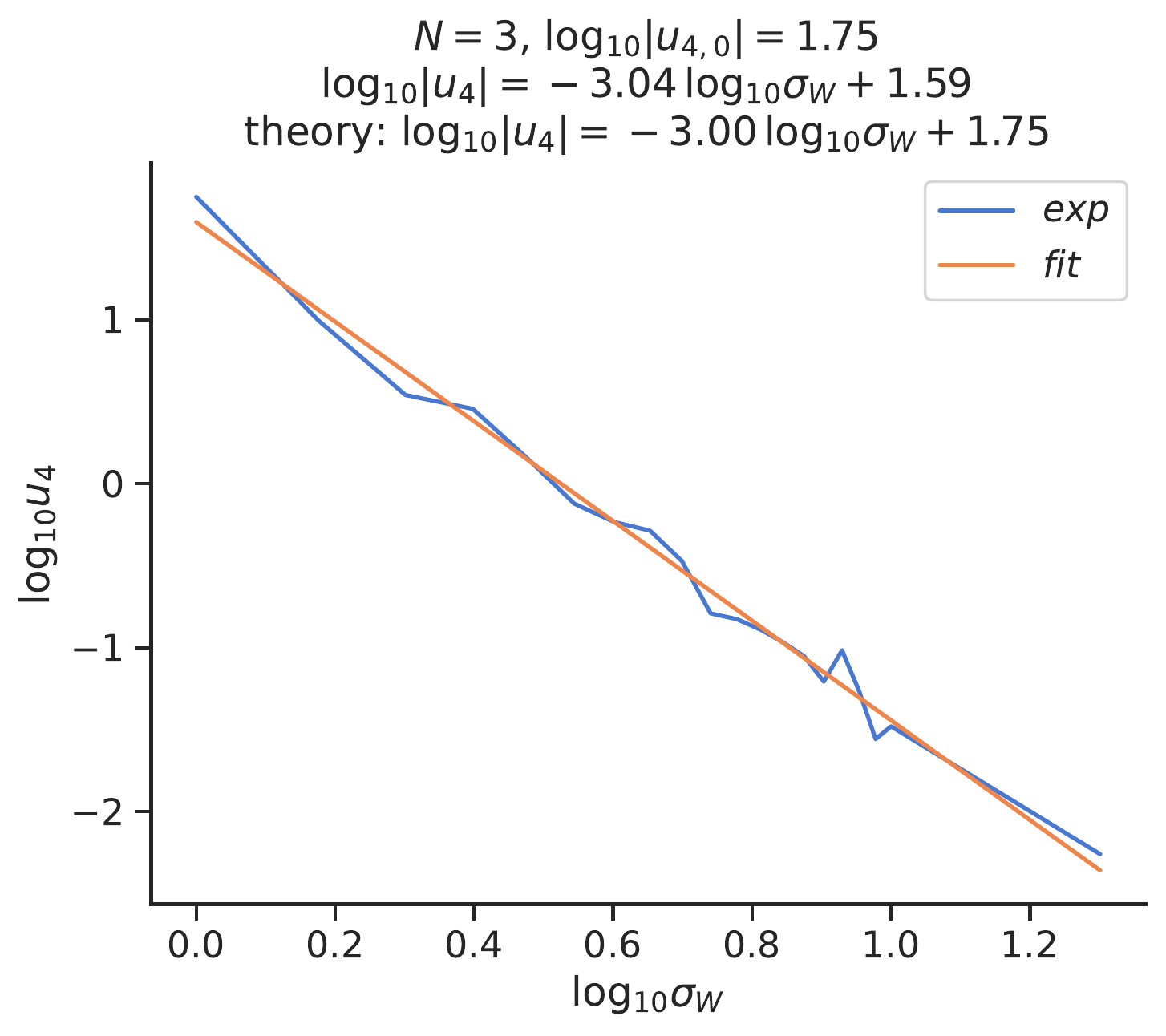}
	}

	\bigskip

	\subcaptionbox{$N = 4$}[.45\linewidth]{%
		\includegraphics[width=\linewidth]{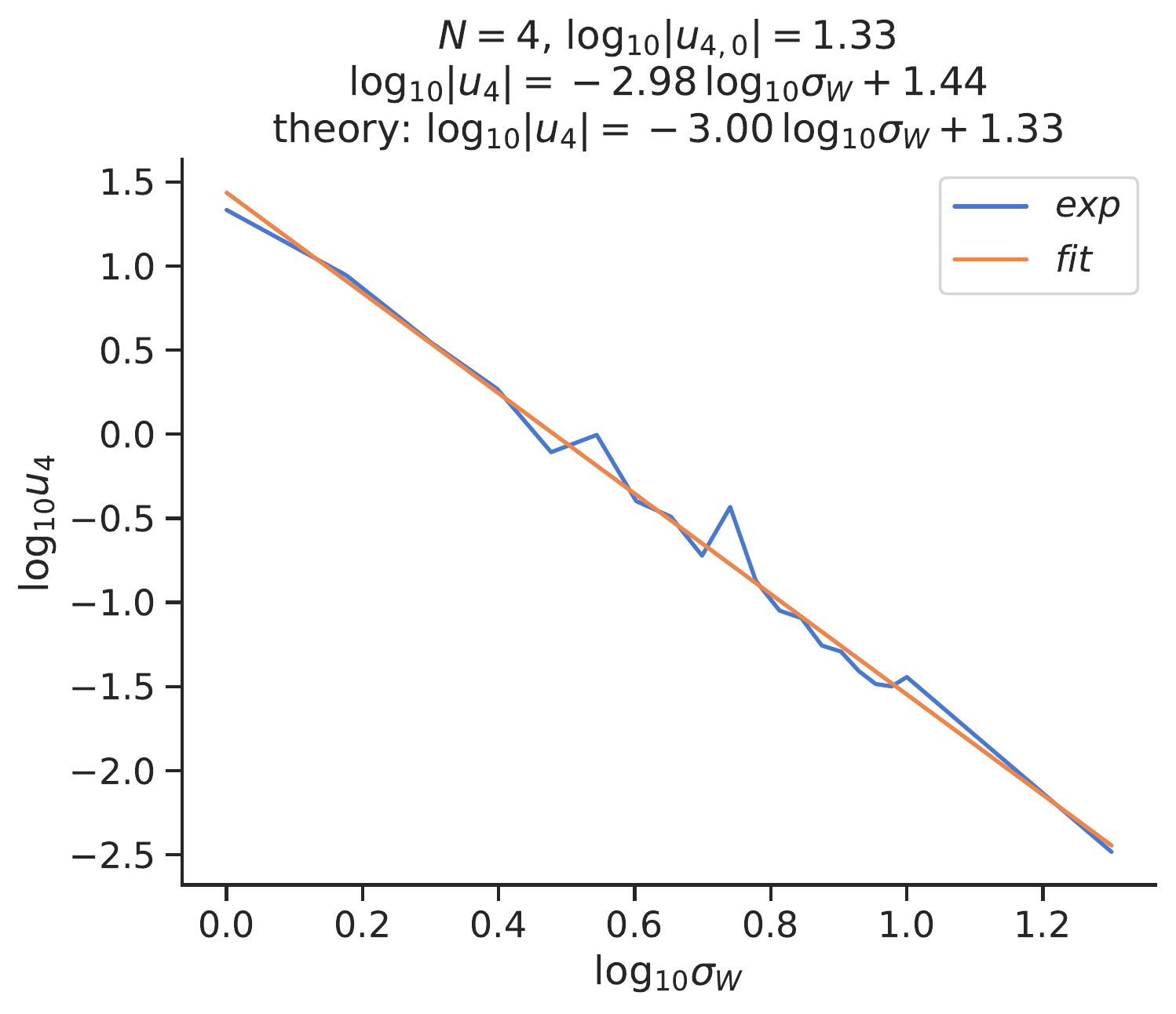}
	}
	\qquad
	\subcaptionbox{$N = 5$}[.45\linewidth]{%
		\includegraphics[width=\linewidth]{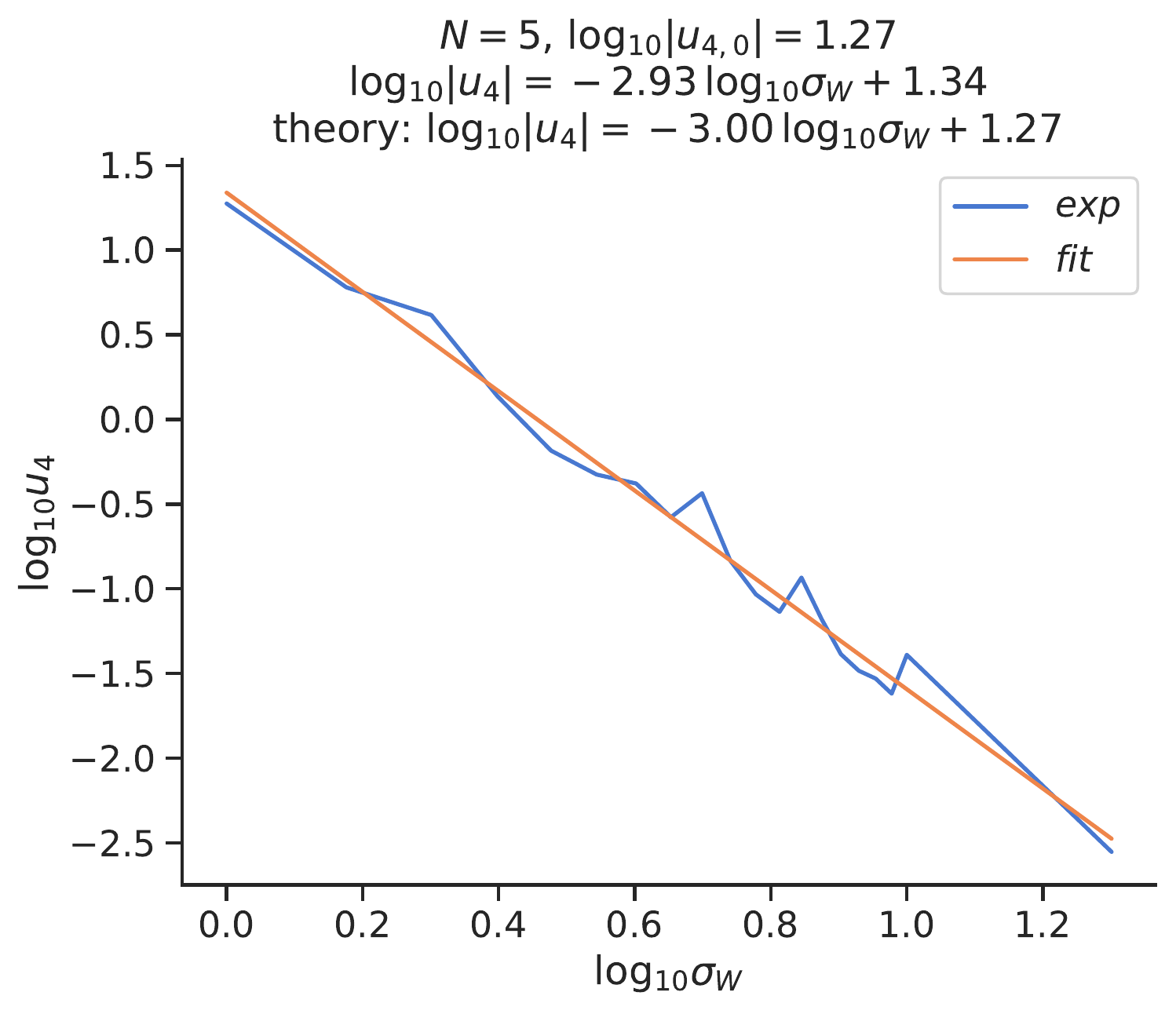}
	}

	\bigskip

	\subcaptionbox{$N = 10$}[.45\linewidth]{%
		\includegraphics[width=\linewidth]{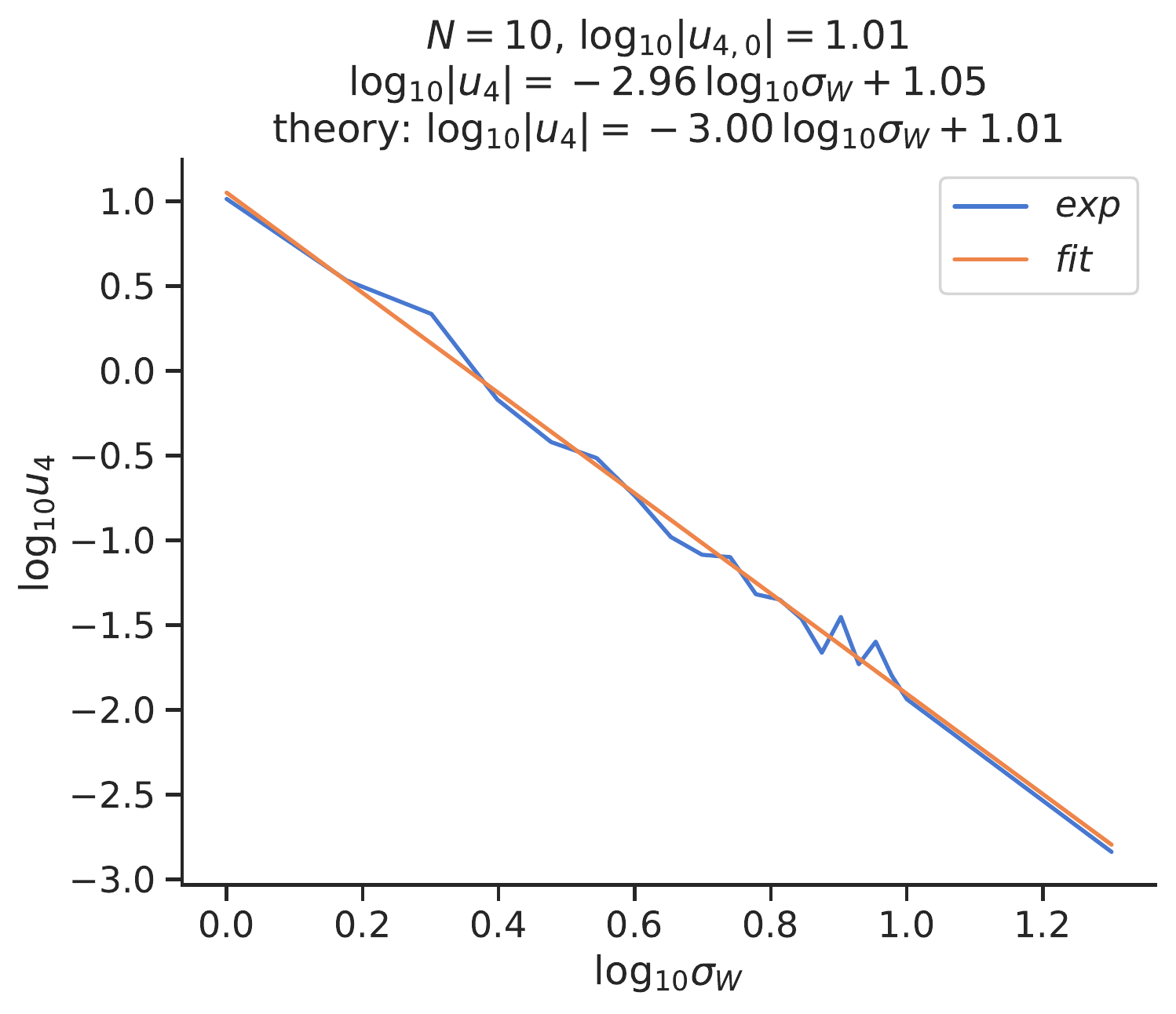}
	}
	\qquad
	\subcaptionbox{$N = 20$}[.45\linewidth]{%
		\includegraphics[width=\linewidth]{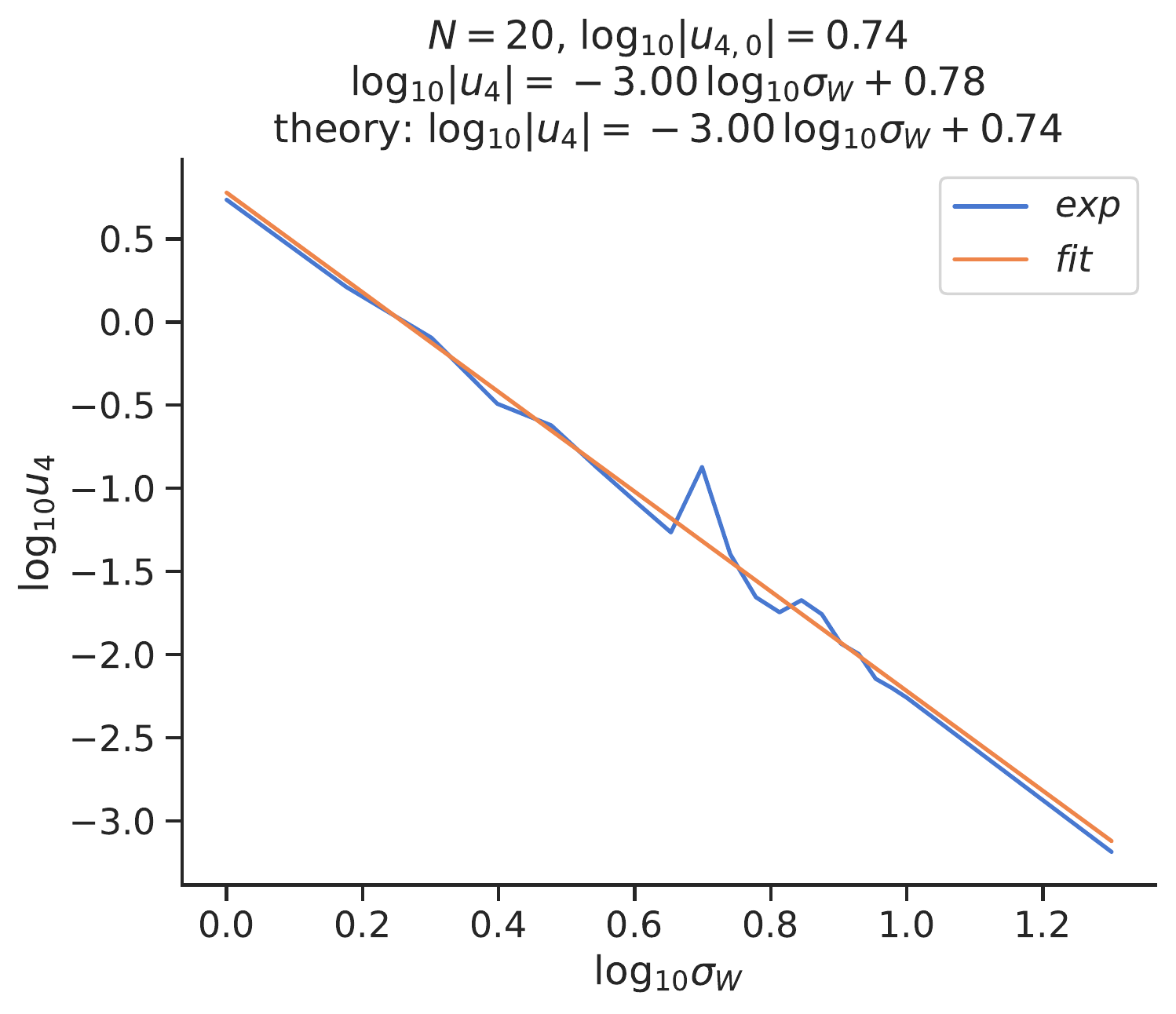}
	}

	\caption{Values of the averaged coupling constant $\langle |u_4| \rangle$ for $N = 2, 3, 4, 5, 10, 20$.}
	\label{fig:u4-fit-1}
\end{figure}

We can use \eqref{eq:relation-k-sigmaw} to write the flow equations in terms of the standard deviation $\sigma_W$:
\begin{equation}
	\label{eq:flow-active-sigmaw}
	\sigma_W \frac{d u_4}{d \sigma_W} = (4 - d_{\text{in}}) u_4 \,,
	\qquad
	\sigma_W \frac{d u_6}{d \sigma_W} = (6 - 2 d_{\text{in}}) u_6 \,,
\end{equation}
where now $u_4$ and $u_6$ are seen as functions of $\sigma_W$.
As displayed in Figures~\ref{fig:u4-fit-1} and~\ref{fig:u4-fit-2}, the numerical simulations match to a good precision with the solution to this equation (see Appendix~\ref{app:NumSim} for the computations of $u_4$).

\begin{remark}
	Finally, let us make a remark in regard to the results obtained in the reference \cite{Halverson:2021:NeuralNetworksQuantum}. Indeed, the authors arrived to the equations \eqref{eq:flow-active-sigmaw} with $\sigma_W$ replaced by an IR cut-off from perturbation theory, whose validity assumes $N$ to be large enough. What is puzzling with this calculation is that the RG predictions work even for small $N$, where we expect that perturbation theory breaks down. Our derivation solves this paradox: using a non-perturbative framework, we are able to show that coupling constants follow scaling laws \eqref{dimensional}, without assumption on the sizes of the coupling constants (however, note that both derivations have been performed with different activation functions, such that it would be interesting to check how general \eqref{eq:flow-active-sigmaw} is).
	\end{remark}

\begin{figure}[p]
	\centering

	\subcaptionbox{$N = 50$}[.45\linewidth]{%
		\includegraphics[width=\linewidth]{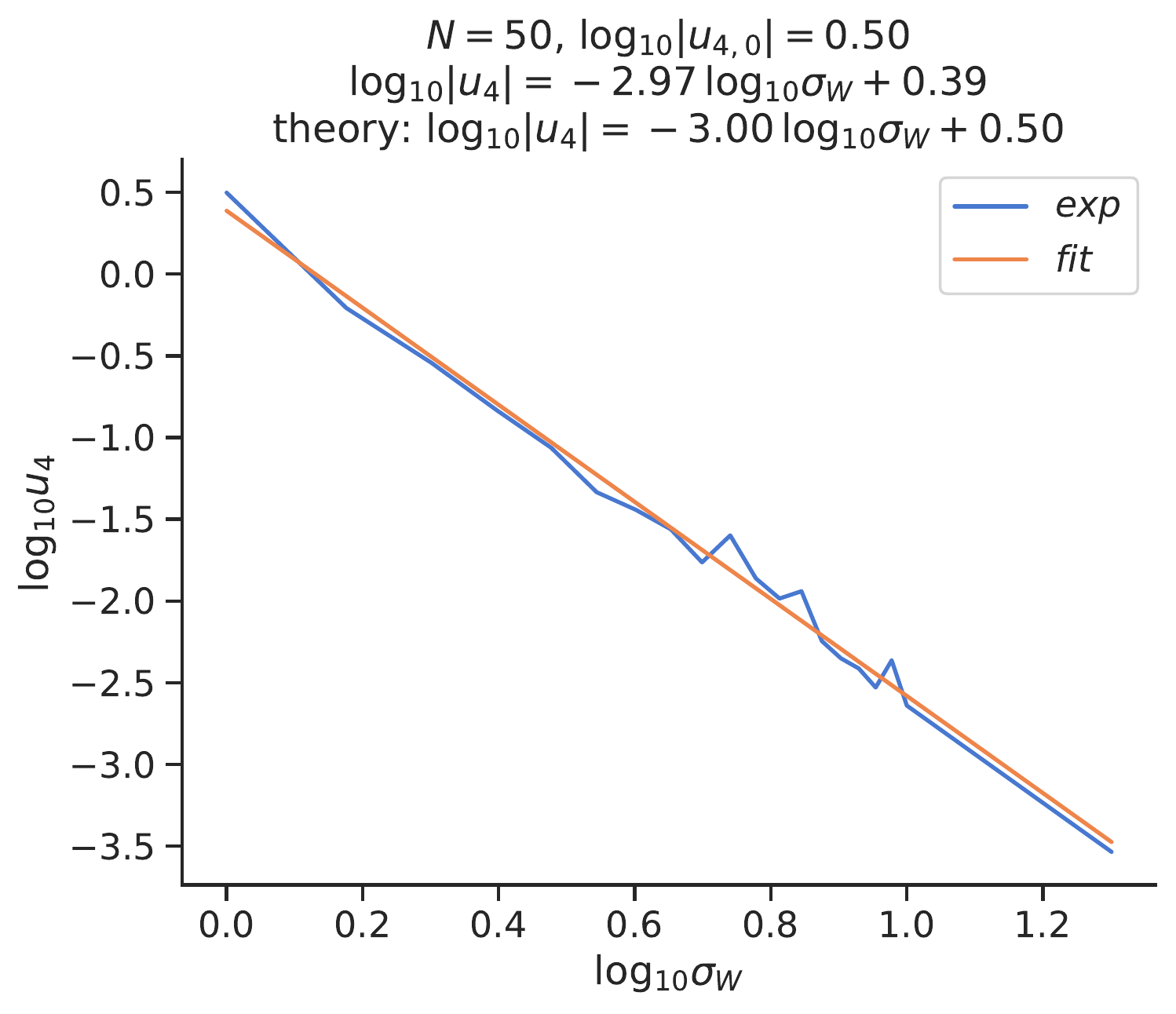}
	}
	\qquad
	\subcaptionbox{$N = 100$}[.45\linewidth]{%
		\includegraphics[width=\linewidth]{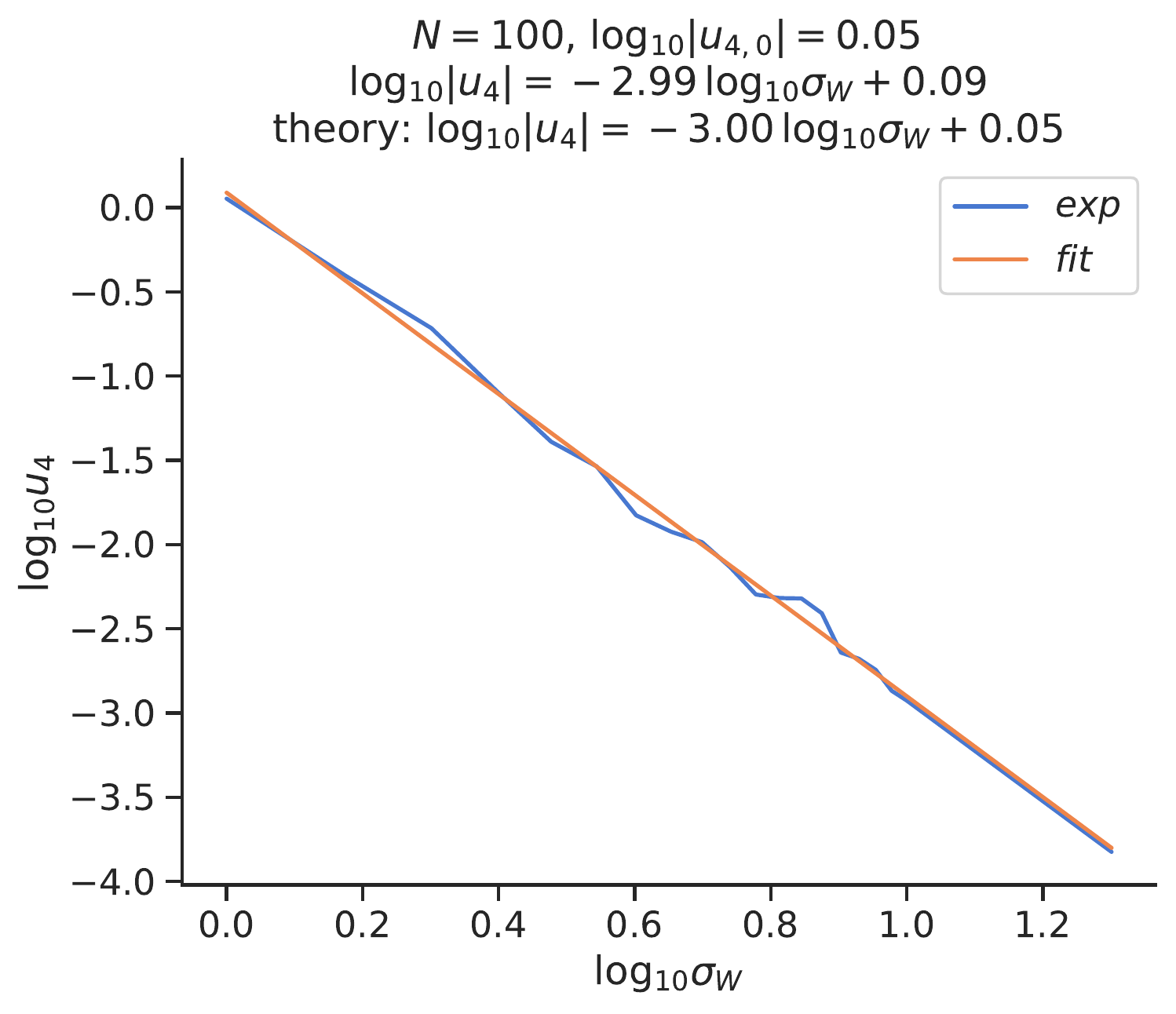}
	}

	\bigskip

	\subcaptionbox{$N = 500$}[.45\linewidth]{%
		\includegraphics[width=\linewidth]{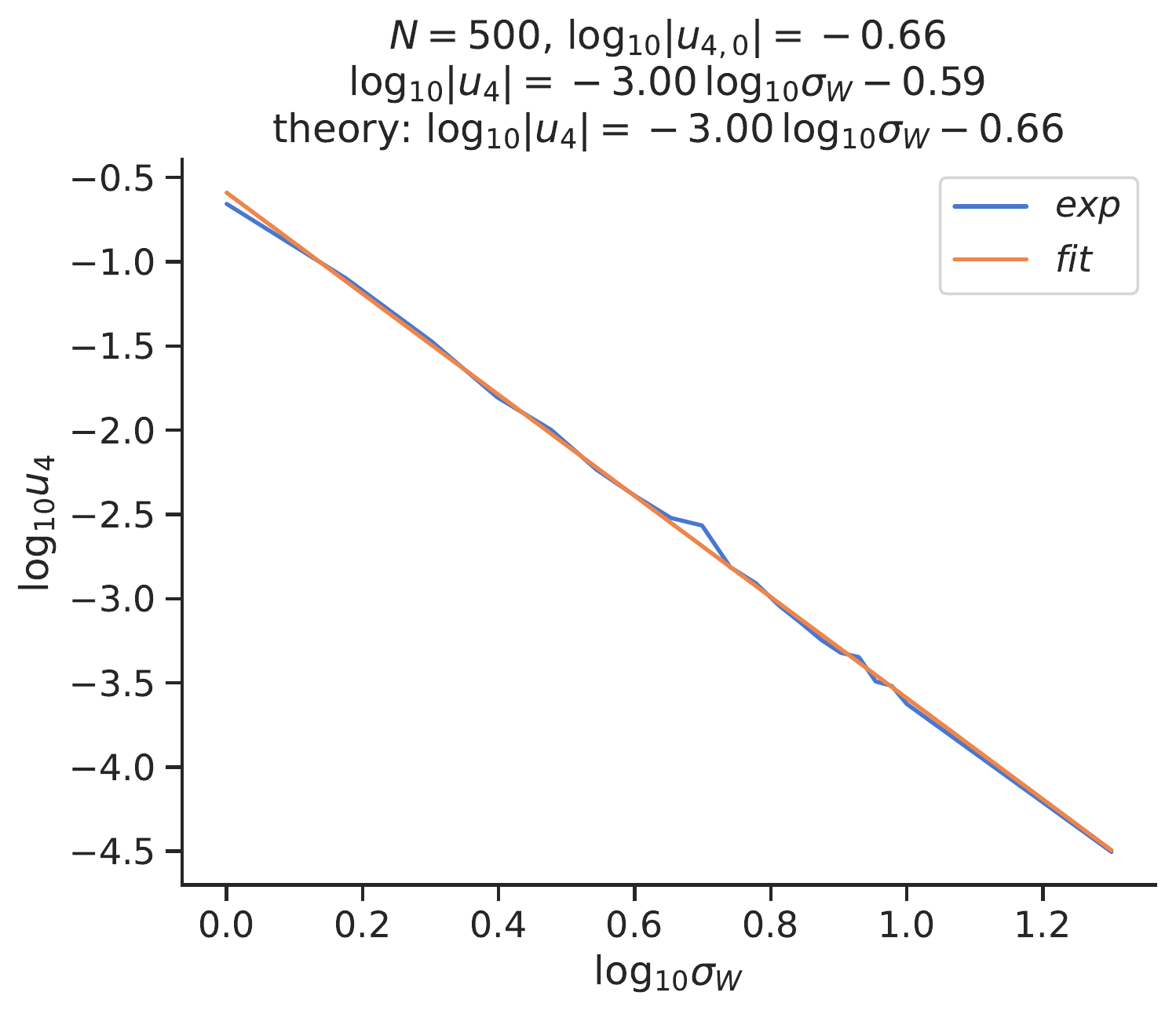}
	}
	\qquad
	\subcaptionbox{$N = 1000$}[.45\linewidth]{%
		\includegraphics[width=\linewidth]{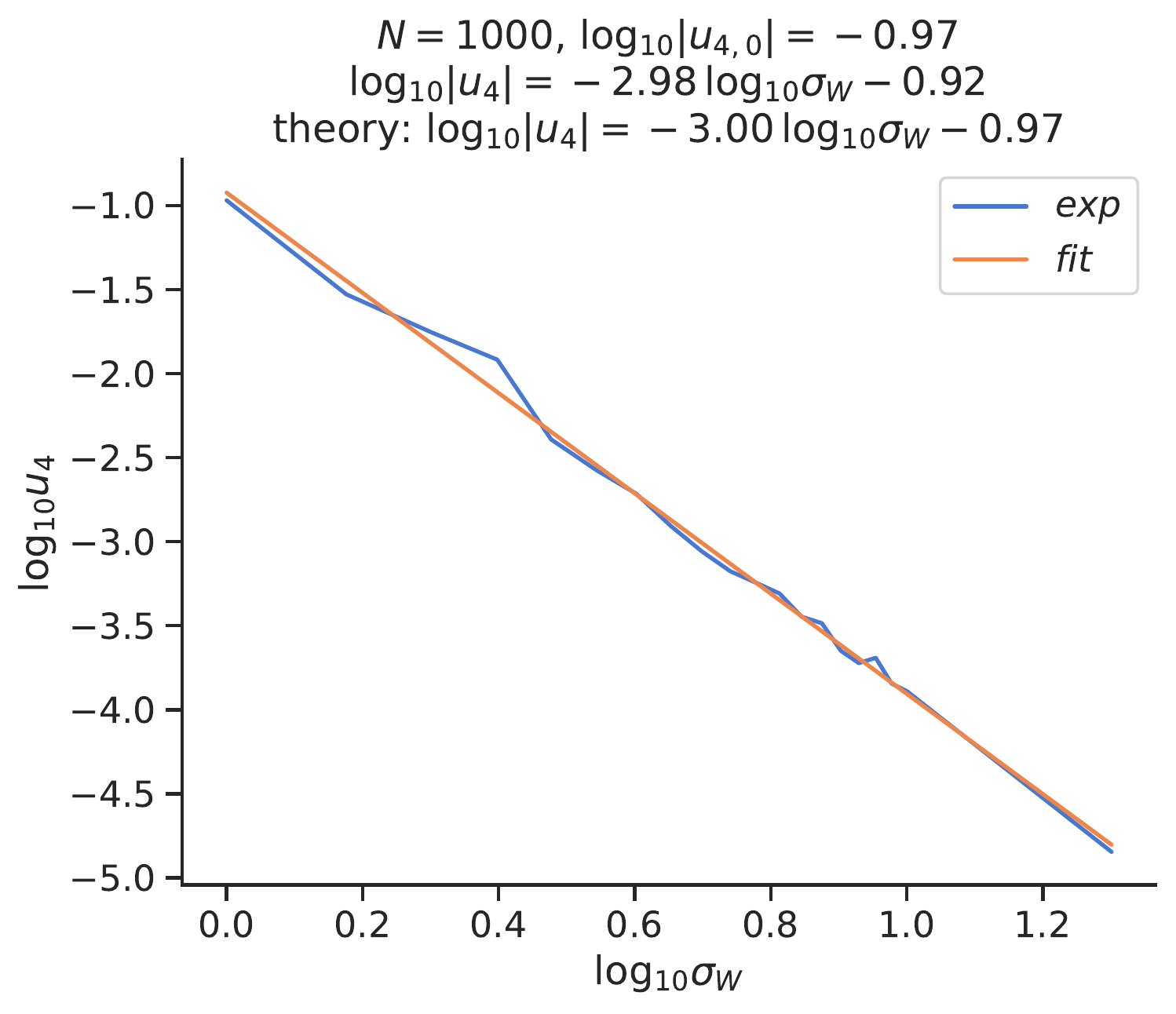}
	}

	\caption{Values of the averaged coupling constant $\langle |u_4| \rangle$ for $N = 50, 100, 500, 1000$.}
	\label{fig:u4-fit-2}
\end{figure}

\section{Conclusion and outlooks}

In this paper, we have pushed further the use of the renormalization group for the NN-QFT correspondence~\cite{Halverson:2021:NeuralNetworksQuantum,Maiti:2021:SymmetryviaDualityInvariantNeural}, which states that a neural network can be represented by a quantum field theory.
In the infinite limit of the hidden layer width $N$, the neural network is described by a Gaussian process and mapped to a free field theory, and interactions translate finite-$N$ corrections.
The main difference with usual QFTs used in physics stems from the choice of the kernel (or propagator), itself inherited from the choice of activation function in the neural network.
Since it encodes important properties on the theory (IR and UV divergences, scaling, etc.), it is important to ask how the data-space of the neural network inputs differ from usual spacetime.
As a consequence, the usual assumptions on interaction locality may not be appropriate.
In the first part of this paper, we have discussed several of these aspects, providing an interpretation slightly different from the one in~\cite{Halverson:2021:NeuralNetworksQuantum}.\footnote{However, let us stress that this divergence in interpretation does not change in any way the computations and numerical results in both sides.}

We have then described how to build a non-perturbative renormalization group flow following Wetterich-Morris formalism.
We introduce two different points of view: in the passive case, the UV cut-off is related to the data resolution, while in the active case, it is given in terms of the standard deviation of the NN weights.
The main difference with~\cite{Halverson:2021:NeuralNetworksQuantum} is that they postulate a global scale invariance with respect to a large volume cut-off (IR, in the language of our paper) on the data-space.
Intriguingly, their results agree strongly with numerical simulations, for the small widths, whereas perturbation theory is expected to fail, and even if a scale invariance with respect to the volume is not expected.
In this paper, we solve this paradox by developing a renormalization group based on an explicit coarse-graining, and derive flow equations from a process of partial integration of the field degrees of freedom.
We find that the active point of view is formally identified with the flow in~\cite{Halverson:2021:NeuralNetworksQuantum}, thus justifying, in an explicitly non-perturbative framework, the agreement between theory and experience found in the paper.
A natural extension of this work is to include non-local interactions using tensor models.
Another possible direction is to generalize the derivation to other networks such as ReLU-net~\cite{Halverson:2021:NeuralNetworksQuantum}.

On the numerical side, the main result of our paper is the flow equation \eqref{eq:flow-active-sigmaw} which shows that the weight standard deviation $\sigma_W$ can be interpreted as a running cut-off in terms of which the couplings of the NN-QFT change.
This means that given the couplings for a specific value of $\sigma_W$, it is possible to compute analytically the couplings for any other value of $\sigma_W$ without doing any numerical simulation.
We have verified this statement using numerical simulations (Figures~\ref{fig:u4-fit-1} and~\ref{fig:u4-fit-2}).
In this paper, we have focused the analysis on the analytical computations in the QFT side: we plan to analyze the equations numerically in future works.

From a function-space perspective, it is natural to understand the learning process as a RG flow induced in a suitable theory space.
It would very interesting to investigate how the notions presented in this paper could generalize to describe this process and how the couplings change under learning.

\section*{Acknowledgments}

We are very grateful for useful discussions with James Halverson, Anindita Maiti, and Keegan Stoner.
We would like to thank particularly Anindita Maiti for help in reproducing the numerical results from~\cite{Halverson:2021:NeuralNetworksQuantum}.
This project has received funding from the European Union's Horizon 2020 research and innovation program under the Marie Skłodowska-Curie grant agreement No 891169.
This work is also supported by the National Science Foundation under Cooperative Agreement PHY-2019786 (The NSF AI Institute for Artificial Intelligence and Fundamental Interactions, \url{http://iaifi.org/}).

\appendix

\section{Numerical simulations}
\label{app:NumSim}

In this appendix, we explain how to compute numerically correlation functions for neural networks defined in Section~\ref{sec:nn-exp} and how to extract relevant information.
In particular, we reproduce the numerical results from~\cite{Halverson:2021:NeuralNetworksQuantum} and provide additional details.
The code is written in Python and is available at \url{https://github.com/melsophos/nnqft}.
Throughout this appendix, we take:
\begin{equation}
	d_{\text{in}} = 1,
	\qquad
	N \in \{ 2, 3, 4, 5, 10, 20, 50, 100, 500, 1000 \}.
\end{equation}

In order to evaluate correlation functions, we consider $n_{\text{nets}}$ neural networks $f_\alpha$.
For each of them, the weights and biases are drawn independently from the distributions $\mathcal N(0, \sigma_W^2 / N)$ and $\mathcal N(0, \sigma_b)$, where $N$ is the width of the hidden layer.
We will take~\cite{Halverson:2021:NeuralNetworksQuantum}:
\begin{equation}
	\sigma_b
		= 1.
\end{equation}
Then, the experimental $n$-point correlation functions are computed as \eqref{eq:exp-green-n-points}:
\begin{equation}
	G^{(n)}_{\text{exp}}(x_1, \ldots, x_n)
		:= \frac{1}{n_{\text{nets}}}
			\sum_{\alpha=1}^{n_{\text{nets}}} f_\alpha(x_1) \cdots f_\alpha(x_n).
\end{equation}
We define the difference with the large $N$ Green functions $G^{(n)}_0$ as (see Section~\ref{sec:finite-N}):
\begin{equation}
	\Delta G^{(n)}_{\text{exp}}(x_1, \ldots, x_n)
		:= G^{(n)}_{\text{exp}}(x_1, \ldots, x_n) - G_0^{(n)}(x_1, \ldots, x_n),
\end{equation}
and the normalized $n$-point functions as:
\begin{equation}
	\label{eq:norm-deviations}
	m_n(x_1, \ldots, x_n)
		:= \frac{\Delta G^{(n)}_{\text{exp}}(x_1, \ldots, x_n)}{G_0^{(n)}(x_1, \ldots, x_n)}.
\end{equation}
Note that no absolute value has been taken until now, and the result can be positive or negative.
The large $N$ Green functions are computed with Wick theorem from the Gauss-net kernel \eqref{eq:kernel}.
For example, the $4$-point function is given by:
\begin{equation}
	G_0^{(4)}(x_1, x_2, x_3, x_4)
		= K(x_1, x_2) K(x_3, x_4)
			+ K(x_1, x_3) K(x_2, x_4)
			+ K(x_1, x_4) K(x_2, x_3).
\end{equation}

In order to reduce variance of the results, we will compute the Green functions by averaging over $n_{\text{bags}}$, each made of $n_{\text{nets}}$ networks:
\begin{equation}
	G^{(n)}_{\text{exp}}(x_1, \ldots, x_n)
		\to \frac{1}{n_{\text{bags}}}
			\sum_{A=1}^{n_{\text{bags}}} G^{(n)}_{\text{exp}}(x_1, \ldots, x_n)|_A\,,
\end{equation}
where $G^{(n)}_{\text{exp}}(x_1, \ldots, x_n)|_A$ means that the correlation functions is computed with the bag $A$.
This also allows extracting standard deviations if needed.

We will compute the correlation functions for the following points~\cite{Halverson:2021:NeuralNetworksQuantum}:
\begin{equation}
	\label{eq:points-exp}
	(x^{(1)}, \ldots, x^{(6)})
		= (-0.01, -0.006, -0.002, +0.002, +0.006, +0.01).
\end{equation}
Given a $n$-point correlation function, we compute it for all the possible combinations of $n$ points $x^{(i)}$ with $i = 1, \ldots, 6$, including identical entries.
Since the experimental Green functions are symmetric by construction (as are the QFT Green functions), we consider only combinations which are inequivalent up to permutations.
For example, we will compute the following $2$-point functions:
\begin{equation}
	\begin{gathered}
		G^{(2)}_{\text{exp}}(x^{(1)}, x^{(1)}),
		\quad
		G^{(2)}_{\text{exp}}(x^{(1)}, x^{(2)}),
		\quad
		\ldots,
		G^{(2)}_{\text{exp}}(x^{(1)}, x^{(6)}),
		\quad
		G^{(2)}_{\text{exp}}(x^{(2)}, x^{(2)}),
		\quad
		\\
		G^{(2)}_{\text{exp}}(x^{(2)}, x^{(3)}),
		\quad
		\ldots,
		\quad
		G^{(2)}_{\text{exp}}(x^{(6)}, x^{(6)})\,.
	\end{gathered}
\end{equation}
For $n = 2, 4, 6$, there are respectively $n_{\text{comb}} = 21, 126, 462$ inequivalent combinations.
We denote by $\langle \cdot \rangle$ the average of a quantity over all possible combinations of points, and by $\langle |\cdot| \rangle$ the average of the absolute value.\footnote{We use the same notation as the average over the networks. However, there is no ambiguity since the latter is used in this section only to compute Green functions and never appears after.}

The numerical Green functions are exact Green functions because they contain already all quantum corrections from loop diagrams.
Hence, it is more natural to write a 1PI effective field theory and determine the coefficients by matching the Green functions computed from 1PI Feynman diagrams.
Moreover, the Wetterich formalism from Sections~\ref{sec3} and~\ref{sec4} gives relations for the 1PI couplings.
We consider the following 1PI interactions to describe the neural network:
\begin{equation}
	\label{eq:Sint}
	\Gamma
		= S_{\text{kin}} + \Gamma_{\text{int}},
	\qquad
	\Gamma_{\text{int}}
		= \frac{u_4}{4!} \int d x \, \phi(x)^4
			+ \frac{u_6}{6!} \int d x \, \phi(x)^6,
\end{equation}
where $S_{\text{kin}}$ is the large $N$ free action \eqref{eq:free-action}.
We consider a local Lagrangian because it turns out that it reproduces well the experimental Green functions for the points considered previously~\cite{Halverson:2021:NeuralNetworksQuantum}.
In the notations of~\cite{Halverson:2021:NeuralNetworksQuantum}, we have $u_4 = 4! \, \lambda$ and $u_6 = 6! \, \kappa$.
However, the interpretation is slightly different compared to~\cite{Halverson:2021:NeuralNetworksQuantum} which writes a microscopic action.
The interactions are associated with the part of the kinetic operator $\Xi_W$ corresponding to the weight only, since the bias part is always Gaussian and independent of $N$~\cite{Halverson:2021:NeuralNetworksQuantum}.
Hence, propagators attached to vertices are $K_W$ instead of $K$: the latter appear only in the disconnected $2$-point propagators.

\begin{figure}[htp]
	\centering
	\subcaptionbox{$n = 2$}[.45\linewidth]{%
		\includegraphics[width=\linewidth]{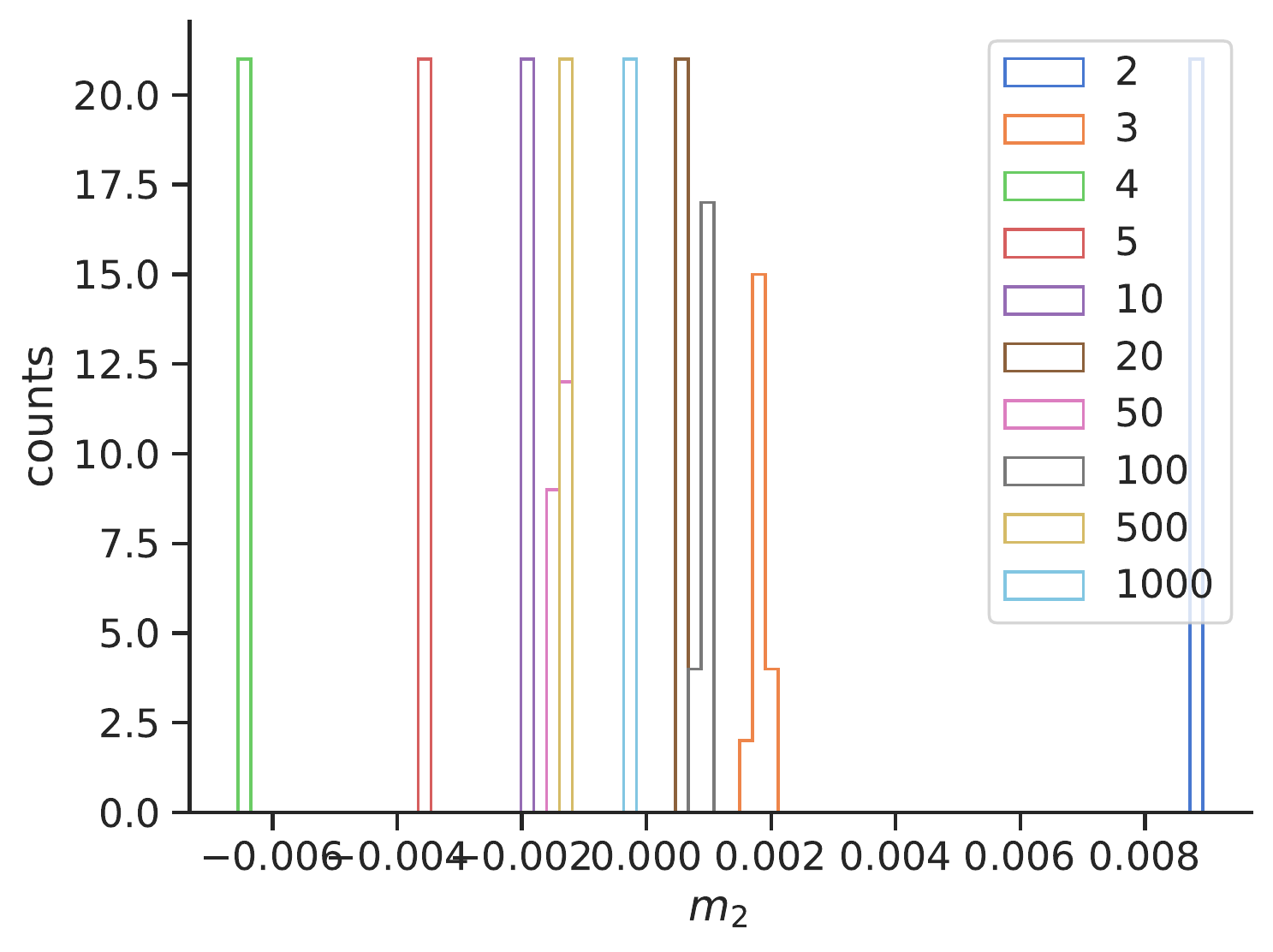}
	}
	\qquad
	\subcaptionbox{$n = 4$}[.45\linewidth]{%
		\includegraphics[width=\linewidth]{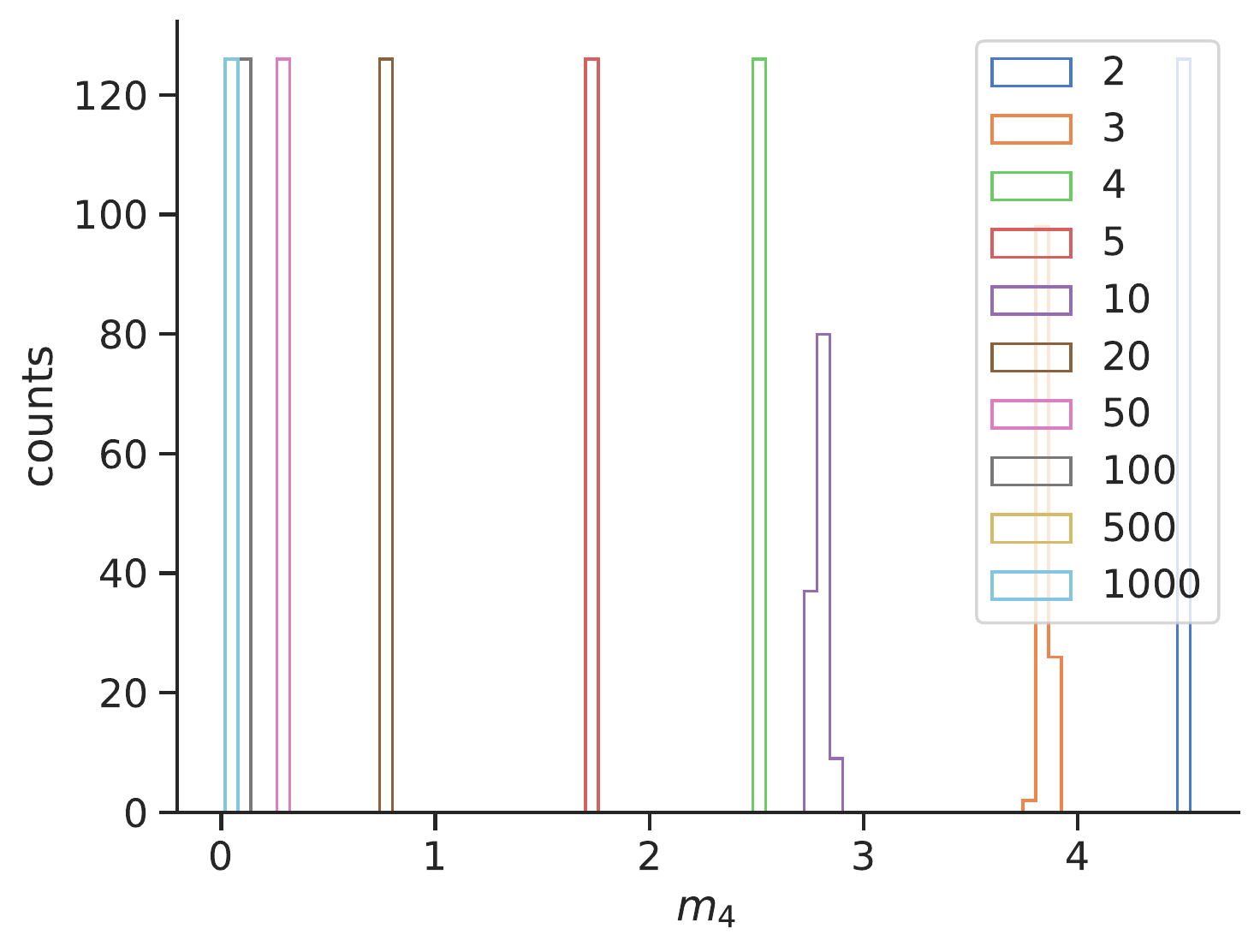}
	}

	\bigskip

	\subcaptionbox{$n = 6$}[.45\linewidth]{%
		\includegraphics[width=\linewidth]{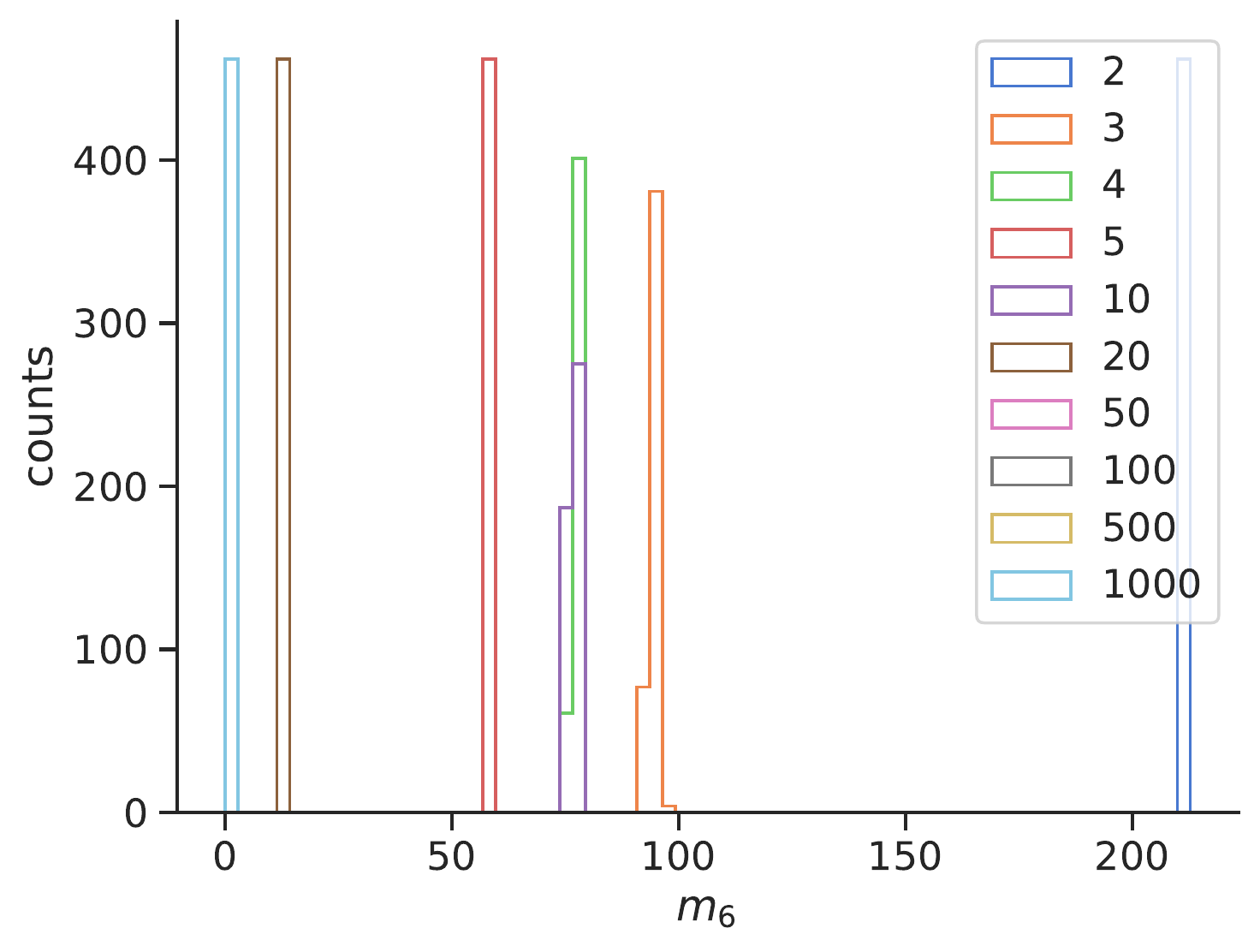}
	}

	\caption{Histograms of the normalized deviations $m_n$ for $n = 2, 4, 6$ for all combinations of points \eqref{eq:points-exp}.}
	\label{fig:mn-hist}
\end{figure}

We now turn our attention to the computation of the experimental Green functions.
We take:
\begin{equation}
	\sigma_W
		= 1,
	\qquad
	n_{\text{bags}}
		= 20,
	\qquad
	n_{\text{nets}}
		= 30000.
\end{equation}
Since we know the exact $2$-point function $G_2 = K$, we must have:
\begin{equation}
	G^{(2)}_{\text{exp}}(x, y)
		\approx K(x, y)
	\quad \Longrightarrow \quad
	m_2(x, y)
		\approx 0\,,
	\qquad
	\forall N.
\end{equation}
Similarly, we know from \eqref{eq:decay-N-Gn} that higher-order Green functions must decrease as $N$ increases:
\begin{equation}
	m_4 = \mathcal{O}(1/N) \,,
	\qquad
	m_6 = \mathcal{O}(1/N) \,.
\end{equation}
We check that it is indeed the case by plotting the values of $m_2$, $m_4$ and $m_6$ for the different combinations of points \eqref{eq:points-exp}.
The figures~\ref{fig:mn-hist} (not present in~\cite{Halverson:2021:NeuralNetworksQuantum}) show that these values go toward $0$ as $N$ increases for $n = 4, 6$.
On the other hand, the values for $n = 2$ don't have any specific pattern, which is expected, since $G^{(2)}_{\text{exp}}$ should be independent of~$N$.

\begin{figure}[htp]
	\centering

	\subcaptionbox{$m = 2$}[.45\linewidth]{%
		\includegraphics[width=\linewidth]{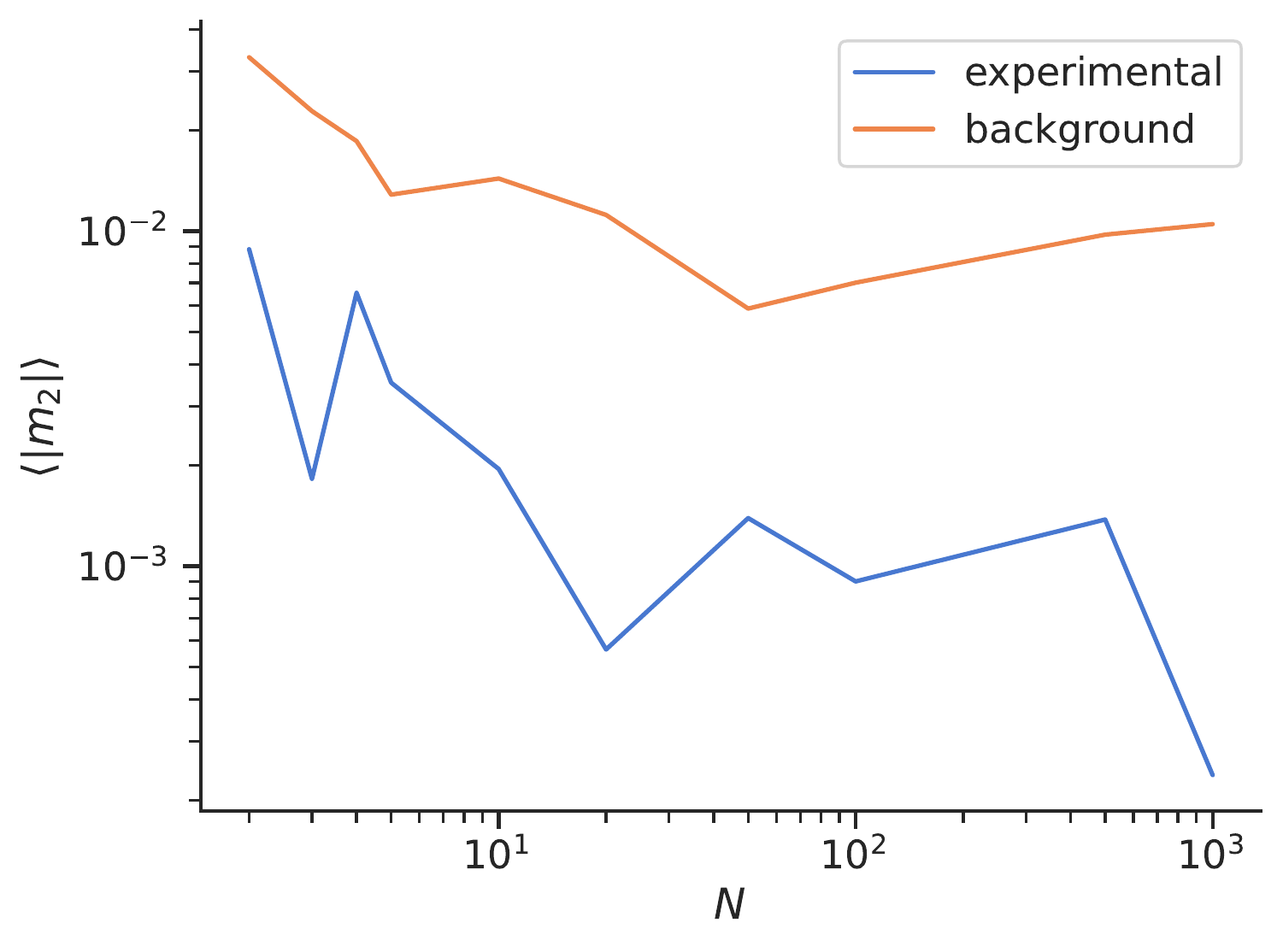}
	}
	\qquad
	\subcaptionbox{$m = 4$}[.45\linewidth]{%
		\includegraphics[width=\linewidth]{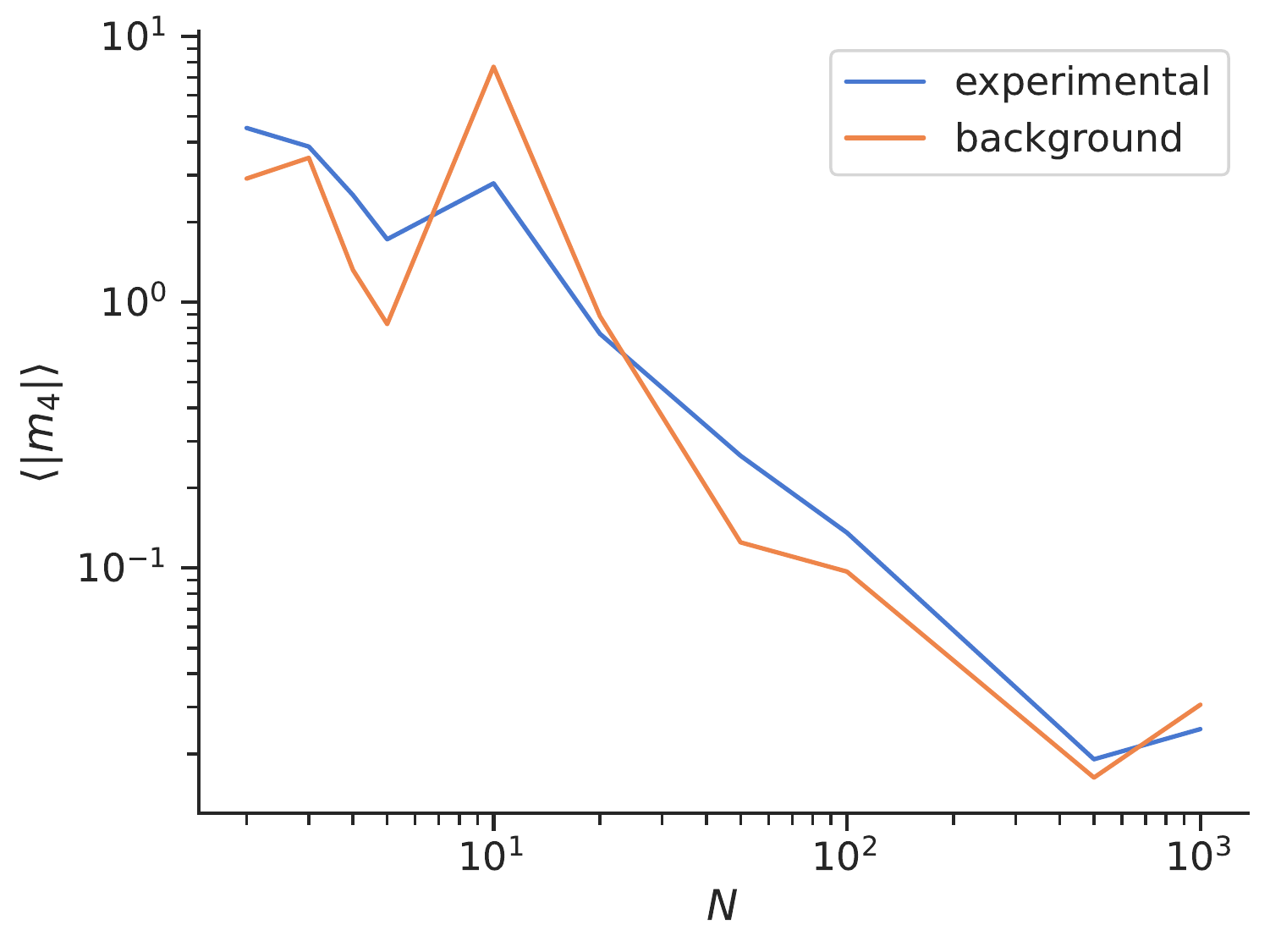}
	}

	\bigskip

	\subcaptionbox{$m = 6$}[.45\linewidth]{%
		\includegraphics[width=\linewidth]{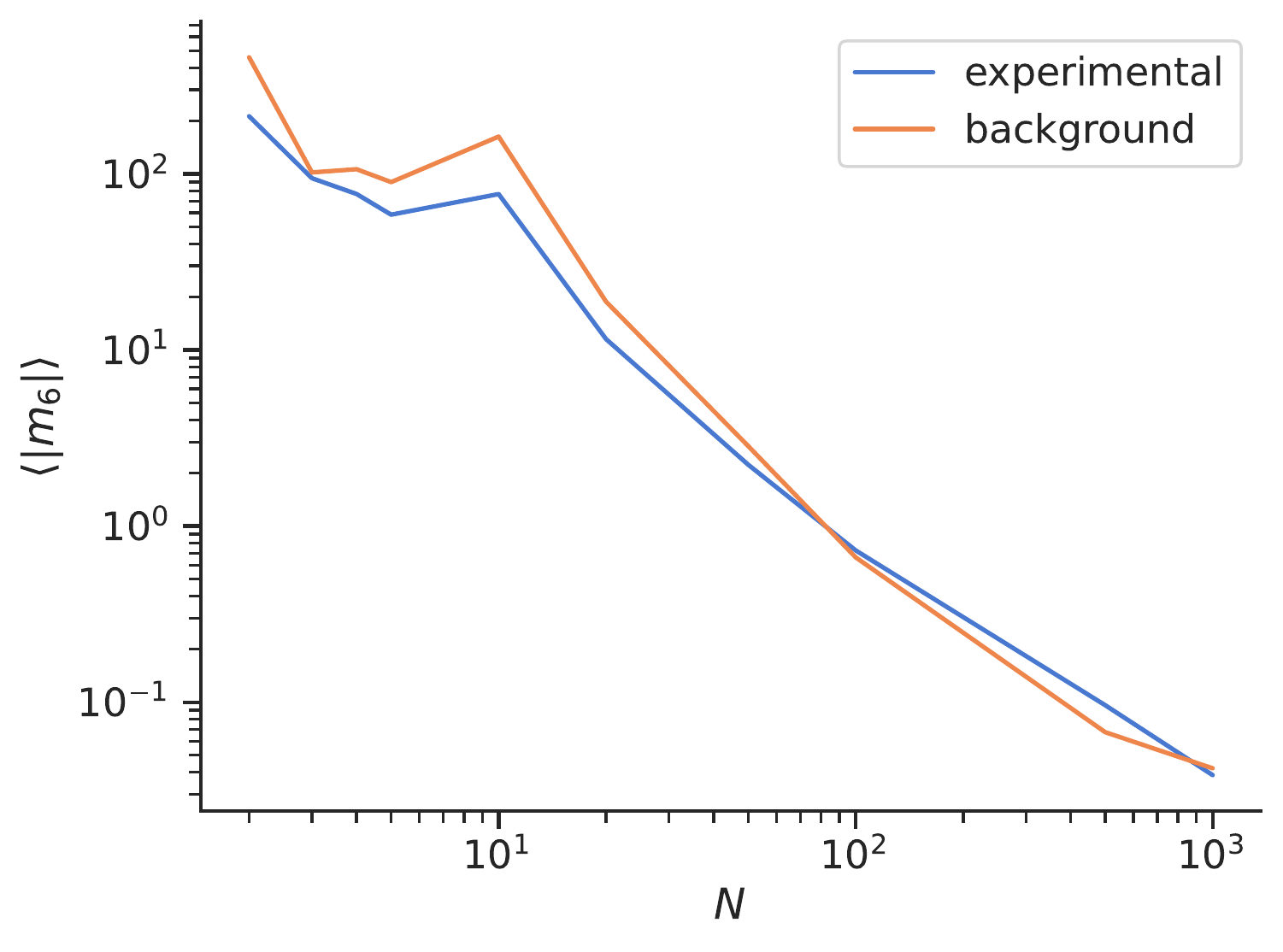}
	}

	\caption{Values of the averaged normalized deviations $\langle |m_n| \rangle$ for $n = 2, 4, 6$.}
	\label{fig:mn-avg}
\end{figure}

We can simplify further this information and extract a single number.
To do this, we take the absolute value of the normalized deviations \eqref{eq:norm-deviations} and average over the different combinations of points \eqref{eq:points-exp} to get $\langle |m_n| \rangle$.
Moreover, to get an idea of how small are the normalized deviations, we define a background as follows: we compute the standard deviation of $m_n$ over all bags of neural networks for all combinations of points \eqref{eq:points-exp}, and then average over the latter.
The idea is to compare the normalized error encoded by $m_n$ with its numerical fluctuations over different bags, represented by the standard variation.
On the figures~\ref{fig:mn-avg}, we reproduce the results from~\cite{Halverson:2021:NeuralNetworksQuantum}: $\langle |m_n| \rangle$ for $n = 4, 6$ is below the background only for small $N$ and for $N = 1000$, and it is always below the background for $n = 2$.
In principle, $\langle |m_n| \rangle$ should always be below the background for higher $N$ (which was not studied in the original paper~\cite{Halverson:2021:NeuralNetworksQuantum} and which we could not reach for computational reasons) so the current test is not very sharp; Figures~\ref{fig:mn-hist} give a cleaner assessment.

Next, we can compute $u_4(x_1, x_2, x_3, x_4)$.
Using Feynman rules, it can be obtained by subtracting the disconnected contributions (equal to $G^{(4)}_0$ and built from the 1PI $2$-point function) from the full $4$-point function to extract the contact interaction
\begin{equation}
	\vcenter{\hbox{\includegraphics[height=3cm]{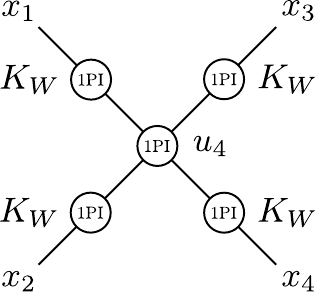}}}
		\quad = \quad
			\vcenter{\hbox{\includegraphics[height=3cm]{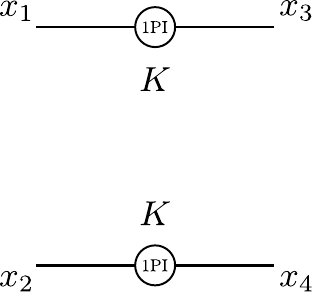}}}
			+ \;\; \text{perms} \;\;
			-
				\vcenter{\hbox{\includegraphics[height=3cm]{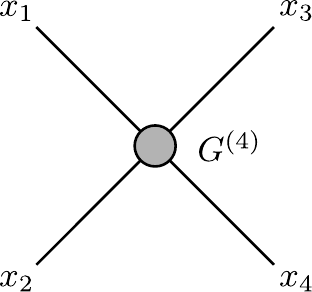}}}
\end{equation}
and truncating the external legs:
\begin{equation}
	\begin{gathered}
	u_4(x_1, x_2, x_3, x_4)
		= - \frac{\Delta G^{(4)}_{\text{exp}}(x_1, x_2, x_3, x_4)}{N_K(x_1, x_2, x_3, x_4)},
	\\\
	N_K(x_1, x_2, x_3, x_4)
		:= \int d x \, K_W(x, x_1) K_W(x, x_2) K_W(x, x_3) K_W(x, x_4),
	\end{gathered}
\end{equation}
where $K_W$ was defined in \eqref{eq:kernel} (see~\cite{Halverson:2021:NeuralNetworksQuantum} for more details).
Importantly, this equation is really an equality and not an approximation as in~\cite{Halverson:2021:NeuralNetworksQuantum}: since we are working with 1PI diagrams, there are no quantum corrections and any $n$-point Green function is built from vertices of order $n' \le n$.
Higher-order vertices $m > n$ appear only in loop diagrams, which are not present.
Hence, this allows determining all 1PI couplings exactly in a recursive way.
Our results agree quantitatively for $u_4$ with those of~\cite{Halverson:2021:NeuralNetworksQuantum} because the loop corrections are subleading in the large $N$ expansion.
However, this may give different results for $u_6$ since the latter receive loop corrections from the microscopic quartic vertex.

\begin{figure}[htp]
	\centering

	\subcaptionbox{$\langle u_4 \rangle$, linear scale.}[.45\linewidth]{%
		\includegraphics[width=\linewidth]{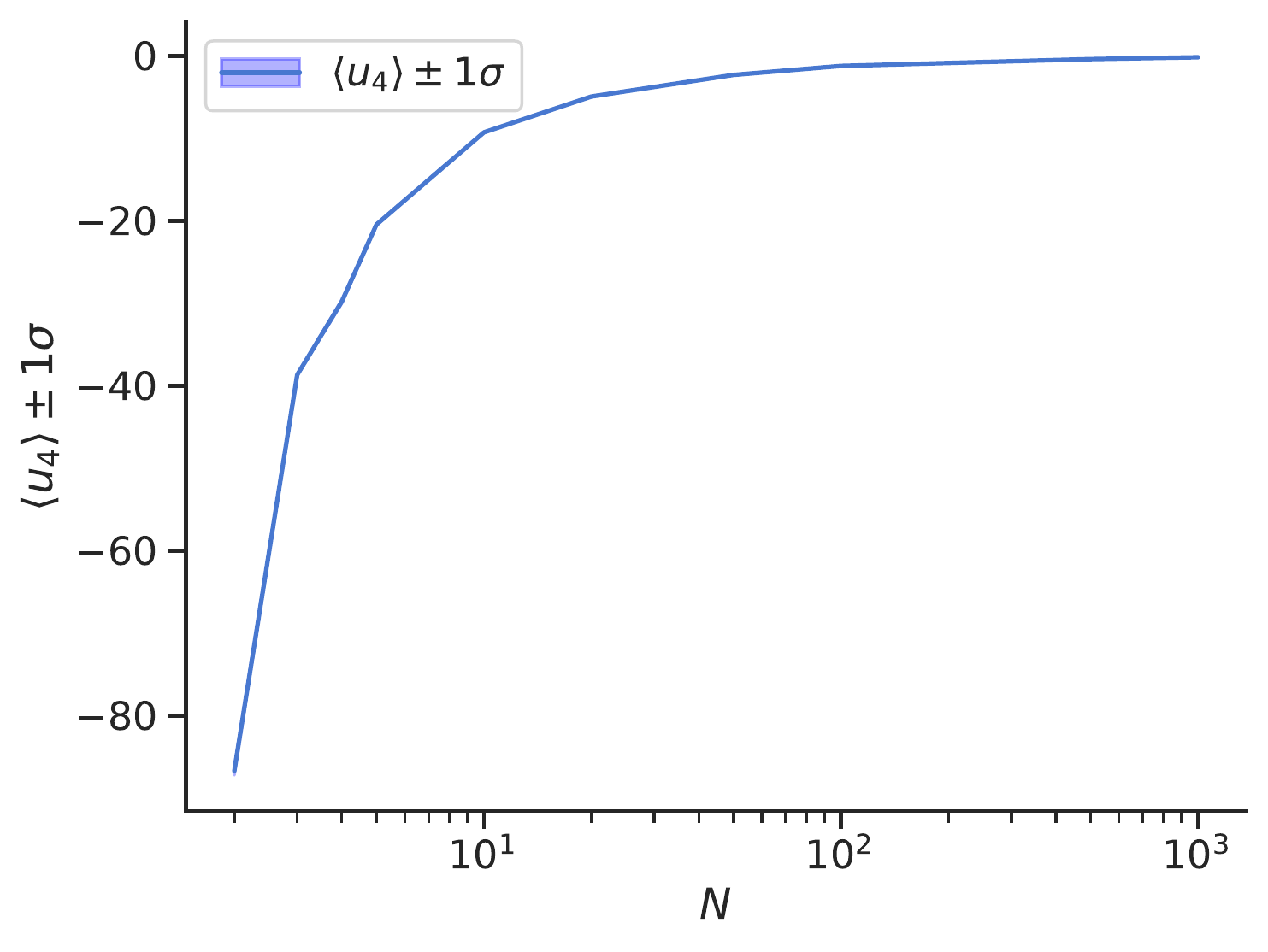}
	}
	\qquad
	\subcaptionbox{$\langle |u_4| \rangle$, log scale.}[.45\linewidth]{%
		\includegraphics[width=\linewidth]{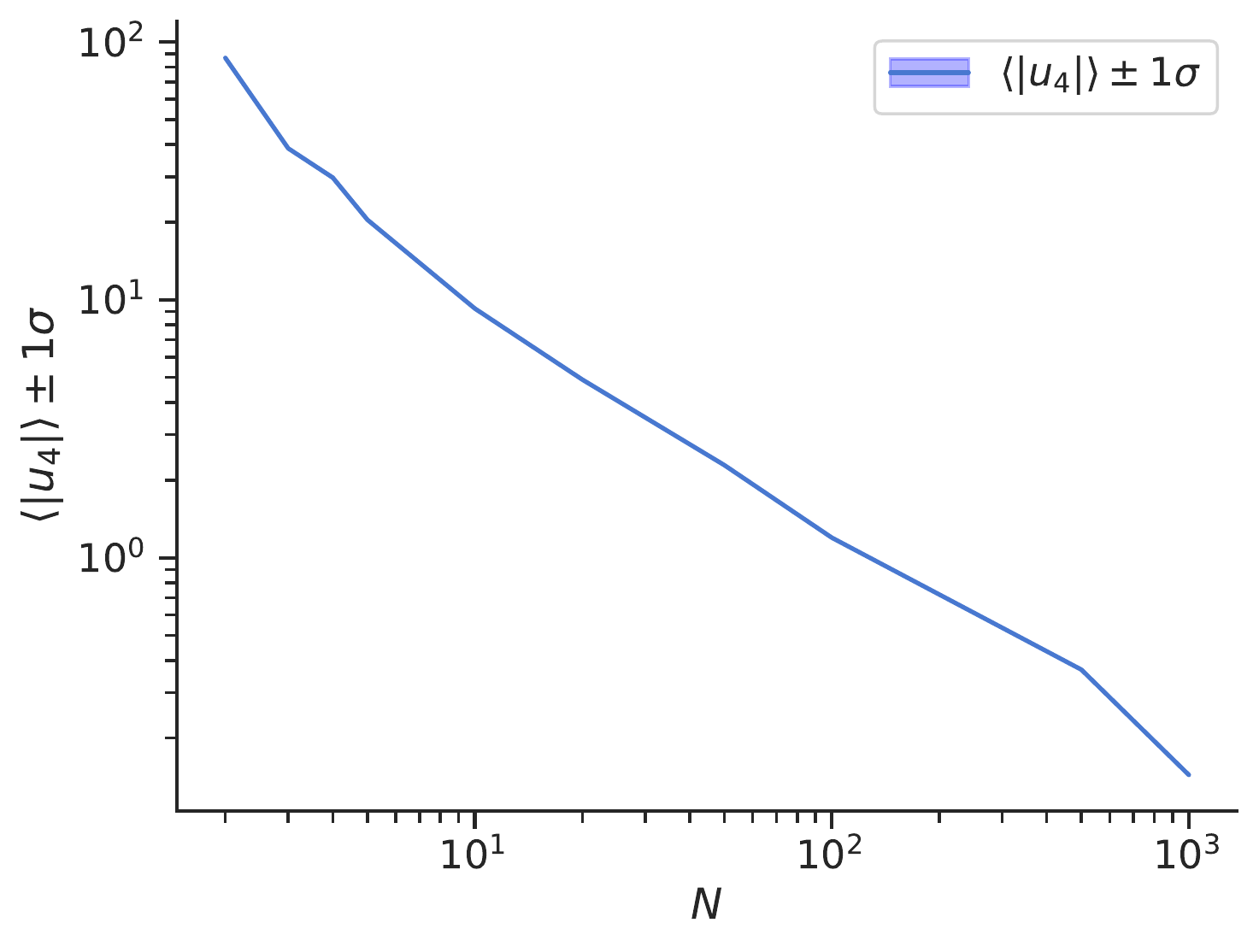}
	}

	\caption{Values of the averaged coupling constant.}
	\label{fig:u4}
\end{figure}

We take:
\begin{equation}
	\sigma_W
		= 1,
	\qquad
	n_{\text{bags}}
		= 30,
	\qquad
	n_{\text{nets}}
		= 30000.
\end{equation}
We find that $u_4$ is constant to a very good precision when evaluated over all combinations of points \eqref{eq:points-exp}.
In Figure~\ref{fig:u4}, we display the values of $u_4$ averaged over all combinations and the corresponding standard deviation and find that its absolute value decreases as $N$ increases, reproducing the results~\cite{Halverson:2021:NeuralNetworksQuantum}.
Importantly, we find that $u_4$ is \emph{negative}, which was not indicated in~\cite{Halverson:2021:NeuralNetworksQuantum} (their figure~4 has an implicit absolute value needed to use the log-scale).
As a consequence, the effective action \eqref{eq:Sint} must include a sixtic contribution, however small, for the path integral to be stable: truncating to quartic interactions as in~\cite{Halverson:2021:NeuralNetworksQuantum} leads to an exponential growth of the weight.
A preliminary analysis of the passive flow equations (Section~\ref{sec3}) indicate that they can be integrated over a large range of $k$ only if the initial conditions satisfy $u_4 < 0$ and $u_6 > 0$, otherwise the flow diverges.

\begin{figure}[htp]
	\centering
	\includegraphics[width=0.7\linewidth]{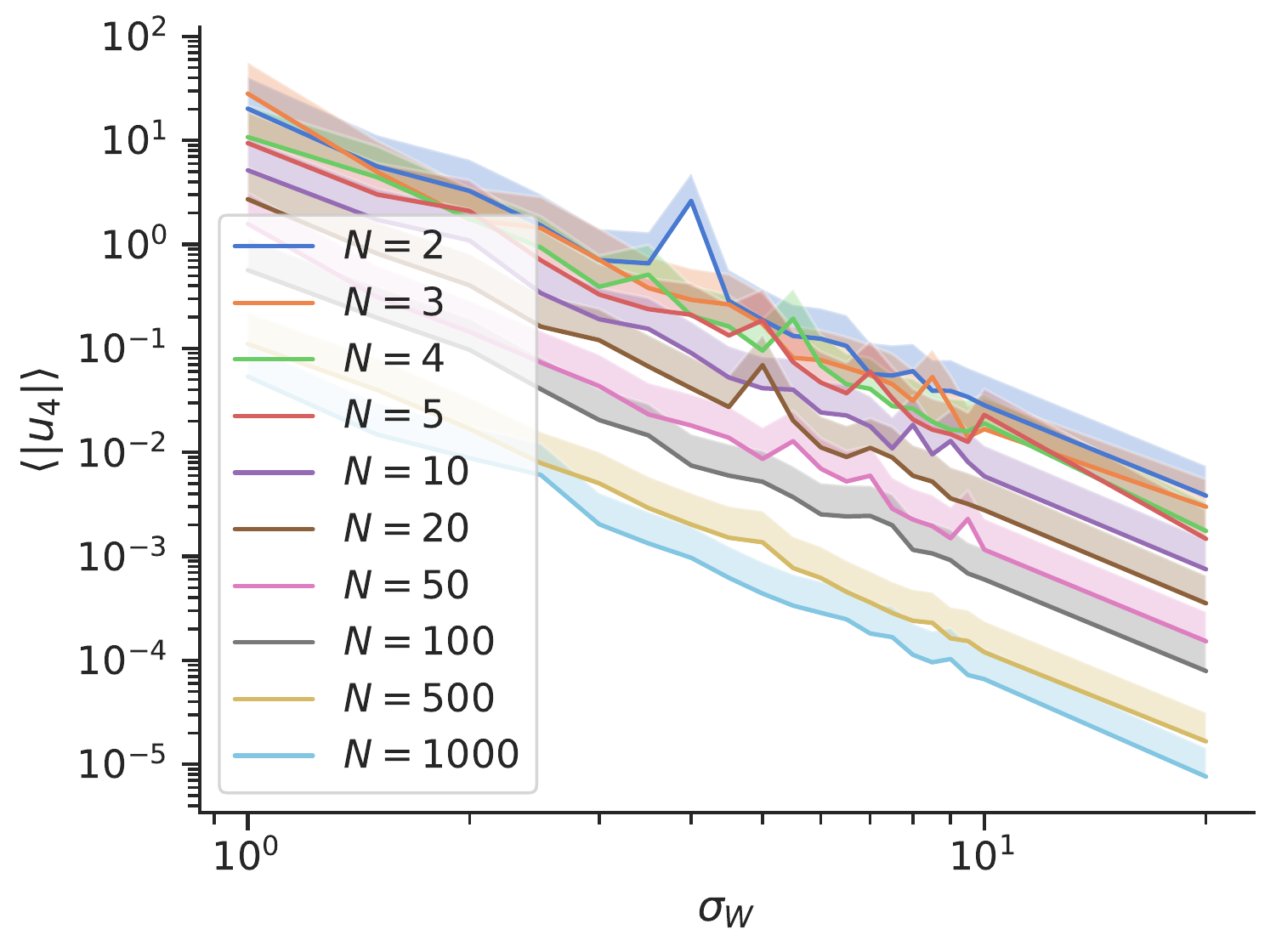}
	\caption{Values of the averaged coupling constant $\langle |u_4| \rangle$ as a function of $\sigma_W$ and $N$. One standard deviation is displayed above the curve.}
	\label{fig:u4-sw-N}
\end{figure}

The final numerical test we perform in this paper is to compute $u_4$ as a function of $N$ and $\sigma_W$ (Figures~\ref{fig:u4-fit-1},~\ref{fig:u4-fit-2} and~\ref{fig:u4-sw-N}).
We consider the following values of $\sigma_W$:
\begin{equation}
	\sigma_W \in \{ 1.0, 1.5, 2.0, \ldots, 9.5, 10, 20 \}.
\end{equation}
We see that $u_4$ decreases as $\sigma_W$ and $N$ increase and that the values are well predicted by the active RG flow equations \eqref{eq:flow-active-sigmaw}.
As such, knowing $u_4$ for a single $\sigma_W$ at fixed $N$ allows computing it for any other $\sigma_W$.

\section{Proofs and technical discussions}
\label{App1}

\subsection{Proof of Proposition \ref{prop1}}

The term involving the field strength $Z$ in the truncation takes the form
\begin{equation}
	\sum_p \, \frac{1}{2}M(-p) (Z[M]p^2) M(p)\,,
\end{equation}
where $Z$ is assumed to depend on $M$. Because we furthermore assume it to be independent of $p^2$, we can define $Z$ operationally as:
\begin{equation}
Z[M=\kappa]\equiv \frac{d}{dp^2}\Gamma^{(2)}_k(p,-p)\bigg\vert_{M=\sqrt{2\kappa}}\,,
\end{equation}
and therefore:
\begin{equation}
\eta_k:= \frac{1}{Z}k\frac{dZ}{dk}=\frac{1}{Z} \frac{d}{dp^2}\dot{\Gamma}^{(2)}_k(p,-p)\,.
\end{equation}
The flow equation for $\Gamma_k^{(2)}$ can be deduced from the Wetterich equation,
\begin{align*}
\dot{\Gamma}^{(2)}_k(p,-p)&=\sum_{q} \dot{r}_k(q^2) \bigg[ G^2(q^2) \Gamma^{(3)}_k(p,0,-p)G((q+p)^2)\Gamma^{(3)}_k(-p,p,0)\\
&\qquad -\frac{1}{2} G(q^2) \Gamma^{(4)}_k(p,-p,0,0)G(q^2) \bigg]\,.
\end{align*}
In the local potential approximation, the vertices are momentum-independent. Therefore, the contribution involving $\Gamma^{(4)}_k$ can be discarded, leading to:
\begin{equation}
\dot{Z}:=( \Gamma^{(3)}_k(0,0,0))^2\bigg\vert_{M=\sqrt{2\kappa}} \frac{d}{dp^2}\sum_{q}\dot{r}_k(q^2) \bigg[ G^2(q^2)G((q+p)^2) \bigg]_{M=\sqrt{2\kappa}, p=0}\,,
\end{equation}
where, according to local potential approximation, we evaluate the RHS over uniform configurations. The derivative is then easy to compute, leading to:
\begin{equation}
\dot{Z}= ( \Gamma^{(3)}_k(0,0,0))^2\bigg\vert_{M=\kappa}  \frac{d}{dp^2} \sum_{q}\dot{r}_k(q^2) G^2(q^2)G((q+p)^2)\bigg\vert_{M=\sqrt{2\kappa}} \,.
\end{equation}
The expression of $ \Gamma^{(3)}_k(0,0,0)$ can be easily obtained by taking the third derivative of the effective potential with respect to $M$:
\begin{equation}
 \Gamma^{(3)}_k(0,0,0)= 3 u_4 \sqrt{2\kappa} + u_6 (2\kappa)^{3/2}\,.
\end{equation}
Note that the renormalized vertex $\bar{\Gamma^{(3)}}_k(0,0,0)$ has to be defined as (the factor $Z$ will be explained below):
\begin{equation}
 \Gamma^{(3)}_k(0,0,0)=Z^{3/2}(k)^{3-d_{\text{in}}/2}\bar{\Gamma}^{(3)}_k(0,0,0)\,. \label{renGamma3}
\end{equation}
Now, we have to compute integrals like:
\begin{equation}
I_n(k,p)=\int \frac{dq}{(2\pi)^{d_{\text{in}}}} (q^2)^n \frac{\theta(k^2-q^2)}{Z(p+q)^2+Z(k^2-(p+q)^2)\theta(k^2-(p+q)^2)+M^2(g,h,\kappa)}\,.
\end{equation}
We focus on small and positive $p$ along axis $1$. The integral decomposes as $I_n(k,p)=I_n^{(+)}(k,p)+I_n^{(-)}(k,p)$, where:
\begin{equation}
I_n^{(+)}(k,p)=\int_{q_1>0} \frac{dq}{(2\pi)^{d_{\text{in}}}}  (q^2)^n \frac{\theta(k^2-q^2)}{Z(p+q)^2+Z(k^2-(p+q)^2)\theta(k^2-(p+q)^2)+M^2(g,h,\kappa)}\,,
\end{equation}
and
\begin{equation}
I_n^{(-)}(k,p)=\int_{q_1<0} \frac{dq}{(2\pi)^{d_{\text{in}}}} (q^2)^n \frac{\theta(k^2-q^2)}{Z(p+q)^2+Z(k^2-(p+q)^2)\theta(k^2-(p+q)^2)+M^2(g,h,\kappa)}\,.
\end{equation}
Because $p>0$, in the negative branch, $(q_1+p)^2<k^2-\sum_{i=2} q_i^2$, we have:
\begin{equation}
I_n^{(-)}(k,p)=\frac{1}{Zk^2+M^2}\times \int_{q_1<0} \frac{dq}{(2\pi)^{d_{\text{in}}}} (q^2)^n \,,
\end{equation}
which is independent of $p$. In the positive branch, in contrast, we get:
\begin{equation}
\begin{aligned}
	I_n^{(+)}(k,p)
		&
		=
			\frac{1}{Zk^2+M^2}\, \int_0^{\sqrt{k^2-q_\bot^2}-p}  \frac{dq}{(2\pi)^{d_{\text{in}}}} (q^2)^n
			\\ & \qquad
			+ \int_{\sqrt{k^2-q_\bot^2}-p}^{\sqrt{k^2-q_\bot^2}}\frac{dq}{(2\pi)^{d_{\text{in}}}}  \frac{(q^2)^n}{Z(q_1+p)^2+\textbf{M}^2(q_{\bot})}\,,
\end{aligned}
\end{equation}
where the bounds in the integrals refer to the integral over coordinate $1$, and $\textbf{M}^2(q_{\bot}):= M^2+Z q_{\bot}^2$, for $q_\bot:=(q_2,\cdots, q_{d_{\text{in}}})$. Note that we omitted the Heaviside functions. Taking the first derivative with respect to $p$, we get:
\begin{align}
\frac{d}{dp}I_n^{(+)}(k,p)= -2Z \int_{\sqrt{k^2-q_\bot^2}-p}^{\sqrt{k^2-q_\bot^2}} \frac{dq}{(2\pi)^{d_{\text{in}}}} (q^2)^n \frac{(q_1+p)}{(Z(q+p)^2+M^2)^2}\theta(k^2-q^2)\,.
\end{align}
Next, we take the second derivative and we set $p=0$. The contribution coming from derivative of the interior of the integral vanishes, because the remaining integration over $q_\bot$ is empty. Thus, only the variation of the bound contributes; assuming $\theta(0)=1$, we get after a tedious calculation:
\begin{equation}
\frac{1}{2}\frac{d^2}{dp^2} I_n^{(+)}(k,0)= -Z \frac{ (k^2)^{n+d/2}}{(2\pi)^{d_{\text{in}}}} H(d_{\text{in}}) \frac{1}{(Zk^2+M^2)^2}\,,
\end{equation}
where:
\begin{equation}
H(d_{\text{in}}):=\frac{\pi^{\frac{d_{\text{in}}+1}{2}}}{d_{\text{in}}+1}\frac{\Gamma\left(\frac{d_{\text{in}}}{2}\right)}{\Gamma \left(\frac{d_{\text{in}}+1}{2}\right)\Gamma \left(\frac{d_{\text{in}}-1}{2}\right)}
\end{equation}
Therefore, we find:
\begin{equation}
Z\eta_k= \frac{(3 u_4 \sqrt{2\kappa} + u_6 (2\kappa)^{3/2})^2}{(Zk^2+M^2)^2} \left(2Z k^2 I_0^{\prime\prime}(k,0)+Z\eta_k (k^2 I_0^{\prime\prime}(k,0)-I_1^{\prime\prime}(k,0))  \right)\,.
\end{equation}
As for the strict LPA, we introduced dimensionless quantities (and then, explain the origin of the factor $Z$ in front of \eqref{renGamma3}). Now, we have to take into account the wave function renormalization. Recovering $1/2$ in front of the kinetic action requires:
\begin{equation}
u_4=\bar{u}_4 Z^2 k^{4-d_{\text{in}}}\,,\qquad u_6=\bar{u}_6 Z^3 k^{6-2d_{\text{in}}}\,,\qquad \bar{u}_2=\bar{u}_2 Z k^2\,,
\end{equation}
where in this expression $\bar{u}_2 = -2u_4 \kappa$ refers to the effective mass. This relation implies $\kappa=\bar{\kappa} k^{d_{\text{in}}-2} Z^{-1}$. After some simplifications, we get:
\begin{equation}
\eta_k=-H(d_{\text{in}})\frac{(3 \bar{u}_4 \sqrt{2\bar{\kappa}} + \bar{u}_6 (2\bar{\kappa})^{3/2})^2}{(1+\bar{M}^2)^4}\,.
\end{equation}
The explicit expression for $\bar{M}^2$ can be easily derived within the local potential approximation:
\begin{equation}
\bar{M}^2=\bar{U}_k^\prime (\bar\kappa)+ 2\bar{\kappa} \bar{U}_k^{\prime \prime}(\bar{\kappa})=2\bar{\kappa} \bar{u}_4\,.
\end{equation}
Then solving for $\eta_k$, we get:
\begin{equation}
\eta_k=-H(d_{\text{in}})\frac{(3 \bar{u}_4 \sqrt{2\bar{\kappa}} + \bar{u}_6 (2\bar{\kappa})^{3/2})^2}{(1+2\bar{\kappa} \bar{u}_4)^4}\,.
\end{equation}
\begin{flushright}
$\square$
\end{flushright}

\subsection{Discussion about Claim \ref{prop2}}

The regulator is obviously positive definite as soon as $m^2>0$. For $p\in [0,k]$, because $m^2(e^{p^2/m^2}-1)\geq p^2$, we must have $r_{k}(p^2) \leq k^2 (1-p^2/k^2)$, therefore:
\begin{equation}
r_{k\to \Lambda}(p^2\ll \Lambda^2) \lesssim \Lambda^2 \,, \qquad r_{k\to \Lambda}(p^2 \sim \Lambda)\sim 0^+\,.
\end{equation}
Thus, low momenta fluctuations are frozen, decoupling from long distance physics, whereas high momenta modes are unchanged and integrated out. Note that the last condition makes $r_k$ an infrared regulator, which prevents infrared divergences along the flow. Finally, $r_{k\to 0} \to 0^+$, meaning that the original model is formally recovered in the deep infrared limit. All these properties ensure that the boundaries interpolation conditions $\Gamma_{k\to \infty} \to S$ and $\Gamma_{k\to 0} \to \Gamma$ holds using $r_k$. Now, let us show that $r_k$ is optimal in the Litim's sense~\cite{Litim:2001:OptimisedRenormalisationGroup}. The Wetterich equation \eqref{Wett} can be singular if the effective propagator diverges, or equivalently, if its inverse $\Gamma^{(2)}_k+r_k$ vanishes. To avoid this difficulty, $r_k$ has to prevent the existence of zero-modes. In other words, the RG requires the existence of a “gap”, and Litim's criterion for optimization is to maximize the gap. This can be done from the observation that only the field-independent part of the inverse $2$-point function is relevant to discuss optimization, i.e.~$F(p^2):=p^2+Z^{-1}_k r_k(p^2)$, in view to establish a (weakly) model-independent criterion. It is suitable to introduce $y=p^2/k^2$. The optimal value for the gap, $C_0$, is:
\begin{equation}
C_0:= \max_{r_k} \min_{y\geq 0} k^2 F(y)\,.
\end{equation}
By construction, the regulator is expected to be efficient for $p^2\sim k^2$, and we fix the normalization such that $F(y=\alpha)= 1$ for some $\alpha \in ]0,1[$. For a large enough family of regulators, this condition imposes that $C_0\leq 1$. If $F(y)$ reaches its absolute minimum for $y=0$, the regulator cannot be optimal from definition. Thus we may have $F(0) \geq C_0$. Without loss of generality we may choose $F(0)\geq 1$. For a regulator which attributes the same size to the IR fluctuations $p^2<k^2$, this reduces to an equality; this is the case for the regulator \eqref{regulatorsuper}. Indeed, in that case $F(y)=\frac{m^2}{k^2}e^{k^2/m^2}$ for $y\leq 1$, $F(y)=\frac{m^2}{k^2} e^{y k^2/m^2}$ for $y\geq 1$: the previous argument holds, up to the normalization factor $\frac{m^2}{k^2}e^{k^2/m^2}$.

\subsection{Proof of proposition \ref{prop3}}

Because we focus on the symmetric phase, odd effective vertices have to vanish identically $\Gamma_k^{(n)}=0$ as $n=2p+1$. Moreover, the effective propagator $G_k(p,p^\prime)$ has to be diagonal: $G_k^{-1}(p,p^\prime)=(g_k(p^2)+r_k(p^2))\delta(p+p^\prime)$. From our ansatz, $g_k(p)$ is given by:
\begin{equation}
g_k(p^2)=m^2(k) e^{p^2/m^2}= m^2+p^2+\mathfrak{K}(p^2)\,,
\end{equation}
where:
\begin{equation}
\mathfrak{K}(p^2):=p^2\sum_{n=1}^{\infty}\,\frac{1}{n!} \frac{(p^2)^n}{(m^2)^{n}}\,. \label{defkernel}
\end{equation}
Taking the second derivative of the exact RG equation \eqref{Wett}, we get for $\dot{g}_k$:
\begin{equation}
\dot{g}_{k}(p^2)= -\frac{1}{2} \int \frac{dq}{(2\pi)^{d_{\text{in}}}}\,\dot{r}_k(q^2)\, (g_k(q^2)+r_k(q^2))^{-2} \Gamma_k^{(4)}(p, -p,q,-q)\,.
\end{equation}
Because the windows of momenta allowed by the function $\dot{r}_k(q^2)$ is limited to the region $q^2<k^2$ by construction, the symmetric function $\Gamma_k^{(4)}(p, -p,q,-q)=:f_k(p,q)$ can be expanded in powers of $q/k$. At leading order, setting $q=0$ and taking into account the definition \ref{prop2}, the equation simplifies as:
\begin{equation}
\dot{g}_{k}(p^2)= -\frac{1}{2} \frac{e^{-2k^2/m^2}}{(m^2)^2}f_k(p,0)  \int \frac{dq}{(2\pi)^{d_{\text{in}}}}\,\dot{r}_k(q^2)\,.\label{flowmass2}
\end{equation}
The derivative of the regulator can be computed from the definition \ref{prop2} as well. We get, for $p^2<k^2$:\footnote{From its definition, the regulator vanishes for $p^2>k^2$, and it is also the case for its derivative.}
\begin{equation}
\dot{r}_k(p^2)= 2k^2 e^{k^2/m^2}+\dot{m}^2 \left[ e^{k^2/m^2}\left(1-\frac{k^2}{m^2}\right)-\left(1-\frac{p^2}{m^2}\right)e^{p^2/m^2} \right]\,.\label{rpoint}
\end{equation}
In the RG transformation, a rescaling of the lattice is required after partial integration of degrees of freedom to ensure preservation of the IR physics. This is equivalent to assuming the existence of a proper rescaling of the coupling, turning the flow equation in an autonomous system. From the equation above, we see in particular that the mass has to be rescaled as:\footnote{See also the discussion at the beginning of the Section \ref{secUV}.}
\begin{equation}
m^2(k)=k^2\bar{u}_2(k)\,,
\end{equation}
and the rescaling for the couplings $f(p,0)$ and $g_k(p^2)$  follows:
\begin{equation}
f_k(p,0)=k^{4-d_{\text{in}}} \bar{f}_k(p/k,0)\,,\qquad g_k(p^2)=k^2\bar{g}_k(p^2/k^2)\,,
\end{equation}
with the conditions:
\begin{equation}
\bar{f}_k(0,0)\equiv\bar{u}_4\,,\quad \bar{g}_k(0)\equiv \bar{u}_2\,,
\end{equation}
respectively defining the local $4$-point coupling and effective mass. Within these dimensionless couplings, equation \eqref{rpoint} becomes:
\begin{equation}
\dot{r}_k(p^2)= 2k^2 e^{1/\bar{u}_2}+k^2(\beta_2+2\bar{u}_2) \left[ e^{1/\bar{u}_2}\left(1-\frac{1}{\bar{u}_2}\right)-\left(1-\frac{p^2/k^2}{\bar{u}_2}\right)e^{p^2/k^2\bar{u}_2} \right]\,,
\end{equation}
where $\beta_{2n}:=\dot{\bar{u}}_{2n}$. Within this approximation, and introducing $x:=p/k$, the flow equation for $\dot{g}_k(p^2)$ takes the form:
\begin{equation*}
\frac{\dot{{g}}_k(x^2)}{k^2}=-\frac{e^{-1/\bar{u}_2}}{\bar{u}_2^2} \bar{f}_k(x,0)\int \frac{dy}{(2\pi)^{d_{\text{in}}}} \left[1+\frac{\beta_2+2\bar{u}_2}{2}\left(\left(1-\frac{1}{\bar{u}_2}\right)-\left(1-\frac{y^2}{\bar{u}_2}\right)e^{\frac{y^2-1}{\bar{u}_2}} \right)  \right]\,,
\end{equation*}
the integral being restricted in the interior of the sphere $y^2<1$. Thus, defining:
\begin{equation}
F_n(\bar{u}_2):=  d_{\text{in}}\int_{0}^{1} r^{d_{\text{in}}+2n-1} e^{\frac{r^2-1}{\bar{u}_2}}dr\,,
\end{equation}
the previous equation reads:
\begin{equation}
\frac{\dot{{g}}_k(x^2)}{k^2}=\frac{e^{-1/\bar{u}_2}}{\bar{u}_2^2} \bar{f}_k(x,0) K_{d_{\text{in}}} \left[\frac{\beta_2+2\bar{u}_2}{2}\left(F_0-\frac{F_1}{\bar{u}_2}\right)-\left(1+  \frac{\beta_2+2\bar{u}_2}{2} \right)\left(1-\frac{1}{\bar{u}_2}\right)\right]\,.\label{equationg}
\end{equation}
From this equation, we easily deduce that $\bar{f}_k(x,0)$ must be a function of $x^2$. Setting $x=0$ on both sides, we get an algebraic closed equation for $\beta_2$:
\begin{equation}
\beta_2+2\bar{u}_2=- K_{d_{\text{in}}} \frac{\bar{u}_4 e^{-1/\bar{u}_2}}{\bar{u}_2^2} \left[\left(1+  \frac{\beta_2+2\bar{u}_2}{2} \right)\left(1-\frac{1}{\bar{u}_2}\right)-\frac{\beta_2+2\bar{u}_2}{2}\left(F_0-\frac{F_1}{\bar{u}_2}\right)\right]\,.\label{equationsuper1}
\end{equation}
Solving it, we get:
\begin{equation}
\begin{boxed}{
\beta_2=-2\bar{u}_2+2K_{d_{\text{in}}} \bar{u}_4 \,\frac{(1-\bar{u}_2)^2+\bar{u}_2(\bar{u}_2F_0-F_1)}{2\bar{u}_2^3e^{1/\bar{u}_2}-K_{d_{\text{in}}}\bar{u}_4 \left((1-\bar{u}_2)+(\bar{u}_2F_0-F_1)\right)}}\end{boxed}\,.
\end{equation}
The solution exhibits a singularity, which has to be taken into account when solving the flow equation. Through the definition \eqref{defkernel}, the solution to this equation provides $g_k(p^2)$. Taking into account that, we find from the chain rule:
\begin{equation}
\frac{\dot{{g}}_k(x^2)}{k^2}=2\bar{g}_k(x^2)-2x^2 \bar{g}_k^\prime(x^2)+\dot{\bar{g}}_k(x^2)\,,
\end{equation}
where $\bar{g}_k^\prime := \partial \bar{g}_k(x^2)/\partial x^2$. We thus obtain for $\bar{f}_k(x,0)$:
\begin{equation}
\begin{boxed}{
\bar{f}_k(x,0)= \frac{\bar{u}_2(\bar{u}_2-x^2)(2\bar{u}_2+\beta_2) e^{(1+x^2)/\bar{u}_2}}{K_{d_{\text{in}}} \left[\frac{\beta_2+2\bar{u}_2}{2}\left(F_0-\frac{F_1}{\bar{u}_2}\right)-\left(1+  \frac{\beta_2+2\bar{u}_2}{2} \right)\left(1-\frac{1}{\bar{u}_2}\right)\right]}}\end{boxed}\,.\label{fexp}
\end{equation}
These equations depend on local couplings $\bar{u}_2$ and $\bar{u}_4$. The flow of $\bar{u}_2$ is fixed by the flow equation \eqref{equationsuper1}, but requires the knowledge of $\bar{u}_4$. It can be obtained using standard LPA, equations \eqref{eqlocal4} and \eqref{eqlocal6} for a sixtic truncation, which discard contributions or order $1/N^3$ from the $N$-scaling. We get, for $\bar{u}_4$ and $\bar{u}_6$:
\begin{align}
\beta_4&=-(4-d_{\text{in}})\bar{u}_4-L_k(\bar{u}_2) \left(\frac{\bar{u}_6 e^{-2/\bar{u}_2}}{\bar{u}_2^2}-\frac{6\bar{u}_4^2e^{-3/\bar{u}_2}}{\bar{u}_2^3}\right)\,,\\
\beta_6&=-(6-2d_{\text{in}})\bar{u}_6-L_k(\bar{u}_2) \left(\frac{90 \bar{u}_4^3 e^{-4/\bar{u}_2}}{\bar{u}_2^4} -\frac{30 \bar{u}_6\bar{u}_4e^{-3/\bar{u}_2}}{\bar{u}_2^3}\right)\,,
\end{align}
where:
\begin{equation}
L_k(\bar{u}_2):=e^{1/\bar{u}_2}K_{d_{\text{in}}} \left[\frac{\beta_2+2\bar{u}_2}{2}\left(F_0-\frac{F_1}{\bar{u}_2}\right)-\left(1+  \frac{\beta_2+2\bar{u}_2}{2} \right)\left(1-\frac{1}{\bar{u}_2}\right)\right]\,.
\end{equation}
Finally, from the flow equation for $\Gamma_k^{(4)}(p_1,p_2,p_3,p_4)$ (equation \eqref{eqlocal4}), setting $p_1=-p_2=p$ and $p_3=p_4=0$, we get:
\begin{equation*}
(4-d_{\text{in}}) \bar{f}_k(x,0)- 2x^2 \bar{f}^\prime_k(x,0)+\dot{\bar{f}}_k(x,0)=L_k(\bar{u}_2) \left(\frac{6(\bar{u}_4  \bar{f}_k(x,0)+R_k(x)) e^{-3/\bar{u}_2}}{\bar{u}_2^3}-\frac{\bar{h}_k(x) e^{-2/\bar{u}_2}}{\bar{u}_2^2}\right)\,,
\end{equation*}
where:
\begin{equation}
h_k(p):= \Gamma_k^{(6)}(p,-p,0,0,0,0)\,,\qquad \bar{h}_k(x):=k^{2d_{\text{in}}-6} h_k(p)\,.
\end{equation}
For reader familiar with QFT, the origin of the function $R_k(x)$ can be traced from the $s$-, $t$- and $u$-channels. In fact, the $(\Gamma_k^{(4)})^2$ contributions have the following structure (the intermediate fat dotted edge materializing the effective loop between effective vertices, see equation \eqref{eqlocal4}):
\begin{equation}
\vcenter{\hbox{\includegraphics[scale=1]{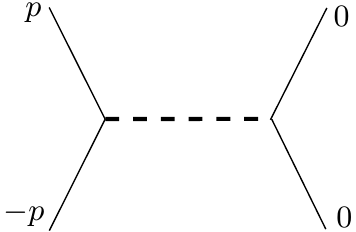} }} +2\,\vcenter{\hbox{\includegraphics[scale=1]{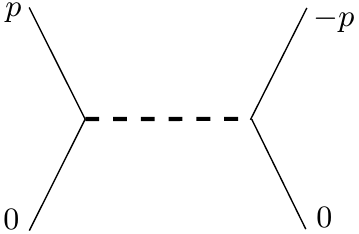} }} \,.
\end{equation}
The first term corresponds to $\bar{u}_4  \bar{f}_k(x,0)$, the second defines $R_k(x)$. A direct inspection shows that $R_k \sim \int dq\, \dot{r}_k(q^2)G_k(-q-p)G_k(q)$, which becomes small for $p$ large enough. We thus obtain the approximation for $\bar{h}_k(x)$ for $x$ large enough:
\begin{equation}\begin{boxed}{
\bar{h}_k(x)\approx \left( \frac{6(\bar{u}_4) e^{-1/\bar{u}_2}}{\bar{u}_2}-\frac{\bar{u}_2^2(4-d_{\text{in}})}{L_k(\bar{u}_2)} e^{2/\bar{u}_2}\right) \bar{f}_k(x,0)+e^{2/\bar{u}_2} \bar{u}_2^2 \frac{2x^2 \bar{f}^\prime_k(x,0)-\dot{\bar{f}}_k(x,0)}{L_k(\bar{u}_2)}\,.}\end{boxed}
\end{equation}
\begin{flushright}
$\square$
\end{flushright}

\subsection{Discussion about Claim \ref{prop4}}

The $4$-point function $\Gamma_k^{(4)}(p_1, p_2,p_3,p_4)$ has to be symmetric under any permutation of the four external momenta $p_1$, $p_2$, $p_3$ and $p_4$. Moreover, because we assume local interactions as building blocks, external momenta have to be conserved: $p_1+p_2+p_3+p_4=0$. Let us assume that $\Gamma_k^{(4)}$ is the analytic continuation with respect to some couplings from a perturbative solution $\Gamma_{k, \text{pert}}^{(4)}$, defined as the formal sum of an asymptotic perturbative series:
\begin{equation}
\Gamma_{k, \text{pert}}^{(4)}(p_1, p_2,p_3,p_4)= \sum_{G\in\mathcal{G}_4}  \left(\prod_{v\in G}  g_{2n(v)}\right) \mathcal{A}_G(p_1, p_2,p_3,p_4)\,,
\end{equation}
where the first sum runs over one-particle irreducible (1PI) Feynman diagrams $\mathcal{G}_4$ having four external points.
The product runs over vertices $v\in G$, $2n(v)$ denoting the number of fields involved in the interaction having coupling constant $g_{2n(v)}$. Finally, $\mathcal{A}_G$ is the Feynman amplitude associated with the graph $G$. Note that all Feynman amplitudes arise with a global Dirac delta $\delta(p_1+p_2+p_3+p_4)$ ensuring momentum conservation. We recall that Feynman diagrams provide a graphical representations of the Wick contractions involved in the perturbative expansion around the Gaussian theory. A typical Feynman graph is a set of vertices and edges, vertices corresponding to interactions and edges to the Wick contractions between pairs of fields. The momentum dependence of the $4$-point function can be investigated from the structure of Feynman graphs labeling its perturbative expansion. First, we assume that the the theory involves only $4$-points vertices. At one loop, $\Gamma_{k, \text{pert}}^{(4)}(p_1, p_2,p_3,p_4)$ has the following structure:
\begin{equation}
\Gamma_{k, \text{1-loop}}^{(4)}(p_1, p_2,p_3,p_4)=\delta(p_1+p_2+p_3+p_4)\sum_{j=2}^4\gamma_{\text{1-loop}}(p_1+p_j)\,,\label{decomp4pts}
\end{equation}
each term corresponding to the allowed permutations of the external momenta\footnote{The so called $s$-, $t$- and $u$-channels.}. Explicitly, the relevant one-loop diagrams have the following structure:
\begin{equation}
\gamma_{\text{one-loop}}(p_1+p_2)=\vcenter{\hbox{\includegraphics[scale=1.2]{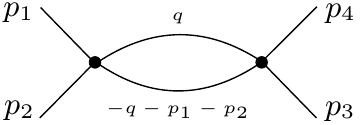} }}\,,
\end{equation}
the loop being proportional to $\int dq \mathcal{K}(q^2)\mathcal{K}((q+p_1+p_2)^2)$. It can be easily checked that the decomposition \eqref{decomp4pts} keeps the same form after including sixtic interactions. Our aim is to prove that such a decomposition remains a suitable approximation for $\Gamma_k^{(4)}$ beyond one-loop. To this end, we will make use of a renormalization group argument, using the explicit expression \eqref{fexp}. Assuming that \eqref{decomp4pts} holds beyond one loop, and there exists a function $\gamma_k(p)$ such that:
\begin{equation}
\Gamma_{k}^{(4)}(p_1, p_2,p_3,p_4)=\delta(p_1+p_2+p_3+p_4)\sum_{j=2}^4\gamma_{k}(p_1+p_j)\,.
\end{equation}
Combining it with the relation \eqref{fexp}, we get:
\begin{equation}
2\gamma_k(p)+\gamma_k(0)=f_k(p)\,,
\end{equation}
and $f_k(0)\equiv u_4= 3\gamma_k(0)$, thus:
\begin{equation}
\gamma_k(p)=\frac{1}{2} f_k(p)-\frac{1}{6}u_4\,.
\end{equation}
The flow equation for $\Gamma_{k}^{(4)}(p_1, p_2,p_3,p_4)$ reads graphically as:
\begin{equation}
\vcenter{\hbox{\includegraphics[scale=0.8]{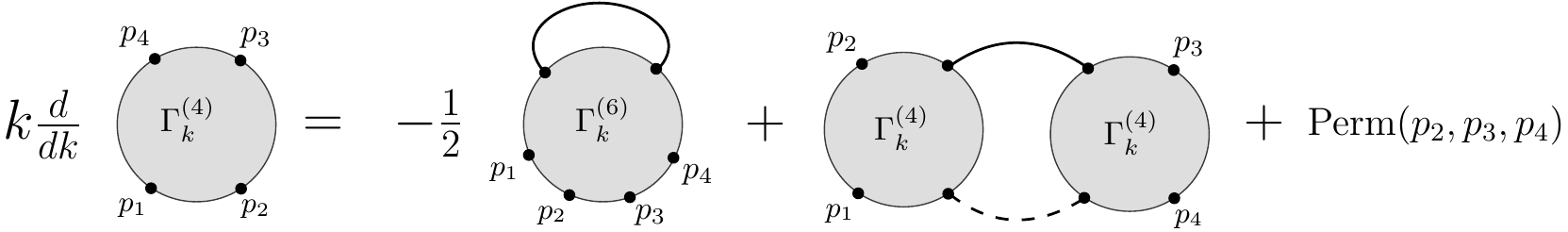} }}\,, \label{eqdiag1}
\end{equation}
the cyclic permutation covering the three pairings $(p_1,p_2)$, $(p_1,p_3)$ and $(p_1,p_4)$, the solid black edge materializing the effective propagator $\dot{r}_k(p^2) \,G_k(p^2)$, whereas the dotted edge corresponds to $G_k(p^2)$. Let us investigate the structure of the $(\Gamma_k^{(4)})^2$ contribution. From our assumption, and neglecting the dependence of the effective vertex on the momentum $q$ running through the effective loop, we get:
\begin{align*}
\vcenter{\hbox{\includegraphics[scale=0.7]{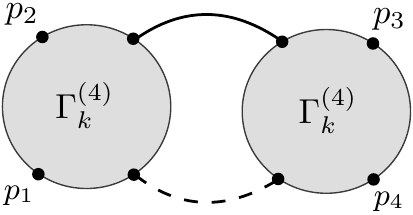} }}
	& \propto (\gamma_k(p_1+p_2)+\gamma_k(p_1)+\gamma_k(p_1))
		\\ & \qquad
		\times (\gamma_k(p_2+p_3)+\gamma_k(p_3)+\gamma_k(p_3)) L_k^{(2)}(p_1+p_2),
\end{align*}
where $L_k^{(2)}(p_1+p_2):=\int dq \dot{r}_k(q^2) G_k((q+p_1+p_2)^2)G_k(q^2)$. For $p_i^2$'s close to the running horizon $\xi^{-2}(k):= k^2 u_2(k)$, $f_k(p)$ becomes small as the explicit expression \eqref{fexp} shows. Thus $\gamma_k(p_i) \sim -u_4/6$,
\begin{equation}
\vcenter{\hbox{\includegraphics[scale=0.7]{images/Graph2} }} \sim (\gamma_k(p_1+p_2)-u_4/3)^2 L_k^{(2)}(p_1+p_2)\,,
\end{equation}
and $(\Gamma_k^{(4)})^2$ contributions ensure stability of the assumption about $\Gamma_k^{(4)}$. To check consistency with $\Gamma_k^{(6)}$, we have to consider the corresponding flow equation:
\begin{equation}
\vcenter{\hbox{\includegraphics[scale=0.8]{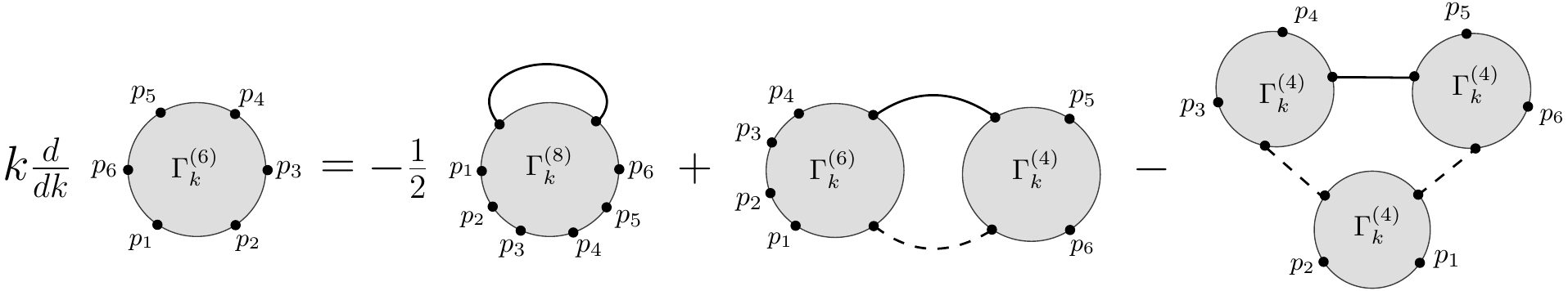} }}\,,
\end{equation}
up to permutation of the external momenta. Assuming we are close to the quartic sector, we focus on the last contribution. Setting $p_5=-p_6=q$, we have two relevant configurations to investigate. The first one is for $p_5$ and $p_6$ hooked to the same vertex. In that case, the loop depends only on two momenta, hooked to another vertex, say $p_1+p_2$ in the following:
\begin{equation}
\vcenter{\hbox{\includegraphics[scale=0.8]{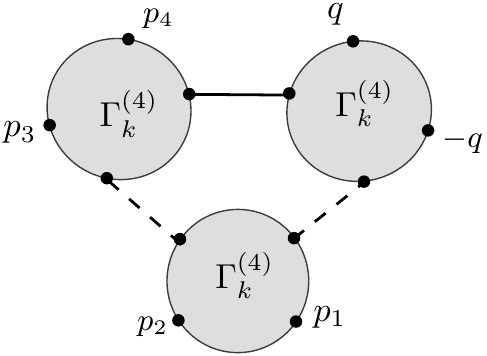}}} \sim (\gamma_k(p_1+p_2)+2\gamma_k(p_1))^2(\gamma_k(0)+2\gamma_k(q)) L_k^{(3)}(p_1+p_2)\,,
\end{equation}
where we discarded the dependence of the effective vertices on the momentum running through the effective loop, and we assumed $\gamma_k(p)=\gamma_k(-p)$. Such a contribution in the first term on the RHS of the equation \eqref{eqdiag1} does not break the ansatz for $\Gamma_k^{(4)}$, the remaining momentum $q$ could be set to zero outside of the tadpole. The second configuration is for $p_5$ and $p_6$ hooked to different effective vertices. It is however easy to check that for large external momenta with respect to the IR cut-off $k$, these contributions are suppressed. For instance, we have:
\begin{align}
\nonumber \vcenter{\hbox{\includegraphics[scale=0.8]{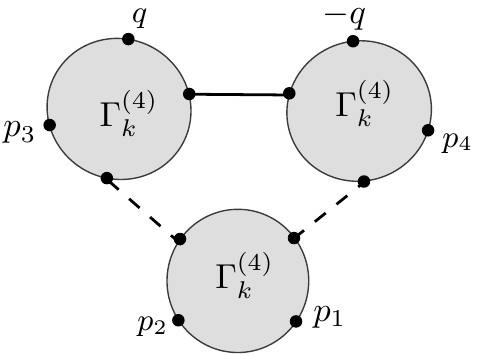}}}& \sim \int dQ \, \dot{r}_k(Q^2) G_k(Q^2)G_k((Q+p_1+p_2)^2)G_k((p_4+q-Q)^2)\,,
\end{align}
and for $p_4^2$ large enough with respect to $k$, this contribution is less relevant than the first in the flow equation for $\Gamma_k^{(4)}$. From the same argument, it is easy to check that the second kind of contributions in the flow of $\Gamma_k^{(6)}$, involving $\Gamma_k^{(6)}$ and $\Gamma_k^{(4)}$ does not break the ansatz for $\Gamma_k^{(4)}$ in the range of momenta that we consider.
\begin{flushright}
$\square$
\end{flushright}

\printbibliography[heading=bibintoc]

\end{document}